\documentclass[11pt, reqno]{amsart}
\usepackage{graphicx,amsmath,amssymb,epsfig}
\usepackage{rotating}
\usepackage{amsfonts}
\usepackage{amsmath}
\usepackage{amssymb}
\usepackage{graphicx}

\theoremstyle{plain}

\newtheorem{theorem}{Theorem}

\newtheorem{corollary}{Corollary}

\newtheorem{algorithm}{Algorithm}

\newtheorem{lemma}{Lemma}

\theoremstyle{definition}
\newtheorem{remark}{Remark}

\numberwithin{equation}{section}
\numberwithin{remark}{section}

\newcommand{\be}{\begin{eqnarray}}
\newcommand{\ee}{\end{eqnarray}}
\newcommand{\ba}{\begin{array}}
\newcommand{\ea}{\end{array}}
\renewcommand{\Pr}{\mathbb{P}}

\numberwithin{theorem}{section}
\numberwithin{lemma}{section}
\numberwithin{corollary}{section}
\input{tcilatex}

\begin{document}
\title[ ]{Inference on Counterfactual Distributions}
\author[ ]{Victor Chernozhukov$^\dag$ \ \ Iv\'an Fern\'andez-Val$^\S $ \ \
Blaise Melly$^\ddag$}
\date{ {\textsc{\today. First ArXiv version: April 6, 2009.} This paper
replaces the earlier independent projects started in 2005 \textquotedblleft
Inference on Counterfactual Distributions Using Conditional Quantile
Models,\textquotedblright\ by Chernozhukov and Fern\'{a}ndez-Val, and
\textquotedblleft Estimation of Counterfactual Distributions Using Quantile
Regression,\textquotedblright\ by Melly. We would like to thank the
co-editors Alberto Abadie, Steve Berry, and James Stock, five anonymous
referees, Isaiah Andrews, Josh Angrist, Manuel Arellano, David Autor,
Alexandre Belloni, Moshe Buchinsky, Arun Chandrasekhar, Mingli Chen, Denis
Chetverikov, Flavio Cunha, Brigham Frandsen, Jerry Hausman, James Heckman,
Michael Jansson, Kengo Kato, Roger Koenker, Joonhwan Lee, Ye Luo,
Pierre-Andre Maugis, Justin McCrary, Miikka Rokkanen, and seminar
participants at Banff International Research Station Conference on
Semiparametric and Nonparametric Methods in Econometrics, Berkeley, Boston
University, CEMFI, Columbia, Harvard/MIT, Michigan, MIT, Ohio State, St.
Gallen, and 2008 Winter Econometric Society Meetings for very useful
comments that helped improve the paper. Companion software developed by the
authors (counterfactual packages for Stata and R) is available from the
authors' web sites. We gratefully acknowledge research support from the
National Science Foundation.}}
\maketitle

\begin{abstract}
{\footnotesize Counterfactual distributions are important ingredients for
policy analysis and decomposition analysis in empirical economics. In this
article we develop modeling and inference tools for counterfactual
distributions based on regression methods. The counterfactual scenarios that
we consider consist of ceteris paribus changes in either the distribution of
covariates related to the outcome of interest or the conditional
distribution of the outcome given covariates. For either of these scenarios
we derive joint functional central limit theorems and bootstrap validity
results for regression-based estimators of the status quo and counterfactual
outcome distributions. These results allow us to construct simultaneous
confidence sets for function-valued effects of the counterfactual changes,
including the effects on the entire distribution and quantile functions of
the outcome as well as on related functionals. These confidence sets can be
used to test functional hypotheses such as no-effect, positive effect, or
stochastic dominance. Our theory applies to general counterfactual changes
and covers the main regression methods including classical, quantile,
duration, and distribution regressions. We illustrate the results with an
empirical application to wage decompositions using data for the United
States. }

{\footnotesize As a part of developing the main results, we introduce
distribution regression as a comprehensive and flexible tool for modeling
and estimating the \textit{entire} conditional distribution. We show that
distribution regression encompasses the Cox duration regression and
represents a useful alternative to quantile regression. We establish
functional central limit theorems and bootstrap validity results for the
empirical distribution regression process and various related functionals. 
\newline
}


{\footnotesize 
}
\end{abstract}

\thispagestyle{empty}

\newpage

\pagestyle{plain}

\section{Introduction}

\setcounter{page}{1}

Counterfactual distributions are important ingredients for policy analysis
(Stock, 1989, Heckman and Vytlacil, 2007) and decomposition analysis (e.g.,
Juhn, Murphy, and Pierce, 1993, DiNardo, Fortin, and Lemieux, 1996, Fortin,
Lemieux, and Firpo, 2011) in empirical economics. For example, we might be
interested in predicting the effect of cleaning up a local hazardous waste
site on the marginal distribution of housing prices (Stock, 1991). Or, we
might be interested in decomposing differences in wage distributions between
men and women into a discrimination effect, arising due to pay differences
between men and women with the same characteristics, and a composition
effect, arising due to differences in characteristics between men and women
(Oaxaca, 1973, and Blinder, 1973). In either example, the key policy or
decomposition effects are differences between observed and counterfactual
distributions. Using econometric terminology, we can often think of a
counterfactual distribution as the result of either a change in the
distribution of a set of covariates $X$ that determine the outcome variable
of interest $Y$, or as a change in the relationship of the covariates with
the outcome, i.e. a change in the conditional distribution of $Y$ given $X$.
Counterfactual analysis consists of evaluating the effects of such changes.%
%
%
%

The main objective and contribution of this paper is to provide estimation
and inference procedures for the entire marginal counterfactual distribution
of $Y$ and its functionals based on regression methods. 
Starting from regression estimators of the conditional distribution of the
outcome given covariates and nonparametric estimators of the covariate
distribution, we obtain uniformly consistent and asymptotically Gaussian
estimators for functionals of the status quo and counterfactual marginal
distributions of the outcome. Examples of these functionals include
distribution functions, quantile functions, quantile effects, distribution
effects, Lorenz curves, and Gini coefficients. We then construct confidence
sets 
that take into account the sampling variation coming from the estimation of
the conditional and covariate distributions. These confidence sets are
uniform in the sense that they cover the entire functional with
pre-specified probability and can be used to test functional hypotheses such
as no-effect, positive effect, or stochastic dominance.

Our analysis specifically targets and covers the regression methods for
estimating conditional distributions most commonly used in empirical work,
including classical, quantile, duration/transformation, and distribution
regressions. We consider simple counterfactual scenarios consisting of
marginal changes in the values of a given covariate, as well as more
elaborate counterfactual scenarios consisting of general changes in the
covariate distribution or in the conditional distribution of the outcome
given covariates. For example, the changes in the covariate and conditional
distributions can correspond to known transformations of these distributions
in a population or to the distributions in different populations. This array
of alternatives allows us to answer a wide variety of counterfactual
questions such as the ones mentioned above.

This paper contains two sets of new theoretical results on counterfactual
analysis. First, we establish the validity of the estimation and inference
procedures under two high-level conditions. The first condition requires the
first stage estimators of the conditional and covariate distributions to
satisfy a functional central limit theorem. The second condition requires
validity of the bootstrap for estimating the limit laws of the first stage
estimators. Under the first condition, we derive functional central limit
theorems for the estimators of the counterfactual functionals of interest,
taking into account the sampling variation coming from the first stage.
Under both conditions, we show that the bootstrap is valid for estimating
the limit laws of the estimators of the counterfactual functionals. The key
new theoretical ingredient to all these results is the Hadamard
differentiability of the counterfactual operator -- that maps the
conditional distributions and covariate distributions into the marginal
counterfactual distributions -- with respect to its arguments, which we
establish in Lemma \ref{lemma: Hadamard dif with estimated cov}. Given this
key ingredient, the other theoretical results above follow from the functional
delta method. A convenient and important feature of these results is that
they automatically imply estimation and inference validity of any existing
or potential estimation method that obeys the two high-level conditions set
forth above.

The second set of results deals with estimation and inference under
primitive conditions in two leading regression methods. Specifically, we
prove that the high-level conditions -- functional central limit theorem and
validity of bootstrap -- hold for estimators of the conditional distribution
based on quantile and distribution regression. In the process of proving
these results we establish also some auxiliary results, which are of
independent interest. In particular, we derive a functional central limit
theorem and prove the validity of exchangeable bootstrap for the (entire)
empirical coefficient process of distribution regression and related
functionals. We also prove the validity of the exchangeable bootstrap for
the (entire) empirical coefficient process of quantile regression and
related functionals. (These are a consequence of a more general result on
functional delta method for Z-processes that we establish in Appendix \ref%
{appendix:delta_zprocesses}).\footnote{%
Prior work by Hahn (1995, 1997) showed empirical and weighted bootstrap
validity for estimating \textit{pointwise} laws of quantile regression
coefficients; see also Chamberlain and Imbens (2003) and Chen and Pouzo
(2009). An important exception is the independent work by Kline and Santos
(2013) that established validity of weighted bootstrap for the entire
coefficient process of Chamberlain (1994) minimum distance estimator in
models with discrete covariates.} Note that the exchangeable bootstrap
covers the empirical, weighted, subsampling, and $m$ out of $n$ bootstraps
as special cases, which gives much flexibility to the practitioner.

This paper contributes to the previous literature on counterfactual analysis
based on regression methods. Stock (1989) introduced integrated kernel
regression-based estimators to evaluate the mean effect of policy
interventions. Gosling, Machin, and Meghir (2000) and Machado and Mata
(2005) proposed quantile regression-based estimators to evaluate
distributional effects, but provided no econometric theory for these
estimators. We also work with quantile regression-based estimators for
evaluating counterfactual effects, though our estimators differ in some
details, and we establish the limit laws as well as inference theory for our
estimators. We also consider distribution regression-based estimators for
evaluating counterfactual effects, and derive the corresponding limit laws
as well as inference theory for our estimators. Moreover, our main results
are generic and apply to any estimator of the conditional and covariate
distributions that satisfy the conditions mentioned above, including
classical regression (Juhn, Murphy and Pierce, 1993), flexible duration
regression (Donald, Green and Paarsch, 2000), and other potential approaches.

Let us comment on the results for distribution regression separately.
Distribution regression, as defined here, consists of the application of a
continuum of binary regressions to the data. We introduce distribution
regression as a comprehensive tool for modeling and estimating the \textit{%
entire} conditional distribution. This partly builds on, but significantly
differs from Foresi and Peracchi (1995), that proposed to use several binary
regressions as a partial description of the conditional distribution.%
\footnote{%
Foresi and Peracchi (1995) considered a fixed number of binary regressions.
In sharp contrast, we consider a continuum of binary regressions, and show
that the \textit{continuum} provides a coherent and flexible model for the 
\textit{entire} conditional distribution. Moreover, we derive the limit
theory for the continuum of binary regressions, establishing functional
central limit theorems and bootstrap functional central limit theorems for
the distribution regression coefficient process and other related
functionals, including the estimators of the entire conditional distribution
and quantile functions.} We show that distribution regression encompasses
the Cox (1972) transformation/duration model as a special case, and
represents a useful alternative to Koenker and Bassett's (1978) quantile
regression.

An alternative approach to counterfactual analysis, which is not covered by
our theoretical results, consists in reweighting the observations using the
propensity score, in the spirit of Horvitz and Thompson (1952). For
instance, DiNardo, Fortin, and Lemieux (1996) applied this idea to estimate
counterfactual densities, Firpo (2007) applied to to quantile treatment effects, and
Donald and Hsu (2013) applied it  to the distribution and quantile functions of
potential outcomes. Under correct specification, the regression and the
weighting approaches are equally valid. In particular, if we use saturated
specifications for the propensity score and conditional distribution, then
both approaches lead to numerically identical results. An advantage of the
regression approach is that the intermediate step---the estimation of the
conditional model---is often of independent economic interest. For example,
Buchinsky (1994) applies quantile regression to analyze the determinants of
conditional wage distributions. This model nests the classical Mincer wage
regression and is useful for decomposing changes in the wage distribution
into factors associated with between-group and within-group inequality.


We illustrate our estimation and inference procedures with a decomposition
analysis of the evolution of the U.S. wage distribution, motivated by the in
influential article by DiNardo, Fortin, and Lemieux (1996). We complement
their analysis by employing a wide range of regression methods (instead of
reweighting methods), providing standard errors for the estimates of the main
effects, and extending the analysis to the entire distribution using
simultaneous confidence bands. The use of standard errors allows us to
disentangle the economic significance of various effects from the
statistical uncertainty, which was previously ignored in most decomposition
analyses in economics. We also compare quantile and distribution regression
as competing models for the conditional distribution of wages and discuss
the different choices that must be made to implement our estimators. Our
empirical results highlight the important role of the decline in the real
minimum wage and the minor role of de-unionization in explaining the
increase in wage inequality during the 80s.

We organize the rest of the paper as follows. Section 2 presents our
setting, the counterfactual distributions and effects of interest, and gives
conditions under which these effects have a causal interpretation. In
Section 3 we describe regression models for the conditional distribution,
introduce the distribution regression method and contrast it with the
quantile regression method, define our proposed estimation and inference
procedures, and outline the main estimation and inference results. Section 4
contains the main theoretical results under simple high-level conditions,
which cover a broad array of estimation methods. In Section 5 we verify the
previous high-level conditions for the main estimators of the conditional
distribution function---quantile and distribution regression---under
suitable primitive conditions. In Section 6 we present the empirical
application, and in Section 7 we conclude with a summary of the main results
and pointing out some possible directions of future research. In the
Appendix, we include all the proofs and additional technical results. We
give a consistency result for bootstrap confidence bands, a numerical
example comparing quantile and distribution regression, and additional
empirical results in the online supplemental material (Chernozhukov,
Fernandez-Val, and Melly, 2013).

\section{The Setting for Counterfactual Analysis}

\subsection{Counterfactual distributions}

In order to motivate the analysis, let us first set up a simple running
example. Suppose we would like to analyze the wage differences between men
and women. Let $0$ denote the population of men and $1$ the population of
women. $Y_{j}$ denotes wages and $X_{j}$ denotes job market-relevant
characteristics affecting wages for populations $j=0$ and $j=1$. The
conditional distribution functions $F_{Y_{0}|X_{0}}(y|x)$ and $%
F_{Y_{1}|X_{1}}(y|x)$ describe the stochastic assignment of wages to workers
with characteristics $x$, for men and women, respectively. Let $F_{Y{\langle
0|0\rangle }}$ and $F_{Y{\langle 1|1\rangle }}$ represent the observed
distribution function of wages for men and women, and $F_{Y{\langle
0|1\rangle }}$ represent the counterfactual distribution function of wages
that would have prevailed for women had they faced the men's wage schedule $%
F_{Y_{0}|X_{0}}$: 
\begin{equation*}
F_{Y{\langle 0|1\rangle }}(y):=\int_{\mathcal{X}%
_{1}}F_{Y_{0}|X_{0}}(y|x)dF_{X_{1}}(x).
\end{equation*}
The latter distribution is called counterfactual, since it does not arise as
a distribution from any observable population. Rather, this distribution is
constructed by integrating the conditional distribution of wages for men
with respect to the distribution of characteristics for women. This quantity
is well defined if $\mathcal{X}_{0},$ the support of men's characteristics,
includes $\mathcal{X}_{1}$, the support of women's characteristics, namely $%
\mathcal{X}_{1}\subseteq \mathcal{X}_{0}.$

The difference in the observed wage distributions between men and women can
be decomposed in the spirit of Oaxaca (1973) and Blinder (1973) as follows: 
\begin{equation*}
F_{Y{\langle 1|1\rangle }}-F_{Y{\langle 0|0\rangle }}=[F_{Y{\langle
1|1\rangle }}-F_{Y{\langle 0|1\rangle }}]+[F_{Y{\langle 0|1\rangle }}-F_{Y{%
\langle 0|0\rangle }}]\text{,}
\end{equation*}%
where the first term in brackets is due to differences in the wage structure
and the second term is a composition effect due to differences in
characteristics. We can decompose similarly any functional of the observed
wage distributions such as the quantile function or Lorenz curve into wage
structure and composition effects. These counterfactual effects are well
defined econometric parameters and are widely used in empirical analysis,
e.g. the first term of the decomposition is a measure of gender
discrimination. It is important to note that these effects do not
necessarily have a causal interpretation without additional conditions.
Section 2.3 provides sufficient conditions for such an interpretation to be
valid. 
Thus, our theory covers both the descriptive decomposition analysis and the
causal policy analysis, because the econometric objects -- the
counterfactual distributions and their functionals -- are the same in either
case.




In what follows we formalize these definitions and treat the more general
case with several populations. We suppose that the populations are labeled
by $k\in \mathcal{K}$, and that for each population $k$ there is a random $%
d_{x}$-vector $X_{k}$ of covariates and a random outcome variable $Y_{k}$.
The covariate vector is observable in all populations, but the outcome is
only observable in populations $j\in \mathcal{J}\subseteq \mathcal{K}$.
Given observability, we can identify the covariate distribution $F_{X_{k}}$
in each population $k\in \mathcal{K},$ and the conditional distribution $%
F_{Y_{j}|X_{j}}$ in each population $j\in \mathcal{J}$, as well as the
corresponding conditional quantile function $Q_{Y_{j}|X_{j}}$.\footnote{%
The inference theory of Section 4 does not rely on observability of $X_k$
and $(X_j,Y_j)$, but it only requires that $F_{X_{k}}$ and $F_{Y_{j}|X_{j}}$
are identified and estimable at parametric rates. In principle, $F_{X_{k}}$
and $F_{Y_{j}|X_{j}}$ can correspond to distributions of latent random
variables. For example, $F_{Y_{j}|X_{j}}$ might be the conditional
distribution of an outcome that we observe censored due to top coding, or it
might be a structural conditional function identified by IV methods in a
model with endogeneity. We focus on the case of observable random variables,
because it is convenient for the exposition and covers our leading examples
in Section 5. We briefly discuss extensions to models with endogeneity in
the conclusion.} Thus, we can associate each $F_{X_{k}}$ with label $k$ and
each $F_{Y_{j}|X_{j}}$ with label $j$. We denote the support of $X_{k}$ by $%
\mathcal{X}_{k}\subseteq \mathbb{R}^{d_{x}}$ and the region of interest for $%
Y_{j}$ by $\mathcal{Y}_{j}\subseteq \mathbb{R}$.\footnote{%
We shall typically exclude tail regions of $Y_{j}$ in estimation, as in
Koenker (2005, p. 148).} We assume for simplicity that the number of
populations, $|\mathcal{K}|$, is finite. Further, we define $\mathcal{Y}_{j}%
\mathcal{X}_{j}=\{(y,x):y\in \mathcal{Y}_{j},x\in \mathcal{X}_{j}\}$, $%
\mathcal{Y}\mathcal{X}\mathcal{J}=\{(y,x,j):(y,x)\in \mathcal{Y}_{j}\mathcal{%
X}_{j},j\in \mathcal{J}\},$ and generate other index sets by taking
Cartesian products, e.g., $\mathcal{J}\mathcal{K}=\{(j,k):j\in \mathcal{J}%
,k\in \mathcal{K}\}$.

Our main interest lies in the counterfactual distribution and quantile
functions created by combining the conditional distribution in population $j$
with the covariate distribution in population $k$, namely: 
\begin{align}
& F_{Y{\langle j|k\rangle }}(y):=\int_{\mathcal{X}%
_{k}}F_{Y_{j}|X_{j}}(y|x)dF_{X_{k}}(x),\ \ y\in \mathcal{Y}_{j},
\label{define: counter} \\
& Q_{Y{\langle j|k\rangle }}(\tau ):=F_{Y{\langle j|k\rangle }}^{\leftarrow
}(\tau ),\ \ \tau \in (0,1),
\end{align}%
where $F_{Y{\langle j|k\rangle }}^{\leftarrow }$ is the left-inverse
function of $F_{Y{\langle j|k\rangle }}$ defined in Appendix A. In the
definition (\ref{define: counter}) we assume the support condition: 
\begin{equation}
\mathcal{X}_{k}\subseteq \mathcal{X}_{j},\ \ \text{ for all }(j,k)\in 
\mathcal{J}\mathcal{K},  \label{support}
\end{equation}%
which ensures that the integral is well defined. This condition is analogous
to the overlap condition in treatment effect models with unconfoundedness
(Rosenbaum and Rubin, 1983). In the gender wage gap example, it means that
every female worker can be matched with a male worker with the same
characteristics. If this condition is not met initially, we need to
explicitly trim the supports and define the parameters relative to the
common support.\footnote{%
Specifically, given initial supports $\mathcal{X}_{j}^{o}$ and $\mathcal{X}%
_{k}^{o}$ such that $\mathcal{X}_{k}^{o}\not\subseteq \mathcal{X}_{j}^{o}$,
we can set $\mathcal{X}_{k}=\mathcal{X}_{j}=(\mathcal{X}_{k}^{o}\cap 
\mathcal{X}_{j}^{o})$. Then the covariate distributions are redefined over
this support. See, e.g., Heckman, Ichimura, Smith, and Todd (1998), and
Crump, Hotz, Imbens, and Mitnik (2009) for relevant discussions.}

The counterfactual distribution $F_{Y{\langle j|k\rangle}}$ is the
distribution function of the counterfactual outcome $Y{\langle j|k\rangle}$
created by first sampling the covariate $X_{k}$ from the distribution $%
F_{X_{k}}$ and then sampling $Y{\langle j|k\rangle}$ from the conditional
distribution $F_{Y_{j}|X_{j}}(\cdot|X_{k})$. This mechanism has a strong
representation in the form\footnote{%
This representation for counterfactuals was suggested by Roger Koenker in
the context of quantile regression, as noted in Machado and Mata (2005).} 
\begin{equation}  \label{eq:shorohod_counterfactual}
Y\langle j|k \rangle= Q_{Y_{j}|X_{j}}(U | X_{k}), \text{ where } U \sim
U(0,1) \text{ independently of } X_{k} \sim F_{X_{k}}.
\end{equation}
This representation is useful for connecting counterfactual analysis with
various forms of regression methods that provide models for conditional
quantiles. In particular, conditional quantile models imply conditional
distribution models through the relation: 
\begin{equation}  \label{eq: conditional via quantiles}
F_{Y_{j}|X_{j}}(y|x) \equiv\int_{(0,1)} 1\{Q_{Y_{j} | X_{j}}(u|x) \leq y \}
du.
\end{equation}

In what follows, we define a \textit{counterfactual effect} as the result of
a shift from one counterfactual distribution $F_{Y{\langle l|m\rangle}}$ to
another $F_{Y{\langle j|k\rangle}},$ for some $j,l \in \mathcal{J}$ and $k,m
\in \mathcal{K}$. Thus, we are interested in estimating and performing
inference on the distribution and quantile effects 
\begin{equation*}
\Delta^{DE}(y) = F_{Y{\langle j|k\rangle}} (y) - F_{Y{\langle l|m\rangle}}
(y) \text{ and } \Delta^{QE}(\tau) = Q_{Y{\langle j|k\rangle}}(\tau) -Q_{Y{%
\langle l|m\rangle}}(\tau),
\end{equation*}
as well as other functionals of the counterfactual distributions. For
example, Lorenz curves, commonly used to measure inequality, are ratios of
partial means to overall means 
\begin{equation*}
L(y, F_{Y{\langle j|k\rangle}}) = \int_{\mathcal{Y}_j} 1(\tilde y \leq y)
\tilde y dF_{Y{\langle j|k\rangle}}(\tilde y)/\int_{\mathcal{Y}_j} \tilde y
dF_{Y{\langle j|k\rangle}}(\tilde y),
\end{equation*}
defined for non-negative outcomes only, i.e. $\mathcal{Y}_j \subseteq
[0,\infty)$. In general, the counterfactual effects take the form 
\begin{equation}  \label{eq: functional_of_interest}
\Delta(w):= \phi\left(F_{Y{\langle j|k\rangle}} : (j,k) \in\mathcal{J}%
\mathcal{K} \right)(w).
\end{equation}
This includes, as special cases, the previous distribution and quantile
effects; Lorenz effects, with $\Delta(y) = L(y,F_{Y{\langle j|k\rangle}}) -
L(y, F_{Y{\langle l|m\rangle}})$; Gini coefficients, with $\Delta= 1 - 2
\int_{\mathcal{Y}_j} L(F_{Y{\langle j|k\rangle}},y) dy =: G_{Y{\langle
j|k\rangle}} $; and Gini effects, with $\Delta= G_{Y{\langle j|k\rangle}} -
G_{Y{\langle l|m\rangle}}$.

\subsection{Types of counterfactuals effects}

Focusing on quantile effects as the leading functional of interest, we can
isolate the following special cases of counterfactual effects (CE): 
\begin{equation*}
\begin{array}{ll}
1)\text{ CE of changing the conditional distribution: } & Q_{Y{\langle
j|k\rangle }}(\tau )-Q_{Y{\langle l|k\rangle }}(\tau ). \\ 
2)\text{ CE of changing the covariate distribution: } & Q_{Y{\langle
j|k\rangle }}(\tau )-Q_{Y{\langle j|m\rangle }}(\tau ). \\ 
3)\text{ CE of changing the conditional and covariate distributions: } & Q_{Y%
{\langle j|k\rangle }}(\tau )-Q_{Y{\langle l|m\rangle }}(\tau ).%
\end{array}%
\end{equation*}%
In the gender wage gap example mentioned at the beginning of the section,
the wage structure effect is a type 1 CE (with $j=1$, $k=1$, and $l=0$),
while the composition effect is an example of a type 2 CE (with $j=0$, $k=1$%
, and $m=0 $). In the wage decomposition application in Section 6 the
populations correspond to time periods, the minimum wage is treated as a
feature of the conditional distribution, and the covariates include union
status and other worker characteristics. We consider type 1 CE by
sequentially changing the minimum wage and the wage structure. We also
consider type 2 CE by sequentially changing the components of the covariate
distribution. The CE of simultaneously changing the conditional and
covariate distributions are also covered by our theoretical results but are
less common in applications.

While in the previous examples the populations correspond to different
demographic groups or time periods, we can also create populations
artificially by transforming status quo populations. This is especially
useful when considering type 2 CE. Formally, we can think of $X_{k}$ as
being created through a known transformation of $X_{0}$ in population $0$: 
\begin{equation}  \label{define:transform}
X_{k}=g_{k}(X_{0}),\ \ \text{ where }g_{k}:\mathcal{X}_{0}\rightarrow 
\mathcal{X}_{k}.
\end{equation}
This case covers, for example, adding one unit to the first covariate, $%
X_{1k}=X_{10}+1,$ holding the rest of the covariates constant. The resulting
effect becomes the \textit{unconditional} quantile regression, which
measures the effect of a unit change in a given covariate component on the
unconditional quantiles of $Y$.\footnote{%
The resulting notion of unconditional quantile regression is related but
strictly different from the notion introduced by Firpo, Fortin and Lemieux
(2009). The latter notion measures a first order approximation to such an
effect, whereas the notion described here measures the exact size of such an
effect on the unconditional quantiles. When the change is small, the two
notions coincide approximately, but generally they can differ substantially.}
For example, this type of counterfactual is useful for estimating the effect
of smoking on the marginal distribution of infant birth weights. Another
example is a mean preserving redistribution of the first covariate
implemented as $X_{1k}=(1-\alpha )E[X_{10}]+\alpha X_{10}.$ These and more
general types of transformation defined in (\ref{define:transform}) are
useful for estimating the effect of a change in taxation on the marginal
distribution of food expenditure, or the effect of cleaning up a local
hazardous waste site on the marginal distribution of housing prices (Stock,
1991).

Even though the previous examples correspond to conceptually different
thought experiments, our econometric analysis covers all of them.

\subsection{When counterfactual effects have a causal interpretation}

Under an assumption called conditional exogeneity, selection on observables
or unconfoundedness (e.g., Rosenbaum and Rubin, 1983, Heckman and Robb,
1985, and Imbens, 2004), CE can be interpreted as causal effects. In order
to explain this assumption and define causal effects, it is convenient to
rely upon the potential outcome notation. Let $(Y_{j}^{\ast } : j \in 
\mathcal{J})$ denote a vector of potential outcome variables for various
values of a policy, $j\in \mathcal{J}$, and let $X$ be a vector of control
variables or, simply, covariates.\footnote{%
We use the term policy in a very broad sense, which could include any
program or treatment. The definition of potential outcomes relies implicitly
on a notion of manipulability of the policy via some thought experiment.
Here there are different views about whether such thought experiment should
be implementable or could be a purely mental act, see e.g., Rubin (1978) and
Holland (1986) for the former view, and Heckman (1998, 2008) and Bollen and
Pearl (2012) for the latter view. Following the treatment effects
literature, we exclude general equilibrium effects in the definition of
potential outcomes.} Let $J$ denote the random variable describing the
realized policy, and $Y:=Y_{J}^{\ast }$ the realized outcome variable. When
the policy $J$ is not randomly assigned, it is well known that the
distribution of the observed outcome $Y$ conditional on $J=j,$ i.e. the
distribution of $Y \mid J =j$, may differ from the distribution of $%
Y_{j}^{\ast }$. However, if $J$ is randomly assigned conditional on the
control variables $X$---i.e. if the conditional exogeneity assumption
holds---then the distributions of $Y\mid X,J=j$ and $Y_{j}^{\ast }\mid X$
agree. In this case the observable conditional distributions have a causal
interpretation, and so do the counterfactual distributions generated from
these conditionals by integrating out $X$.

To explain this point formally, let $F_{Y_{j}^{\ast }\mid J}(y\mid k)$
denote the distribution of the potential outcome $Y_{j}^{\ast }$ in the
population with $J=k\in \mathcal{J}$. The causal effect of exogenously
changing the policy from $l$ to $j$ on the distribution of the potential
outcome in the population with realized policy $J=k$, is 
\begin{equation*}
F_{Y_{j}^{\ast }\mid J}(y\mid k)-F_{Y_{l}^{\ast }\mid J}(y\mid k).
\end{equation*}%
%
%
%
%
%
%
%
%
%
In the notation of the previous sections, the policy $J$ corresponds to an
indicator for the population labels $j \in \mathcal{J}$, and the observed
outcome and covariates are generated as $Y_j = Y \mid J = j,$ and $X_k = X
\mid J = k.$\footnote{%
The notation $Y_{j} = Y\mid J=j$ designates that $Y_{j} = Y$ if $J=j$, and $%
X_k = X \mid J = k$ designates that $X_k = X$ if $J=k$.} The lemma given
below shows that under conditional exogeneity, for any $j,k\in \mathcal{J}$
the counterfactual distribution $F_{Y{\langle j|k\rangle }}\left( y\right) $
exactly corresponds to $F_{Y_{j}^{\ast }\mid J}(y\mid k)$, and hence the
causal effect of exogenously changing the policy from $l$ to $j$ in the
population with $J=k$ corresponds to the CE of changing the conditional
distribution from $l$ to $j,$ i.e., 
\begin{equation*}
F_{Y_{j}^{\ast }\mid J}(y\mid k)-F_{Y_{l}^{\ast }\mid J}(y\mid k)=F_{Y{%
\langle j|k\rangle }}\left( y\right) -F_{Y{\langle l|k\rangle }}\left(
y\right) .
\end{equation*}

\begin{lemma}[Causal interpretation for counterfactual distributions]
\label{causal effect} Suppose that 
\begin{equation}  \label{exog}
(Y^*_j:j\in \mathcal{J}) \perp \!\!\!\perp J\mid X, \ \text{a.s.,}
\end{equation}
where $\perp \!\!\!\perp $ denotes independence. Under (\ref{support}) and (%
\ref{exog}), 
\begin{equation*}
F_{Y{\langle j|k\rangle }}(\cdot ) = F_{Y^*_j\mid J}(\cdot \mid k),\ \
j,k\in \mathcal{J}.
\end{equation*}
\end{lemma}

The CE of changing the covariate distribution, $F_{Y{\langle j|k\rangle }%
}(y)-F_{Y{\langle j|m\rangle }}(y)$, also has a causal interpretation as the
policy effect of changing exogenously the covariate distribution from $%
F_{X_{m}}$ to $F_{X_{k}}$ under the assumption that the policy does not
affect the conditional distribution. Such a policy effect arises, for
example, in Stock (1991)'s analysis of the impact of cleaning up a hazardous
site on housing prices. Here, the distance to the nearest hazardous site is
one of the characteristics, $X$, that affect the price of a house, $Y$, and
the cleanup changes the distribution of $X$, say, from $F_{X_{m}}$ to $%
F_{X_{k}}$. The assumption for causality is that the cleanup does not alter
the hedonic pricing function $F_{Y_{m}|X_{m}}(y|x)$, which describes the
stochastic assignment of prices $y$ to houses with characteristics $x$. We
do not discuss explicitly the potential outcome notation and the formal
causal interpretation for this case.


\section{Modeling Choices and Inference Methods for Counterfactual Analysis}

In this section we discuss modeling choices, introduce our proposed
estimation and inference methods, and outline our results, without
submersing into mathematical details. Counterfactual distributions in our
framework have the form (\ref{define: counter}), so we need to model and
estimate the conditional distributions $F_{Y_{j}|X_{j}}$ and covariate
distributions $F_{X_{k}}$. As leading approaches for modeling and estimating 
$F_{Y_{j}|X_{j}}$ we shall use semi-parametric quantile and distribution
regression methods. As the leading approach to estimating $F_{X_k}$ we shall
consider an unrestricted nonparametric method. Note that our proposal of
using distribution regressions is new for counterfactual analysis, while our
proposal of using quantile regressions builds on earlier work by Machado and
Mata (2005), though differs in algorithmic details.

\subsection{Regression models for conditional distributions}

\label{sec:cond_models} The counterfactual distributions of interest depend
on either the underlying conditional distribution, $F_{Y_{j}|X_{j}} $, or
the conditional quantile function, $Q_{Y_{j}|X_{j}},$ through the relation (%
\ref{eq: conditional via quantiles}). Thus, we can proceed by modeling and
estimating either of these conditional functions. There are several
principal approaches to carry out these tasks, and our theory covers these
approaches as leading special cases. In this section we drop the dependence
on the population index $j$ to simplify the notation.

\textbf{1. Conditional quantile models.} Classical regression is one of the
principal approaches to modeling and estimating conditional quantiles. The
classical location-shift model takes the linear-in-parameters form: $%
Y=P(X)^{\prime }\beta +V,\ \ V=Q_{V}(U),$ where $U\sim U(0,1)$ is
independent of $X$, $P(X)$ is a vector of transformations of $X$ such as
polynomials or B-splines, and $P(X)^{\prime }\beta $ is a location function
such as the conditional mean. The additive disturbance $V$ has unknown
distribution and quantile functions $F_{V}$ and $Q_{V}$. The conditional
quantile function of $Y$ given $X$ is $Q_{Y|X}(u|x)=P(X)^{\prime }\beta
+Q_{V}(u),$ and the corresponding conditional distribution is $%
F_{Y|X}(y|x)=F_{V}(y-P(X)^{\prime }\beta ).$ This model, used in Juhn,
Murphy and Pierce (1993), is parsimonious but restrictive, since no matter
how flexible $P(X)$ is, the covariates impact the outcome only through the
location. In applications this model as well as its location-scale
generalizations are often rejected, so we cannot recommend its use without
appropriate specification checks. 

A major generalization and alternative to classical regression is quantile
regression, which is a rather complete method for modeling and estimating
conditional quantile functions (Koenker and Bassett, 1978, Koenker, 2005).%
\footnote{%
Quantile regression is one of most important methods of regression analysis
in economics. For applications, including to counterfactual analysis, see,
e.g., Buchinsky (1994), Chamberlain (1994), Abadie (1997), Gosling, Machin,
and Meghir (2000), Machado and Mata (2005), Angrist, Chernozhukov, and
Fern\'andez-Val (2006), and Autor, Katz, and Kearney (2006b).} In this
approach, we have the general non-separable representation: $Y = Q_{Y|X}(U |
X) = P(X)^{\prime}\beta(U),$ where $U \sim U(0,1)$ is independent of $X$%
(Koenker, 2005, p. 59). We can back out the conditional distribution from
the conditional quantile function through the integral transform: 
\begin{equation*}
F_{Y|X}(y|x) = \int_{(0,1)} 1 \{ P(x)^{\prime}\beta(u) \leq y\} d u, \ y \in 
\mathcal{Y}.
\end{equation*}
The main advantage of quantile regression is that it permits covariates to
impact the outcome by changing not only the location or scale of the
distribution but also its entire shape. Moreover, quantile regression is
flexible in that by considering $P(X)$ that is rich enough, one could
approximate the true conditional quantile function arbitrarily well, when $Y$
has a smooth conditional density (Koenker, 2005, p. 53). 

\textbf{2. Conditional distribution models.} A common approach to model
conditional distributions 
is through the Cox (1972) transformation (duration regression) model: $%
F_{Y|X}(y | x) = 1 - \exp(-\exp(t(y) - P(x)^{\prime}\beta)),$ where $%
t(\cdot) $ is an unknown monotonic transformation. This conditional
distribution corresponds to the following location-shift representation: $%
t(Y) = P(X)^{\prime}\beta+ V, $ where $V$ has an extreme value distribution
and is independent of $X$. In this model, covariates impact an unknown
monotone transformation of the outcome only through the location. The role
of covariates is therefore limited in an important way. Note, however, that
since $t(\cdot)$ is unknown this model is not a special case of quantile
regression. 

Instead of restricting attention to the transformation model for the
conditional distribution, we advocate modelling $F_{Y|X}(y|x)$ separately at
each threshold $y \in \mathcal{Y}$, building upon related, but different,
contributions by Foresi and Peracchi (1995) and Han and Hausman (1990). 
Namely, we propose considering the \textit{distribution regression} model 
\begin{equation}
F_{Y|X}(y|x)=\Lambda (P(x)^{\prime }\beta (y)), \ \text{ for all } y\in 
\mathcal{Y},  \label{dr}
\end{equation}%
where $\Lambda $ is a known link function and $\beta (\cdot )$ is an unknown
function-valued parameter. This specification includes the Cox (1972) model
as a strict special case, but allows for a much more flexible effect of the
covariates. Indeed, to see the inclusion, we set the link function to be the
complementary log-log link, $\Lambda (v)=1-\exp (-\exp (v))$, take $P(x)$ to
include a constant as the first component, and let $P(x)^{\prime }\beta
(y)=t(y)-P(x)^{\prime }\beta $, so that only the first component of $\beta
(y)$ varies with the threshold $y$. To see the greater flexibility of (\ref%
{dr}), we note that (\ref{dr}) allows all components of $\beta (y)$ to vary
with $y$.

The fact that distribution regression with a complementary log-log link
nests the Cox model leads us to consider this specification as an important
reference point. Other useful link functions include the logit, probit,
linear, log-log, and Gosset functions (see Koenker and Yoon, 2009, for the
latter). We also note that the distribution regression model is \textit{%
flexible} in the sense that, for any given link function $\Lambda $, we can
approximate the conditional distribution function $F_{Y|X}(y|x)$ arbitrarily
well by using a rich enough $P(X)$.\footnote{%
Indeed, let $P(X)$ denote the first $p$ components of a basis in $L^{2}(%
\mathcal{X},P)$. Suppose that $\Lambda ^{-1}(F_{Y|X}(y|X))\in L^{2}(\mathcal{%
X},P)$ and $\lambda(t) =\partial \Lambda(t) / \partial t$ is bounded above
by $\bar{\lambda}$. Then, there exists $\beta (y)$ depending on $p$, such
that $\delta _{p}=E\left[ \Lambda ^{-1}(F_{Y|X}(y|X))-P(X)^{\prime }\beta (y)%
\right] ^{ 2}\rightarrow 0$ as $p$ grows, so that $E\left[
F_{Y|X}(y|X)-\Lambda \left( P(X)^{\prime }\beta \left( y\right) \right) %
\right] ^{2}\leq \bar{\lambda}\delta _{p}\rightarrow 0$.} Thus, the choice
of the link function is not important for sufficiently rich $P(X)$.

\textbf{3. Comparison of distribution regression vs. quantile regression.} It
is important to compare and contrast the quantile regression and
distribution regression models. Just like quantile regression generalizes
location regression by allowing all the slope coefficients $\beta (u) $ to
depend on the quantile index $u$, distribution regression generalizes
transformation (duration) regression by allowing all the slope coefficients $%
\beta (y)$ to depend on the threshold index $y$. Both models therefore
generalize important classical models and are semiparametric because they
have infinite-dimensional parameters $\beta (\cdot )$. When the
specification of $P(X)$ is saturated, the quantile regression and
distribution regression models coincide.\footnote{%
For example, when $P(X)$ contains indicators of all points of support of $X$%
, if the support of $X$ is finite.} When the specification of $P(X)$ is not
saturated, distribution and quantile regression models may differ
substantially and are not nested. Accordingly, the model choice cannot be
made on the basis of generality.

Both models are flexible in the sense that by allowing for a sufficiently
rich $P(X)$, we can approximate the conditional distribution arbitrarily
well. However, linear-in-parameters quantile regression is only flexible if $%
Y$ has a smooth conditional density, and may provide a poor approximation to
the conditional distribution otherwise, e.g. when $Y$ is discrete or has
mass points, as it happens in our empirical application. In sharp contrast,
distribution regression does not require smoothness of the conditional
density, since the approximation is done pointwise in the threshold $y$, and
thus handles continuous, discrete, or mixed $Y$ without any special
adjustment. Another practical consideration is determined by the functional
of interest. For example, we show in Remark 3.1 that the algorithm to
compute estimates of the counterfactual distribution involves simpler steps
for distribution regression than for quantile regression, whereas this
computational advantage does not carry over the counterfactual quantile
function. Thus, in practice, we recommend researchers to choose one method
over the other on the basis of empirical performance, specification testing,
ability to handle complicated data situations, or the functional of
interest. In section 6 we explain how these factors influence our decision
in a wage decomposition application.

\subsection{Estimation of counterfactual distributions and their functionals}

The estimator of each counterfactual distribution is obtained by the
plug-in-rule, namely integrating an estimator of the conditional
distribution $\widehat{F}_{Y_{j}|X_{j}}$ with respect to an estimator of the
covariate distribution $\widehat{F}_{X_{k}}$, 
\begin{equation}
\widehat{F}_{Y{\langle j|k\rangle }}(y)=\int_{\mathcal{X}_{k}}\widehat{F}%
_{Y_{j}|X_{j}}(y|x)d\widehat{F}_{X_{k}}(x),\ y\in \mathcal{Y}_{j},\ (j,k)\in 
\mathcal{J}\mathcal{K}.  \label{eq: estimate cfdf}
\end{equation}%
For counterfactual quantiles and other functionals, we also obtain the
estimators via the plug-in rule: 
\begin{equation}
\widehat{Q}_{Y{\langle j|k\rangle }}(\tau )=\widehat{F}_{Y{\langle
j|k\rangle }}^{r\leftarrow }(\tau )\text{ and }\widehat{\Delta }(w)=\phi (%
\widehat{F}_{Y{\langle j|k\rangle }}:(j,k)\in \mathcal{J}\mathcal{K})(w),
\label{eq: plugins}
\end{equation}%
where $\widehat{F}_{Y{\langle j|k\rangle }}^{r}$ denotes the rearrangement
of $\widehat{F}_{Y{\langle j|k\rangle }}$ if $\widehat{F}_{Y{\langle
j|k\rangle }}$ is not monotone (see Chernozhukov, Fernandez-Val, and
Galichon, 2010).\footnote{%
If a functional $\phi _{0}$ requires proper distribution functions as
inputs, we assume that the rearrangement is applied before applying $\phi
_{0}$. Hence formally, to keep notation simple, we interpret the final
functional $\phi $ as the composition of the original functional $\phi _{0}$
with the rearrangement.}


Assume that we have samples $\{ (Y_{ki}, X_{ki}): i=1,...,n_{k}\}$ composed
of i.i.d. copies of $(Y_{k},X_{k})$ for all populations $k \in \mathcal{K}$,
where $Y_{ji}$ is observable only for $j \in\mathcal{J} \subseteq \mathcal{K}
$. We estimate the covariate distribution $F_{X_{k}} $ using the empirical
distribution function 
\begin{equation}  \label{eq:covariate_distribution}
\widehat F_{X_{k}}(x) = n^{-1}_{k} \sum _{i=1}^{n_{k}} 1\{X_{ki} \leq x \},
\ \ k \in\mathcal{K}.
\end{equation}
To estimate the conditional distribution $F_{Y_{j}|X_{j}},$ we develop
methods based on the regression models described in Section 3.1. The
estimator based on distribution regression (DR) takes the form: 
\begin{eqnarray}  \label{eq:conditional_dr}
&& \widehat F_{Y_{j}|X_{j}}(y|x) = \Lambda(
P(x)^{\prime}\widehat\beta_{j}(y)), \ \ (y,x) \in \mathcal{Y}_j\mathcal{X}%
_j, \ \ j \in\mathcal{J}, \ \   \label{est dr} \\
&& \widehat\beta_{j}(y) = \arg\max_{b \in\mathbb{R}^{d_p}} \sum
_{i=1}^{n_{j}} \Big[ 1\{ Y_{ji} \leq y\} \ln[\Lambda( P(X_{ji})^{\prime}b) ]
+ 1\{Y_{ji}>y\} \ln[1- \Lambda(P(X_{ji})^{\prime}b)] \Big],  \label{est dr2}
\end{eqnarray}
where $d_p = \dim P(X_j)$. The estimator based on quantile regression (QR)
takes the form: 
\begin{eqnarray}
&& \widehat F_{Y_{j}|X_{j}}(y|x) = \varepsilon +
\int_{\varepsilon}^{1-\varepsilon} 1 \{ P(x)^{\prime}\widehat \beta_{j}(u)
\leq y\} d u, \ \ (y,x) \in \mathcal{Y}_j\mathcal{X}_j, \ \ j \in\mathcal{J}%
, \ \   \label{est qr} \\
&& \widehat\beta_{j}(u) = \arg\min_{b \in\mathbb{R}^{d_p}} \sum
_{i=1}^{n_{j}} [u - 1\{Y_{ji} \leq P(X_{ji})^{\prime}b\}][Y_{ji}
-P(X_{ji})^{\prime}b],  \label{est qr2}
\end{eqnarray}
for some small constant $\varepsilon > 0$. The trimming by $\varepsilon$
avoids estimation of tail quantiles (Koenker, 2005, p. 148), and is valid
under the conditions set forth in Theorem \ref{theorem:main}.\footnote{%
In our empirical example, we use $\varepsilon=.01$. Tail trimming seems
unavoidable in practice, unless we impose stringent tail restrictions on the
conditional density or use explicit extrapolation to the tails as in
Chernozhukov and Du (2008).}

We provide additional examples of estimators of the conditional distribution
function in the working paper version (Chernozhukov, Fernandez-Val and
Melly, 2009). Also our conditions in Section 4 allow for various additional
estimators of the covariate distribution.

To sum-up, our estimates are computed using the following algorithm:

\begin{algorithm}[Estimation of counterfactual distributions and their
functionals]
(i) Obtain estimates $\widehat F_{X_{k}}$ of the covariate distributions $%
F_{X_{k}}$ using (\ref{eq:covariate_distribution}). (ii) Obtain estimates $%
\widehat F_{Y_{j}|X_{j}}$ of the conditional distribution using (\ref{est dr}%
)--(\ref{est dr2}) for DR or (\ref{est qr})--(\ref{est qr2}) for QR. (iii)
Obtain estimates of the counterfactual distributions, quantiles and other
functionals via (\ref{eq: estimate cfdf}) and (\ref{eq: plugins}).\qed
\end{algorithm}

\begin{remark}
In practice, the quantile regression coefficients can be estimated on a fine
mesh $\varepsilon = u_{1} < ... < u_{S} = 1 - \varepsilon$, with meshwidth $%
\delta$ such that $\delta\sqrt{n_{j}} \to 0 $. In this case the final
counterfactual distribution estimator is computed as: $\widehat F_{Y{\langle
j|k\rangle}}(y)= \varepsilon + n_{k}^{-1} \delta\sum_{i=1}^{n_{k}}
\sum_{s=1}^{S} 1\{P(X_{ki})^{\prime}\widehat\beta(u_{s}) \leq y \}. $ For
distribution regression, the counterfactual distribution estimator takes the
computationally convenient form $\widehat F_{Y{\langle j|k\rangle}}(y) =
n_{k}^{-1} \sum_{i=1}^{n_{k}} \Lambda(
P(X_{ki})^{\prime}\widehat\beta_{j}(y))$ that does not involve inversion nor
trimming 
\qed
\end{remark}

\subsection{Inference}

The estimators of the counterfactual effects follow functional central limit
theorems under conditions that we will make precise in the next section. For
example, the estimators of the counterfactual distributions satisfy 
\begin{equation*}
\sqrt{n} (\widehat F_{Y{\langle j|k\rangle}} - F_{Y{\langle j|k\rangle}})
\rightsquigarrow\bar Z_{jk}, \ \ \text{ jointly in } (j,k) \in\mathcal{J}%
\mathcal{K},
\end{equation*}
where $n$ is a sample size index (say, $n$ denotes the sample size of
population $0$) and $\bar Z_{jk}$ are zero-mean Gaussian processes. We
characterize the limit processes for our leading examples in Section 5, so
that we can perform inference using standard analytical methods. However,
for ease of inference, we recommend and prove the validity of a general
resampling procedure called the \textit{exchangeable bootstrap} (e.g.,
Praestgaard and Wellner, 1993, and van der Vaart and Wellner, 1996). This
procedure incorporates many popular forms of resampling as special cases,
namely the empirical bootstrap, weighted bootstrap, $m$ out of $n$
bootstrap, and subsampling. It is quite useful for applications to have all
of these schemes covered by our theory. For example, in small samples with
categorical covariates, we might want to use the weighted bootstrap to gain
accuracy and robustness to ``small cells", whereas in large samples, where
computational tractability can be an important consideration, we might
prefer subsampling.

In the rest of this section we briefly describe the exchangeable bootstrap
method and its implementation details, leaving a more technical discussion
of the method to Sections 4 and 5. Let $(w_{k1},...,w_{kn_{k}}),$ $k\in 
\mathcal{K},$ be vectors of nonnegative random variables that are
independent of data, and satisfy Condition EB in Section 5. For example, $%
(w_{k1},...,w_{kn_{k}})$ are multinomial vectors with dimension $n_{k}$ and
probabilities $(1/n_{k},\ldots ,1/n_{k})$ in the empirical bootstrap. The
exchangeable bootstrap uses the components of $(w_{k1},...,w_{kn_{k}})$ as
random sampling weights in the construction of the bootstrap version of the
estimators. Thus, the bootstrap version of the estimator of the
counterfactual distribution is 
\begin{equation}
\widehat{F}_{Y{\langle j|k\rangle }}^{\ast }(y)=\int_{\mathcal{X}_{k}}%
\widehat{F}_{Y_{j}|X_{j}}^{\ast }(y|x)d\widehat{F}_{X_{k}}^{\ast }(x),\ \
y\in \mathcal{Y}_{j},\ \ (j,k)\in \mathcal{J}\mathcal{K}.
\label{eq:bs_estimate cfdf}
\end{equation}%
The component $\widehat{F}_{X_{k}}^{\ast }$ is a bootstrap version of
covariate distribution estimator. For example, if using the estimator of $%
F_{X_{k}}$ in (\ref{eq:covariate_distribution}), set 
\begin{equation}
\widehat{F}_{X_{k}}^{\ast }(x)=(n_{k}^{\ast
})^{-1}\sum_{i=1}^{n_{k}}w_{ki}1\{X_{ki}\leq x\},\ \ x\in \mathcal{X}_{k},\
\ k\in \mathcal{K},  \label{eq:bs_covariate_distribution}
\end{equation}%
for $n_{k}^{\ast }=\sum_{i=1}^{n_{k}}w_{ki}$. The component $\widehat{F}%
_{Y_{j}|X_{j}}^{\ast }$ is a bootstrap version of the conditional
distribution estimator. For example, if using DR, set $\widehat{F}%
_{Y_{j}|X_{j}}^{\ast }(y|x)=\Lambda (P(x)^{\prime }\widehat{\beta }%
_{j}^{\ast }(y)),\ (y,x)\in \mathcal{Y}_{j}\mathcal{X}_{j}$, $j\in \mathcal{J%
}$, for 
\begin{equation*}
\widehat{\beta }_{j}^{\ast }(y)=\arg \max_{b\in \mathbb{R}%
^{d_p}}\sum_{i=1}^{n_{j}}w_{ji}\Big[1\{Y_{ji}\leq y\}\ln [\Lambda
(P(X_{ji})^{\prime }b)]+1\{Y_{ji}>y\}\ln [1-\Lambda (P(X_{ji})^{\prime }b)]%
\Big].
\end{equation*}%
If using QR, set $\widehat{F}_{Y_{j}|X_{j}}^{\ast }(y|x)=\varepsilon
+\int_{\varepsilon }^{1-\varepsilon }1\{P(x)^{\prime }\widehat{\beta }%
_{j}^{\ast }(u)\leq y\}du$, $(y,x)\in \mathcal{Y}_{j}\mathcal{X}_{j}$, $j\in 
\mathcal{J}$, for 
\begin{equation*}
\widehat{\beta }_{j}^{\ast }(u)=\arg \min_{b\in \mathbb{R}%
^{d_p}}\sum_{i=1}^{n_{j}}w_{ji}[u-1(Y_{ji}\leq P(X_{ji})^{\prime
}b)][Y_{ji}-P(X_{ji})^{\prime }b)].
\end{equation*}

Bootstrap versions of the estimators of the counterfactual quantiles and
other functionals are obtained by monotonizing $\widehat F^{*}_{Y{\langle
j|k\rangle}}$ using rearrangement if required and setting 
\begin{equation}  \label{eq:pluginsboot}
\widehat Q^{*}_{Y{\langle j|k\rangle}}(\tau) = \widehat F^{* \leftarrow}_{Y{%
\langle j|k\rangle}}(\tau) \text{ and } \widehat\Delta^{*}(w) = \phi \left
(\widehat F^{*}_{Y{\langle j|k\rangle}} : (j,k) \in\mathcal{J}\mathcal{K}
\right )(w).
\end{equation}

The following algorithm describes how to obtain an exchangeable bootstrap
draw of a counterfactual estimator.

\begin{algorithm}[Exchangeable bootstrap for estimators of counterfactual
functionals]
\label{algorithm: bs} (i) Draw a realization of the vectors of weights $%
(w_{k1},...,w_{kn_{k}}),$ $k\in \mathcal{K},$ that satisfy Condition EB in
Section 5. (ii) Obtain a bootstrap draw $\widehat{F}_{X_{k}}^{\ast }$ of the
covariate distribution estimator $\widehat{F}_{X_{k}}$ using (\ref%
{eq:bs_covariate_distribution}). (iii) Obtain a bootstrap draw $\widehat{F}%
_{Y_{j}|X_{j}}^{\ast }$ of the conditional distribution estimator $\widehat{F%
}_{Y_{j}|X_{j}}$ using the same regression method as for the estimator. (iv)
Obtain the bootstrap draws of the estimators of the counterfactual
distribution, quantiles, and other functionals via (\ref{eq:bs_estimate cfdf}%
) and (\ref{eq:pluginsboot}). \qed
\end{algorithm}

The exchangeable bootstrap distributions are useful to perform
asymptotically valid inference on the counterfactual effects of interest. We
focus on uniform methods that cover standard pointwise methods for
real-valued parameters as special cases, and that also allow us to consider
richer functional parameters and hypotheses. For example, an asymptotic
simultaneous $(1-\alpha )$-confidence band for the counterfactual
distribution $F_{Y{\langle j|k\rangle }}(y)$ over the region $y\in \mathcal{Y%
}_{j}$ is defined by the end-point functions 
\begin{equation}
\widehat{F}_{Y{\langle j|k\rangle }}^{\pm }(y)=\widehat{F}_{Y{\langle
j|k\rangle }}(y)\pm \widehat{t}_{1-\alpha}\widehat{\Sigma }_{jk}(y)^{1/2}/ 
\sqrt{n},  \label{eq: bands}
\end{equation}%
such that 
\begin{equation}
\lim_{n\rightarrow \infty }\Pr \left\{ F_{Y{\langle j|k\rangle }}(y)\in
\lbrack \widehat{F}_{Y{\langle j|k\rangle }}^{-}(y),\widehat{F}_{Y{\langle
j|k\rangle }}^{+}(y)]\text{ for all }y\in \mathcal{Y}_j\right\} =1-\alpha .
\label{eq: bands cover}
\end{equation}%
Here, $\widehat{\Sigma }_{jk}(y)$ is a uniformly consistent estimator of $%
\Sigma_{jk} (y)$, the asymptotic variance function of $\sqrt{n}(\widehat{F}%
_{Y{\langle j|k\rangle }}(y)-F_{Y{\langle j|k\rangle }}(y))$. In order to
achieve the coverage property (\ref{eq: bands cover}), we set the critical
value $\widehat{t}_{1-\alpha}$ as a consistent estimator of the $(1-\alpha )$%
-quantile of the Kolmogorov-Smirnov maximal $t$-statistic: 
\begin{equation*}
t=\sup_{y\in \mathcal{Y}_j}\sqrt{n}\widehat{\Sigma }_{jk}(y)^{-1/2}|\widehat{%
F}_{Y{\langle j|k\rangle }}(y)-F_{Y{\langle j|k\rangle }}(y)|.
\end{equation*}

The following algorithm describes how to obtain uniform bands using
exchangeable bootstrap:

\begin{algorithm}[Uniform inference for counterfactual analysis]
\label{algorithm: inference} (i) Using Algorithm \ref{algorithm: bs}, draw $%
\{\widehat{Z}_{jk,b}^{\ast }:1\leq b\leq B\}$ as i.i.d. realizations of $%
\widehat{Z}_{jk}^{\ast }(y)=\sqrt{n}(\widehat{F}_{Y{\langle j|k\rangle }%
}^{\ast }(y)-\widehat{F}_{Y{\langle j|k\rangle }}(y)),$ for $y\in \mathcal{Y}%
_j,$ $(j,k)\in \mathcal{J}\mathcal{K}$. (ii) Compute a bootstrap estimate of 
$\Sigma_{jk}(y)^{1/2}$ such as the bootstrap interquartile range rescaled
with the normal distribution: $\widehat{\Sigma }%
_{jk}(y)^{1/2}=(q_{.75}(y)-q_{.25}(y))/(z_{.75}-z_{.25})$ for $y\in \mathcal{%
Y}_j$, where $q_{p}(y)$ is the p-th quantile of $\{\widehat{Z}_{jk,b}^{\ast
}(y):1\leq b\leq B\}$ and $z_p$ is the p-th quantile of $N(0,1)$. (3)
Compute realizations of the maximal t-statistic $\widehat{t}_{b}=\sup_{y\in 
\mathcal{Y}_j}\widehat{\Sigma }_{jk}(y)^{-1/2}|\widehat{Z}_{jk,b}^{\ast
}(y)| $ for $1 \leq b \leq B.$ (iii) Form a $(1-\alpha ) $-confidence band
for $\{F_{Y{\langle j|k\rangle }}(y):y\in \mathcal{Y}_j\}$ using (\ref{eq:
bands}) setting $\widehat{t}_{1-\alpha}$ to the $(1-\alpha ) $-sample
quantile of $\{\widehat{t}_{b}:1\leq b\leq B\}$. \qed
\end{algorithm}

We can obtain similar uniform bands for the counterfactual quantile
functions and other functionals replacing $\widehat{F}_{Y{\langle j|k\rangle 
}}^{\ast }$ by $\widehat{Q}_{Y{\langle j|k\rangle }}^{\ast }$ or $\widehat{%
\Delta }^{\ast }$ and adjusting the indexing sets accordingly. If the sample
size is large, we can reduce the computational complexity of step (i) of the
algorithm by resampling the first order approximation to the estimators of
the conditional distribution, by using subsampling, or by simulating the
limit process $\bar{Z}_{jk}$ using multiplier methods (Barrett and Donald,
2003).

Our confidence bands can be used to test functional hypotheses about
counterfactual effects. For example, it is straightforward to test
no-effect, positive effect or stochastic dominance hypotheses by verifying
whether the entire null hypothesis falls within the confidence band of the
relevant counterfactual functional, e.g., as in Barrett and Donald (2003)
and Linton, Song, and Whang (2010).\footnote{%
For further references and other approaches, see McFadden (1989), Klecan,
McFadden, and McFadden (1991), Anderson (1996), Davidson and Duclos (2000),
Abadie (2002), Chernozhukov and Fernandez-Val (2005), Linton, Massoumi, and
Whang (2005), Chernozhukov and Hansen (2006), or Maier (2011), among others.}


\begin{remark}[On Validity of Confidence Bands]
\label{remark:confidence} Algorithm \ref{algorithm: inference} uses the
rescaled bootstrap interquartile range $\widehat{\Sigma }_{jk}(y)$ as a
robust estimator of $\Sigma_{jk} (y)$. Other choices of quantile spreads are
also possible. If $\Sigma_{jk} (y)$ is bounded away from zero on the region $%
y\in\mathcal{Y}_j,$ uniform consistency of $\widehat{\Sigma }_{jk}(y)$ over $%
y\in \mathcal{Y}_j$ and consistency of the confidence bands follow from the
consistency of bootstrap for estimating the law of the limit Gaussian
process $\bar{Z}_{jk},$ shown in Sections 4 and 5, and Lemma 1 in
Chernozhukov and Fernandez-Val (2005). Appendix A of the Supplemental
Material provides a formal proof for the sake of completeness. The bootstrap
standard deviation is a natural alternative estimator for $\Sigma_{jk}(y),$
but its consistency requires the uniform integrability of $\{\widehat{Z}%
_{jk}^{\ast }(y)^2 : y \in \mathcal{Y}_j\}$ , which in turn requires
additional technical conditions that we do not impose (see Kato, 2011). 
\qed
\end{remark}


\section{Inference Theory for Counterfactual Analysis under General
Conditions}

This section contains the main theoretical results of the paper. We state
the results under simple high-level conditions, which cover a broad array of
estimation methods. We verify the high-level conditions for the principal
approaches -- quantile and distribution regressions -- in the next section.
Throughout this section, $n$ denotes a sample size index and all limits are
taken as $n \to \infty$. We refer to Appendix A for additional notation.

\subsection{Theory under general conditions}

We begin by gathering the key modeling conditions introduced in Section 2.

\textbf{Condition S.} \textit{\ (a) The condition (\ref{support}) on the
support inclusion holds, so that the counterfactual distributions (\ref%
{define: counter}) are well defined. (b) The sample size $n_k$ for the $k$%
-th population is nondecreasing in the index $n$ and $n/n_{k} \to s_{k}
\in[0, \infty),$ for all $k \in\mathcal{K}$, as $n \to \infty$.}

We impose high-level regularity conditions on the following empirical
processes: 
\begin{equation*}
\widehat{Z}_{j}(y,x) := \sqrt{n_{j}} ( \widehat{F}_{Y_{j}|X_{j}}(y| x) -
F_{Y_{j}|X_{j}}(y| x) ) \text{ and } \widehat{G}_{{k}}(f) := \sqrt{n_{k}}
\int f d( \widehat{F}_{X_{k}} - F_{X_{k}} ),
\end{equation*}
indexed by $( y,x,j,k, f) \in\mathcal{Y}\mathcal{X}\mathcal{J}\mathcal{K}%
\mathcal{F}$, where $\widehat{F}_{Y_{j}|X_{j}}$ is the estimator of the
conditional distribution $F_{Y_{j}|X_{j}}$, $\widehat{F}_{X_{k}} $ is the
estimator of the covariate distribution $F_{X_{k}}$, and $\mathcal{F}$ is a
function class specified below.\footnote{%
Throughout the paper we assume that $\widehat{F}_{Y_{j}|X_{j}}$ takes values
in $[0,1]$. This can always be imposed in practice by truncating negative
values to $0$ and values above $1$ to $1$.} We require that these empirical
processes converge to well-behaved Gaussian processes. In what follows, we
consider $\mathcal{Y}_j\mathcal{X}_j$ as a subset of $\overline{\mathbb{R}}%
^{1+d_x}$ with topology induced by the standard metric $\rho$ on $\overline{%
\mathbb{R}}^{1+d_x}$, where $\overline{\mathbb{R}} = \mathbb{R} \cup \{ +
\infty, - \infty\}$. We also let $\lambda_k(f,\tilde{f}) = [\int (f - \tilde
f)^2 dF_{X_k}]^{1/2}$ be a metric on $\mathcal{F} $.

\textbf{Condition D.} \textit{\ Let $\mathcal{F}$ be a class of measurable
functions that includes $\{ F_{Y_{j}|X_{j}}(y| \cdot): y \in\mathcal{Y}_j, j
\in\mathcal{J}\} $ as well as the indicators of all the rectangles in $%
\overline{\mathbb{R}}^{d_x}$, such that $\mathcal{F}$ is totally bounded
under $\lambda_k$. (a) In the metric space $\ell^{\infty}(\mathcal{YXJKF}%
)^{2}$, 
\begin{equation*}
(\widehat Z_{j}(y,x), \widehat{G}_{{k}}(f) ) \rightsquigarrow ( Z_{j}(y,x),
G_{{k}}(f) ),
\end{equation*}
as stochastic processes indexed by $(y,x,j,k,f) \in\mathcal{YXJKF}.$ The
limit process is a zero-mean tight Gaussian process, where $Z_{j}$ a.s. has
uniformly continuous paths with respect to $\rho$, and $G_{{k}}$ a.s. has
uniformly continuous paths with respect to the metric $\lambda_k$ on $%
\mathcal{F}$. } \textit{(b) The map $y \mapsto F_{Y_{j}|X_{j}}(y| \cdot)$ is
uniformly continuous with respect to the metric $\lambda_k$ for all $(j,k)
\in\mathcal{J}\mathcal{K}$, namely as $\delta \to 0$, $\sup_{ \rho(y,\bar y)
\leq \delta} \lambda_k ( F_{Y_{j}|X_{j}}(y| \cdot), F_{Y_{j}|X_{j}}(\bar y|
\cdot) ) \to 0$ uniformly in $(j,k) \in\mathcal{J}\mathcal{K}$.}

Condition D requires that a uniform central limit theorem holds for the
estimators of the conditional and covariate distributions. We verify
Condition D for semi-parametric estimators of the conditional distribution
function, such as quantile and distribution regression, under i.i.d.
sampling assumption. For the case of duration/transformation regression,
this condition follows from the results of Andersen and Gill (1982) and Burr
and Doss (1993). For the case of classical (location) regression, this
condition follows from the results reported in the working paper version
(Chernozhukov, Fernandez-Val and Melly, 2009). We expect Condition D to hold
in many other applied settings. The requirement $\widehat {G}_{{k}}
\rightsquigarrow{G}_{{k}}$ on the estimated measures is weak and is
satisfied when $\widehat F_{X_{k}}$ is the empirical measure based on a
random sample, as in the previous section. Finally, we note that Condition D
does not even impose the i.i.d sampling conditions, only that a functional
central limit theorem is satisfied. Thus, Condition D can be expected to
hold more generally, which may be relevant for time series applications.%

\begin{remark}
Condition D does not require that the regions $\mathcal{Y}_j$ and $\mathcal{X%
}_{k}$ are compact subsets of $\mathbb{R}$ and $\mathbb{R}^{d_x}$, but we
shall impose compactness when we provide primitive conditions. The
requirement $\widehat {G}_{{k}} \rightsquigarrow{G}_{{k}} $ holds not only
for empirical measures but also for various smooth empirical measures; in
fact, in the latter case the indexing class of functions $\mathcal{F}$ can
be much larger than Glivenko-Cantelli or Donsker; see Radulovic and Wegkamp
(2003) and Gine and Nickl (2008). \qed
\end{remark}

\begin{theorem}[Uniform limit theory for counterfactual distributions and
quantiles]
\label{theorem:main} Suppose that Conditions S and D hold. (1) Then, 
\begin{equation}  \label{eq:marg_process_cdf_cov}
\sqrt{n} \left( \widehat{F}_{Y\langle j|k\rangle}(y) - F_{Y\langle
j|k\rangle}(y) \right) \rightsquigarrow\bar Z_{jk}(y)
\end{equation}
as a stochastic process indexed by $(y,j,k) \in\mathcal{Y}\mathcal{J}%
\mathcal{K}$ in the metric space $\ell^{\infty}( \mathcal{Y}\mathcal{J}%
\mathcal{K})$, where $\bar Z_{jk}$ is a tight zero-mean Gaussian process
with a.s. uniformly continuous paths on $\mathcal{Y}_j$, defined by 
\begin{equation}  \label{eq:marg_process_df}
\bar Z_{jk}(y) := \sqrt{s_{j}} \int Z_{j}(y,x) d F_{X_{k}}(x) + \sqrt{s_{k}}
G_{{k}}(F_{Y_{j}|X_{j}}(y| \cdot)).
\end{equation}
(2) If in addition $F_{Y{\langle j|k\rangle}}$ admits a positive continuous
density $f_{Y{\langle j|k\rangle}}$ on an interval $[a,b]$containing an $%
\epsilon $-enlargement of the set $\{Q_{Y{\langle j|k\rangle}}(\tau): \tau
\in{\mathcal{T}}\}$ in $\mathcal{Y}_j$, where ${\mathcal{T}} \subset (0,1)$,
then 
\begin{equation}  \label{eq:marg_process_qf}
\sqrt{n} \left( \widehat{Q}_{Y{\langle j|k\rangle}}(\tau) - Q_{Y{\langle
j|k\rangle}}(\tau) \right) \rightsquigarrow- \bar Z_{jk}(Q_{Y{\langle
j|k\rangle}}(\tau)) / f_{Y{\langle j|k\rangle}}( Q_{Y_{\langle
j|k\rangle}}(\tau)) =: V_{jk}(\tau),
\end{equation}
as a stochastic process indexed by $(\tau,j,k) \in{\mathcal{T}}\mathcal{J}%
\mathcal{K}$ in the metric space $\ell^{\infty}( {\mathcal{T}}\mathcal{J}%
\mathcal{K})$, where $V_{jk}$ is a tight zero mean Gaussian process with
a.s. uniformly continuous paths on ${\mathcal{T}}$.
\end{theorem}

This is the first main and new result of the paper. It shows that if the
estimators of the conditional and covariate distributions satisfy a
functional central limit theorem, then the estimators of the counterfactual
distributions and quantiles also obey a functional central limit theorem.
This result forms the basis of all inference results on counterfactual
estimators.

As an application of the result above, we derive functional central limit
theorems for distribution and quantile effects. Let $\mathcal{Y} \subseteq 
\mathcal{Y}_j \cap \mathcal{Y}_l,$ ${\mathcal{T}} \subset (0,1),$ and 
\begin{align*}
& \Delta^{DE} (y) = F_{Y{\langle j|k\rangle}}(y) -F_{Y{\langle l|m\rangle}%
}(y), \ \ \widehat\Delta^{DE}(y) = \widehat F_{Y{\langle j|k\rangle}}(y) -
\widehat F_{Y{\langle l|m\rangle}}(y), \\
& \Delta^{QE} (\tau) = Q_{Y{\langle j|k\rangle}}(\tau) -Q_{Y{\langle
l|m\rangle}}(\tau), \ \ \widehat\Delta^{QE}(\tau) = \widehat Q_{Y{\langle
j|k\rangle}}(\tau) - \widehat Q_{Y{\langle l|m\rangle}}(\tau).
\end{align*}

\begin{corollary}[Limit theory for quantile and distribution effects]
\label{cor1} Under the conditions of Theorem \ref{theorem:main}, part 1, 
\begin{equation}  \label{eq:marg_process_de}
\sqrt{n} \left( \widehat\Delta^{DE} (y) - \Delta^{DE} (y) \right)
\rightsquigarrow \bar Z_{jk}(y) - \bar Z_{lm}(y) =: S_{t}(y),
\end{equation}
as a stochastic process indexed by $y \in\mathcal{Y}$ in the space $%
\ell^{\infty}( \mathcal{Y} )$, where $S_{t}$ is a tight zero-mean Gaussian
process with a.s. uniformly continuous paths on $\mathcal{Y}$. Under
conditions of Theorem \ref{theorem:main}, part 2, 
\begin{equation}  \label{eq:marg_process_qe}
\sqrt{n} \left( \widehat\Delta^{QE} (\tau) - \Delta^{QE} (\tau) \right)
\rightsquigarrow V_{jk}(\tau) - V_{lm}(\tau) =: W_{t}(\tau),
\end{equation}
as a stochastic process indexed by $\tau \in{\mathcal{T}}$ in the space $%
\ell^{\infty}( {\mathcal{T}} )$, where $W_{t}$ is a tight zero-mean Gaussian
process with a.s. uniformly continuous paths on $\mathcal{T}$.
\end{corollary}

The following corollary is another application of the result above. It shows
that plug-in estimators of Hadamard-differentiable functionals also satisfy
functional central limit theorems. Examples include Lorenz curves and Lorenz
effects, as well as real-valued parameters, such as Gini coefficients and
Gini effects. Regularity conditions for Hadamard-differentiability of Lorenz
and Gini functionals are given in Bhattacharya (2007).

\begin{corollary}[Limit theory for smooth functionals]
\label{cor2} Consider the parameter $\theta$ as an element of a parameter
space $\mathbb{D}_{\theta} \subset \mathbb{D}= \times_{(jk) \in \mathcal{J}%
\mathcal{K}} \ell^{\infty}(\mathcal{Y}_j)$, with $\mathbb{D}_{\theta}$
containing the true value $\theta_0 = (F_{Y\langle j|k\rangle} : (j,k) \in%
\mathcal{J}\mathcal{K}) $. Consider the plug-in estimator $\widehat \theta =
(\widehat F_{Y\langle j|k\rangle} : (j,k) \in\mathcal{J} \mathcal{K})$.
Suppose $\phi(\theta) $, a functional of interest mapping $\mathbb{D}%
_{\theta}$ to $\ell^{\infty}( \mathcal{W})$, is Hadamard differentiable in $%
\theta$ at $\theta_0$ tangentially to $\times_{(jk) \in \mathcal{J}\mathcal{K%
}} C(\mathcal{Y}_j)$ with derivative $(\phi^{\prime}_{jk}: (j,k) \in\mathcal{%
J}\mathcal{K})$. Let $\Delta = \phi(\theta_0)$ and $\widehat\Delta =
\phi(\widehat\theta)$. Then, under the conditions of Theorem \ref%
{theorem:main}, part 1, 
\begin{equation}  \label{eq:marg_process_diff_func}
\sqrt{n} \left( \widehat{\Delta}(w) - \Delta(w) \right)
\rightsquigarrow\sum_{(j,k) \in\mathcal{J}\mathcal{K}} (\phi^{\prime}_{jk}
\bar Z_{jk})(w) =: T (w),
\end{equation}
as a stochastic processes indexed by $w \in\mathcal{W}$ in $\ell^{\infty}( 
\mathcal{W})$, where $T$ is a tight zero-mean Gaussian process.
\end{corollary}

\subsection{Validity of resampling and other simulation methods for
counterfactual analysis}

As we discussed in Section 3.3, Kolmogorov-Smirnov type procedures offer a
convenient and computationally attractive approach for performing inference
on function-valued parameters using functional central limit theorems. A
complication in our case is that the limit processes in (\ref%
{eq:marg_process_df})--(\ref{eq:marg_process_diff_func}) are non-pivotal, as
their covariance functions depend on unknown, though estimable, nuisance
parameters.\footnote{%
Similar non-pivotality issues arise in a variety of goodness-of-fit problems
studied by Durbin and others, and are referred to as the Durbin problem by
Koenker and Xiao (2002).} We deal with this non-pivotality by using
resampling and simulation methods. An attractive result shown as part of our
theoretical analysis is that the counterfactual operator is Hadamard
differentiable with respect to the underlying conditional and covariate
distributions. As a result, if bootstrap or any other method consistently
estimates the limit laws of the estimators of the conditional and covariate
distributions, it also consistently estimates the limit laws of the
estimators of the counterfactual distributions and their smooth functionals.
This convenient result follows from the functional delta method for
bootstrap of Hadamard differentiable functionals.

In order to state the results formally, we follow the notation and
definitions in van der Vaart and Wellner (1996). Let $D_{n}$ denote the data
vector and $M_{n}$ be the vector of random variables used to generate
bootstrap draws or simulation draws given $D_{n}$ (this may depend on the
particular resampling or simulation method). Consider the random element $%
\mathbb{Z}^{*}_{n} = \mathbb{Z}_{n}(D_{n}, M_{n})$ in a normed space $%
\mathbb{D}$. We say that the bootstrap law of $\mathbb{Z}^{*}_{n}$
consistently estimates the law of some tight random element $\mathbb{Z}$ and
write $\mathbb{Z}^{*}_{n} \rightsquigarrow_{\Pr} \mathbb{Z} $ in $\mathbb{D}$
if 
\begin{equation}  \label{boot1}
\begin{array}{r}
\sup_{h \in\text{BL}_{1}(\mathbb{D})} \left| E_{M_{n}} h \left( \mathbb{Z}%
^{*}_{n}\right) - E h(\mathbb{Z})\right| \rightarrow_{\Pr} 0,%
\end{array}%
\end{equation}
where $\text{BL}_{1}(\mathbb{D})$ denotes the space of functions with
Lipschitz norm at most 1 and $E_{M_{n}}$ denotes the conditional expectation
with respect to $M_{n}$ given the data $D_{n}$; $\rightarrow_{\Pr}$ denotes
convergence in (outer) probability.

Next, consider the processes $\widehat\vartheta(t)$ $=$ $(\widehat
F_{Y_{j}|X_{j}}(y|x),$ $\int f d \widehat F_{X_{k}})$ and $\vartheta(t)$ $=$ 
$(F_{Y_{j}|X_{j}}(y|x), \int f d F_{X_{k}}),$ indexed by $t=(y,x,j,k,f) \in
T =\mathcal{Y}\mathcal{X}\mathcal{J}\mathcal{K}\mathcal{F}$, as elements of $%
\mathbb{E}_{\vartheta} = \ell^{\infty} (T)^{2}$. Condition D(a) can be
restated as $\sqrt{n} (\widehat\vartheta - \vartheta) \rightsquigarrow 
\mathbb{Z}_{\vartheta}$ in $\mathbb{E}_{\vartheta}$, where $\mathbb{Z}%
_{\vartheta}$ denotes the limit process in Condition D(a). Let $%
\widehat\vartheta^{*}$ be the bootstrap draw of $\widehat\vartheta$.
Consider the functional of interest $\phi=\phi(\vartheta)$ in the normed
space $\mathbb{E}_{\phi}$, which can be either the counterfactual
distribution and quantile functions considered in Theorem \ref{theorem:main}%
, the distribution or quantile effects considered in Corollary \ref{cor1},
or any of the functionals considered in Corollary \ref{cor2}. Denote the
plug-in estimator of $\phi$ as $\widehat\phi= \phi(\widehat\vartheta)$ and
the corresponding bootstrap draw as $\widehat \phi^{*} = \phi(
\widehat\vartheta^{*})$. Let $\mathbb{Z}_{\phi}$ denote the limit law of $%
\sqrt{n}(\widehat\phi- \phi)$, as described in Theorem \ref{theorem:main},
Corollary \ref{cor1}, and Corollary \ref{cor2}.

\begin{theorem}[Validity of resampling and other simulation methods for
counterfactual analysis]
\label{theorem:bs} Assume that the conditions of Theorem \ref{theorem:main}
hold. If $\sqrt{n}(\widehat{\vartheta }^{\ast }-\widehat{\vartheta }%
)\rightsquigarrow _{\Pr }\mathbb{Z}_{\vartheta }$ in $\mathbb{E}_{\vartheta
} $, then $\sqrt{n}(\widehat{\phi }^{\ast }-\widehat{\phi})\rightsquigarrow
_{\Pr }\mathbb{Z}_{\phi } $ in $\mathbb{E}_{\phi }$. In words, if the
exchangeable bootstrap or any other simulation method consistently estimates
the law of the limit stochastic process in Condition D, then this method
also consistently estimates the laws of the limit stochastic processes (\ref%
{eq:marg_process_df})--(\ref{eq:marg_process_diff_func}) for the estimators of
counterfactual distributions, quantiles, distribution effects, quantile
effects, and other functionals.
\end{theorem}

This is the second main and new result of the paper. Theorem \ref{algorithm:
bs} shows that any resampling method is valid for estimating the limit laws
of the estimators of the counterfactual effects, provided this method is
valid for estimating the limit laws of the (function-valued) estimators of
the conditional and covariate distributions. We verify the latter condition
for our principal estimators in Section 5, where we establish the validity
of exchangeable bootstrap methods for estimating the laws of function-valued
estimators of the conditional distribution based on quantile regression and
distribution regression processes. As noted in Remark \ref{remark:confidence}%
, this result also implies the validity of the Kolmogorov-Smirnov type
confidence bands for counterfactual effects under non-degeneracy of the
variance function of the limit processes for the estimators of these
effects; see Appendix A of the supplemental material for details. 


\section{Inference Theory for Counterfactual Analysis under Primitive
Conditions}

\label{sec:qr&dr}

We verify that the high-level conditions of the previous section hold for
the principal estimators of the conditional distribution functions, and so
the various conclusions on inference methods also apply to this case. We
also present new results on limit distribution theory for distribution
regression processes and exchangeable bootstrap validity for quantile and
distribution regression processes, which may be of a substantial independent
interest. Throughout this section, we re-label $P(X)$ to $X$ to simplify the
notation. This entails no loss of generality when $P(X)$ includes $X$ as one
of the components.

\subsection{Preliminaries on sampling.}

We assume there are samples $\{ (Y_{ki}, X_{ki}): i=1,...,n_{k}\}$ composed
of i.i.d. copies of $(Y_{k},X_{k})$ for all populations $k \in \mathcal{K}$.
The samples are independent across $k \in \mathcal{K}_0 \subseteq \mathcal{K}
$. We assume that $Y_{ji}$ is observable only for $j \in\mathcal{J}
\subseteq \mathcal{K}_0$. We shall call the case with $\mathcal{K} = 
\mathcal{K}_0$ the \emph{independent samples} case. The independent samples
case arises, for example, in the wage decomposition application of Section
6. In addition, we may have transformation samples indexed by $k \in 
\mathcal{K}_t$ created via transformation of some ``originating" samples $l
\in \mathcal{K}_0$. For example, in the unconditional quantile regression
mentioned in Section 2, we create a transformation sample by shifting one of
the covariates in the original sample up by a unit.

We need to account for the dependence between the transformation and
originating samples in the limit theory for counterfactual estimators. In
order to do so formally, we specify the relation of each transformation
sample, with index $k \in \mathcal{K}_t$, to an originating sample, with
index $l(k) \in \mathcal{K}_0$, as follows: $(Y_{k i}, X_{k i}) = g_{l(k),k}
(Y_{l(k) i}, X_{l(k)i}), i = 1,..., n_k$, where $g_{l(k),k}$ is a known
measurable transformation map, and $l: \mathcal{K}_{t} \to\mathcal{K}_0 $ is
the indexing function that gives the index $l(k)$ of the sample from which
the transformation sample $k$ is originated. We also let $\mathcal{K} = 
\mathcal{K}_t \cup \mathcal{K}_0$. The main requirement on the map $%
g_{l(k),k}: \mathbb{R}^{d_x+1} \mapsto \mathbb{R}^{d_x+1} $ is that it
preserves the Dudley-Koltchinskii-Pollard's (DKP) sufficient condition for
universal Donskerness: given a class $\mathcal{F}$ of suitably measurable
and bounded functions mapping a measurable subset of $\mathbb{R}^{d_x+1}$ to 
$\mathbb{R} $ that obeys Pollard's entropy condition, the class $\mathcal{F}%
\circ g_{l(k),k} $ continues to contain bounded and suitably measurable
functions, and obeys Pollard's entropy condition.\footnote{%
The definitions of suitably measurable and Pollard's entropy condition are
recalled in Appendix A. Together with boundedness, these are well-known
sufficient conditions for a function class to be universal Donsker (Dudley,
1987, Koltchinskii, 1981, and Pollard, 1982).} For example, this holds if $%
g_{l(k),k}$ is an affine or a uniformly Lipschits map. The following
condition states formally the sampling requirements.

\textbf{Condition SM.} \textit{\ The samples $D_{k} = \{(Y_{ki}, X_{ki}): 1
\leq i \leq n_{k} \}$, $k \in \mathcal{K},$ are generated as follows: (a)
For each population $k \in \mathcal{K}_0$, $D_{k}$ contains i.i.d. copies of
the random vector $(Y_{k}, X_{k})$ that has probability law $P_{k}$, and $%
D_k $ are independent across $k \in \mathcal{K}_0$. (b) For each population $%
k \in \mathcal{K}_t$, the samples $D_{k}$ are created by transformation, $%
D_k = \{g_{l(k),k} (Y_{l(k)i}, X_{l(k)i}): 1 \leq i \leq n_{l(k)} \}$ for $%
l(k) \in \mathcal{K}_0$, where the maps $g_{l(k),k} $ preserve the DKP
condition.}

Lemma \ref{lemma:empirical measures} in Appendix D shows the following
result under Condition SM: As $n \to\infty$ the empirical processes 
\begin{equation}  \label{eq: define empirical process}
\widehat{\mathbb{G}}_{k}(f) := \frac{1}{\sqrt{n_{k}}} \sum_{i=1}^{n_k} \left
( f (Y_{ki},X_{ki}) - \int f d P_{k} \right)
\end{equation}
converge weakly, 
\begin{equation}  \label{eq: cov_proc1}
\widehat{{\mathbb{G}}}_{k}(f) \rightsquigarrow{\mathbb{G}}_{k}(f),
\end{equation}
as stochastic processes indexed by $(k,f) \in\mathcal{K}\mathcal{F}$ in $%
\ell^{\infty }(\mathcal{K}\mathcal{F})$. The limit processes $\mathbb{G}_{k}$
are tight $P_{k} $-Brownian bridges, which are independent across $k \in 
\mathcal{K}_0$,\footnote{%
A zero-mean Gaussian process ${\mathbb{G}}_{k}$ is a $P_k$-Brownian bridge
if its covariance function takes the form $E[{\mathbb{G}}_k(f) {\mathbb{G}}%
_k(l) ] = \int f l d P_k - \int f d P_k \int l d P_k$, for any $f$ and $l$
in $L^2(F_{X_k}) $; see van der Vaart (1998).} and for $k \in \mathcal{K}_t$
are defined by: 
\begin{equation}  \label{eq: cov_proc2}
{\mathbb{G}}_{k} (f) = {\mathbb{G}}_{l(k)} (f \circ g_{l(k),k}), \ \ \forall
f \in \mathcal{F}.
\end{equation}


\subsection{Exchangeable bootstrap}

The following condition specifies how we should draw the bootstrap weights
to mimic the dependence between the samples in the exchangeable bootstrap
version of the estimators of counterfactual functionals described in Section
3.

\textbf{Condition EB.} \textit{\ For each $n_{k}$ and $k \in\mathcal{K}_{0}$%
, $(w_{k1}, ..., w_{kn_{k}})$ is an exchangeable,\footnote{%
A sequence of random variables $X_1, X_2, ..., X_n$ is exchangeable if for
any finite permutation $\sigma$ of the indices $1,2, ..., n$ the joint
distribution of the permuted sequence $X_{\sigma(1)}, X_{\sigma(2)},
...,X_{\sigma(n)} $ is the same as the joint distribution of the original
sequence.} nonnegative random vector, which is independent of the data $%
(D_k)_{k \in \mathcal{K}}$, such that for some $\epsilon> 0$ 
\begin{equation}  \label{eq: assumptions weighted bootstrap}
\begin{split}
\sup_{n_{k}} E[w_{k1}^{2+\epsilon}] < \infty, \ \ {n_{k}}^{-1}\sum
_{i=1}^{n_{k}} \left( w_{ki} - \bar{w}_{k} \right) ^{2} \to_{\Pr} 1, \ \
\bar w_{k} \to_{\Pr} 1,
\end{split}%
\end{equation}
where $\bar w_{k} = {n_{k}}^{-1} \sum_{i=1}^{n_{k}} w_{ki} $. Moreover, the
vectors $(w_{k1}, ..., w_{kn_{k}})$ are independent across $k \in \mathcal{K}%
_0$. For each $k \in\mathcal{K}_{t}$, 
\begin{equation}  \label{eq:bs_weights}
w_{ki} = w_{l(k)i}, \ \ k \in\mathcal{K}_t.
\end{equation}%
}

\begin{remark}[Common bootstrap schemes]
As pointed out in van der Vaart and Wellner (1996), by appropriately
selecting the distribution of the weights, exchangeable bootstrap covers the
most common bootstrap schemes as special cases. The empirical bootstrap
corresponds to the case where $(w_{k1},...,w_{kn_{k}})$ is a multinomial
vector with parameter $n_{k}$ and probabilities $(1/n_{k},...,1/n_{k})$. The
weighted bootstrap corresponds to the case where $w_{k1},...,w_{kn_{k}}$ are
i.i.d. nonnegative random variables with $E[w_{k1}]=Var[w_{k1}]=1$, e.g.
standard exponential. 
The $m$ out of $n$ bootstrap corresponds to letting $(w_{k1},...,w_{kn_{k}})$
be equal to $\sqrt{n_{k}/m_{k}}$ times multinomial vectors with parameter $%
m_{k}$ and probabilities $(1/n_{k},...,1/n_{k})$. The subsampling bootstrap
corresponds to letting $(w_{k1},...,w_{kn_{k}})$ be a row in which the
number $n_{k}(n_{k}-m_{k})^{-1/2}m_{k}^{-1/2}$ appears $m_{k}$ times and 0
appears $n_{k}-m_{k}$ times ordered at random, independent of the data. 
\qed
\end{remark}

\subsection{Inference theory for counterfactual estimators based on quantile
regression}

We proceed to impose the following conditions on $(Y_j,X_j)$ for each $j \in 
\mathcal{J}$.

\textbf{Condition QR.} \textit{\ (a) The conditional quantile function takes
the form $Q_{Y_{j}|X_{j}}(u|x)= x^{\prime}\beta_{j}(u)$ for all $u \in 
\mathcal{U}=[\varepsilon, 1- \varepsilon]$ with $0< \varepsilon< 1/2$, and $%
x \in\mathcal{X}_{j}$. (b) The conditional density function $%
f_{Y_{j}|X_{j}}(y|x)$ exists, is uniformly continuous in $(y,x)$ in the
support of $(Y_j, X_j)$, and is uniformly bounded. (c) The minimal
eigenvalue of $J_{j}(u)$ $=$ $E[f_{Y_{j}|X_{j}}(
X_{j}^{\prime}\beta_j(u)|X_{j}) X_{j}X_{j}^{\prime}]$ is bounded away from
zero uniformly over $u \in{\mathcal{U}}$. (d) $E\|X_{j}\|^{2+\epsilon} <
\infty$ for some $\epsilon>0$. }

In order to state the next result, let us define 
\begin{eqnarray*}
&& \ell_{j,y,x}(Y_{j},X_{j}) = - f_{Y_{j}|X_{j}}(y|x) x^{\prime}\psi_{j,
F_{Y_{j}|X_{j}}(y|x)}(Y_{j},X_{j}), \\
&& \psi_{j,u} (Y_{j},X_{j})= -J_{j}(u)^{-1} \{1(Y_{j} \leq
X_{j}^{\prime}\beta_{j}(u)) - u\} X_{j}, \\
&& \kappa_{jk,y} (Y_{j},X_{j},X_{k}) = \sqrt{s_j} \int
\ell_{j,y,x}(Y_{j},X_{j}) dF_{X_k}(x) + \sqrt{s_k} F_{Y_{j}|X_{j}}(y|X_k).
\end{eqnarray*}

\begin{theorem}[Validity of QR based counterfactual analysis]
\label{theorem:qr} Suppose that for each $j \in\mathcal{J},$ Conditions S,
SM, and QR hold, the region of interest $\mathcal{Y}_{j}\mathcal{X}_{j}$ is
a compact subset of $\mathbb{R}^{1+ d_x},$ and $\mathcal{U}_j := \{ u :
x^{\prime }\beta_j (u) \in \mathcal{Y}_{j}, \text{ for some } x \in \mathcal{%
X}_j\} \subseteq \mathcal{U}$. Then, (1) Condition D holds for the quantile
regression estimator (\ref{est qr}) of the conditional distribution and the
empirical distribution estimator (\ref{eq:covariate_distribution}) of the
covariate distribution. The limit processes are given by 
\begin{equation*}
Z_{j}(y,x) = {\mathbb{G}}_{j}( \ell_{j,y,x}), \ \ G_{{k}}(f) = {\mathbb{G}}%
_{k}(f), \ \ (j,k) \in \mathcal{J}\mathcal{K},
\end{equation*}
where ${\mathbb{G}}_{k}$ are the $P_{k}$-Brownian bridges defined in (\ref%
{eq: cov_proc1}) and (\ref{eq: cov_proc2}). In particular, $%
\{F_{Y_{j}|X_{j}}(y|\cdot): y \in\mathcal{Y}_{j}\}$ is a universal Donsker
class. (2) Exchangeable bootstrap consistently estimates the limit law of
these processes under Condition EB. (3) Therefore, all conclusions of
Theorems \ref{theorem:main}- \ref{theorem:bs} and Corollaries \ref{cor1} - %
\ref{cor2} apply. In particular, the limit law for the estimated
counterfactual distribution is given by $\bar Z_{jk}(y) := {\mathbb{G}}_{j}
( \kappa_{jk,y}), $ with covariance function $E[\bar Z_{jk}(y) \bar
Z_{lm}(\bar y)] = E[ \kappa_{jk,y} \kappa_{ lm,\bar y} ] - E[ \kappa_{jk, y}
] E[ \kappa_{lm,\bar y}]$.
\end{theorem}

This is the third main and new result of the paper. It derives the joint
functional central limit theorem for the quantile regression estimator of
the conditional distribution and the empirical distribution function
estimator of the covariate distribution. It also shows that exchangeable
bootstrap consistently estimates the limit law. Moreover, the result
characterizes the limit law of the estimator of the counterfactual
distribution in Theorem 4.1, which in turn determines the limit laws of the
estimators of the counterfactual quantile functions and other functionals,
via Theorem 4.1 and Corollaries 4.1 and 4.2. Note that $\mathcal{U}_j
\subseteq \mathcal{U}$ is the condition that permits the use of trimming in (%
\ref{est qr}), since it says that the conditional distribution of $Y_j$
given $X_j$ on the region of interest $\mathcal{Y}_j\mathcal{X}_j$ is not
determined by the tail conditional quantiles.

While proving Theorem \ref{theorem:qr}, we establish the following
corollaries that may be of independent interest. 

\begin{corollary}[Validity of exchangeable bootstrap for QR coefficient
process]
\label{corQR} Let $\{(Y_{ji},X_{ji}): 1\leq i \leq n_{j}\}$ be a sample of
i.i.d. copies of the random vector $(Y_{j},X_{j})$ that has probability law $%
P_{j}$ and obeys Condition QR. (1) As $n_{j} \to\infty$, the QR coefficient
process possesses the following first order approximation and limit law: $%
\sqrt{n_{j}}(\widehat\beta_{j}(\cdot) - \beta_{j}(\cdot)) = \widehat{\mathbb{%
G}}_j(\psi_{j,\cdot}) + o_{\Pr}(1) \rightsquigarrow {\mathbb{G}}%
_{j}(\psi_{j,\cdot} ) $ in $\ell^{\infty}({\mathcal{U}})^{d_x}$, where ${%
\mathbb{G}}_j$ is a $P_{j}$- Brownian Bridge. (2) The exchangeable bootstrap
law is consistent for the limit law, namely, as $n_{j} \to\infty$, 
\begin{equation*}
\sqrt{n_{j}}(\widehat\beta^{*}_{j}(\cdot) - \widehat\beta_{j}(\cdot))
\rightsquigarrow _{\Pr} {\mathbb{G}}_{j}(\psi_{j,\cdot} ) \text{ in }
\ell^{\infty}({\mathcal{U}})^{d_x}.
\end{equation*}
\end{corollary}

The result (2) is new and shows that exchangeable bootstrap (which includes
empirical bootstrap, weighted bootstrap, $m$ out of $n$ bootstrap, and
subsampling) is valid for estimating the limit law of the entire QR
coefficient process. Previously, such a result was available only for
pointwise cases (e.g. Hahn, 1995, 1997, and Feng, He, and Hu, 2011), and the
process result was available only for subsampling (Chernozhukov and
Fernandez-Val, 2005, and Chernozhukov and Hansen, 2006). 

Let $\widehat Q_{Y_j|X_j}(u|x):= x^{\prime }\widehat \beta_j(u)$ be the QR
estimator of the conditional quantile function, and $u \mapsto \widehat
Q^r_{Y_j|X_j}(u|x)$ be the non-decreasing rearrangement of $u \mapsto
\widehat Q_{Y_j|X_j}(u|x)$ over the region $\mathcal{U}_j$. 
Let $\widehat F_{Y_j|X_j}(y|x) $ be the QR estimator of the conditional
distribution function defined in (\ref{est qr}). Also, we use the star
superscript to denote the bootstrap versions of all these estimators, and
define 
\begin{equation*}
\bar \ell_{j,u,x} (Y_j,X_j):= x^{\prime }\psi_{j,u} (Y_{j},X_{j}).
\end{equation*}

\begin{corollary}[Limit law and exchangeable bootstrap for QR-based
estimators of conditional distribution and quantile functions]
\label{corQR2} Suppose that the conditions of Theorem \ref{theorem:qr} hold.
Then, (1) As $n_{j} \to\infty$, in $\ell^{\infty}({\mathcal{U}}_{j}\mathcal{X%
}_{j})$, $\sqrt{n_{j}}(\widehat Q_{Y_j|X_j}(u|x) - Q_{Y_j|X_j}(u|x) ) = 
\widehat { \mathbb{G}}_{j}( \bar \ell_{j,u,x}) + o_{\Pr}(1) \rightsquigarrow{%
\mathbb{G}}_{j}( \bar \ell_{j,u,x}),$ and $\sqrt{n_{j}}(\widehat
Q^r_{Y_j|X_j}(u|x) - Q_{Y_j|X_j}(u|x) ) = \widehat {\mathbb{G}}_{j}( \bar
\ell_{j,u,x}) + o_{\Pr}(1) \rightsquigarrow{\mathbb{G}}_{j}( \bar
\ell_{j,u,x}),$ as stochastic processes indexed by $(u,x) \in {\mathcal{U}}%
_{j}\mathcal{X}_{j}$. In $\ell^{\infty}({\mathcal{Y}}_{j}\mathcal{X}_{j})$, $%
\sqrt{n_{j}}(\widehat F_{Y_j|X_j}(y|x) - F_{Y_j|X_j}(y|x) ) = \widehat {%
\mathbb{G}}_{j}( \ell_{j,y,x} ) + o_{\Pr}(1) \rightsquigarrow{\mathbb{G}}%
_{j}(\ell_{j,y,x}),$ as a stochastic process indexed by $(y,x) \in {\mathcal{%
Y}}_{j}\mathcal{X}_{j}$. (2) The exchangeable bootstrap law is consistent
for the limit laws, namely, as $n_{j} \to\infty$, in $\ell^{\infty}({%
\mathcal{U}}_{j}\mathcal{X}_{j})$, $\sqrt{n_{j}}(\widehat Q^*_{Y_j|X_j}(u|x)
- \widehat Q_{Y_j|X_j}(u|x) ) \rightsquigarrow_{\Pr} {\mathbb{G}}_{j}( \bar
\ell_{j,u,x}),$ and $\sqrt{n_{j}}(\widehat Q^{r*}_{Y_j|X_j}(u|x) - \widehat
Q^{r}_{Y_j|X_j}(u|x) ) \rightsquigarrow_{\Pr} {\mathbb{G}}_{j}( \bar
\ell_{j,u,x}),$ as stochastic processes indexed by $(u,x) \in {\mathcal{U}}%
_{j}\mathcal{X}_{j}$. In $\ell^{\infty}({\mathcal{Y}}_{j}\mathcal{X}_{j})$, $%
\sqrt{n_{j}}(\widehat F^*_{Y_j|X_j}(y|x) - \widehat F_{Y_j|X_j}(y|x) )
\rightsquigarrow_{\Pr} {\mathbb{G}}_{j}( \ell_{j,y,x}),$ as a stochastic
process indexed by $(y,x) \in {\mathcal{Y}}_{j}\mathcal{X}_{j}$.
\end{corollary}

Corollary \ref{corQR2} establishes first order approximations, functional
central limit theorems and exchangeable bootstrap validity for QR-based
estimators of the conditional distribution and quantile functions. The two
estimators of the conditional quantile function -- $\widehat Q _{Y_j|X_j}$
and $\widehat Q^r_{Y_j|X_j}$ -- are asymptotically equivalent. However, $%
\widehat Q_{Y_j|X_j}$ is not necessarily monotone, while $\widehat
Q^r_{Y_j|X_j}$ is monotone and has better finite sample properties
(Chernozhukov, Fernandez-Val, and Galichon, 2009).

\subsection{Inference Theory for Counterfactual Estimators based on
Distribution Regression}

We shall impose the following conditions on $(Y_{j},X_{j})$ for each $j \in%
\mathcal{J}$.

\textbf{Condition DR.} \textit{\ (a) The conditional distribution function
takes the form $F_{Y_{j}|X_{j}}(y|x)=\Lambda(x^{\prime}\beta_j(y))$ for all $%
y \in \mathcal{Y}_j$ and $x \in \mathcal{X}_{j}$, where $\Lambda$ is either
the probit or logit link function. (b) The region of interest $\mathcal{Y}_j$
is either a compact interval in $\mathbb{R}$ or a finite subset of $\mathbb{R%
}$. In the former case, the conditional density function $%
f_{Y_{j}|X_{j}}(y|x)$ exists, is uniformly bounded and uniformly continuous
in $(y,x)$ in the support of $(Y_j,X_j)$. (c) $E\|X_{j}\|^{2} < \infty$ and
the minimum eigenvalue of 
\begin{equation*}
J_{j}(y) := E \left[ \frac{\lambda(X_{j}^{\prime}\beta_{j}(y))^{2}}{%
\Lambda(X_{j}^{\prime}\beta_{j}(y))[1 - \Lambda(X_{j}^{\prime}\beta_{j}(y))]}
X_{j}X_{j}^{\prime}\right] ,
\end{equation*}
is bounded away from zero uniformly over $y \in \mathcal{Y}_{j}$, where $%
\lambda$ is the derivative of $\Lambda$.}

In order to state the next result, we define 
\begin{eqnarray*}
&& \ell_{j,y,x}(Y_{j},X_{j}) = \lambda(x^{\prime}\beta_j(y))x^{\prime}\psi
_{j,y}(Y_{j},X_{j}) , \\
&& \psi_{j,y}(Y_{j},X_{j})= -J_{j}^{-1}(y) \frac{\Lambda(X_{j}^{\prime}%
\beta_j(y)) - 1 \{Y_{j} \leq y\}}{ \Lambda(X_{j}^{\prime}\beta_j(y))(1-
\Lambda(X_{j}^{\prime }\beta_j(y)))}\lambda(X_{j}^{\prime}\beta_{j}(y))X_{j},
\\
&& \kappa_{jk,y} (Y_{j},X_{j},X_{k}) = \sqrt{s_j} \int
\ell_{j,y,x}(Y_{j},X_{j}) dF_{X_k}(x) + \sqrt{s_k} F_{Y_{j}|X_{j}}(y|X_k).
\end{eqnarray*}

\begin{theorem}[Validity of DR based counterfactual analysis]
\label{theorem:dr} Suppose that for each $j \in\mathcal{J}$, Conditions S,
SM, and DR hold, and the region $\mathcal{Y}_{j}\mathcal{X}_{j}$ is a
compact subset of $\mathbb{R}^{1 + d_x}$. Then, (1) Condition D holds for
the distribution regression estimator (\ref{est dr}) of the conditional
distribution and the empirical distribution estimator (\ref%
{eq:covariate_distribution}) of the covariate distribution, with limit
processes given by 
\begin{equation*}
Z_{j}(y,x) = {\mathbb{G}}_{j}( \ell_{j,y,x}), \ \ G_{{k}}(f) = {\mathbb{G}}%
_{k}(f), \ \ (j,k) \in \mathcal{J}\mathcal{K},
\end{equation*}
where ${\mathbb{G}}_{k}$ are the $P_{k}$-Brownian bridges defined in (\ref%
{eq: cov_proc1}) and (\ref{eq: cov_proc2}). In particular, $%
\{F_{Y_{j}|X_{j}}(y|\cdot) : y \in\mathcal{Y}_{j}\}$ is a universal Donsker
class. (2) Exchangeable bootstrap consistently estimates the limit law of
these processes under Condition EB. (c) Therefore, all conclusions of
Theorem \ref{theorem:main} and \ref{theorem:bs}, and of Corollaries \ref%
{cor1} and \ref{cor2} apply to this case. In particular, the limit law for
the estimated counterfactual distribution is given by $\bar Z_{jk}(y) := {%
\mathbb{G}}_{j} ( \kappa_{jk,y}), $ with covariance function $E [\bar
Z_{jk}(y) \bar Z_{lm}(\bar y)] = E[ \kappa_{jk,y} \kappa_{ lm,\bar y} ] - E[
\kappa_{jk, y} ] E[ \kappa_{lm,\bar y}]$.
\end{theorem}

This is the fourth main and new result of the paper. It derives the joint
functional central limit theorem for the distribution regression estimator
of the conditional distribution and the empirical distribution function
estimator of the covariate distribution. It also shows that bootstrap
consistently estimates the limit law. Moreover, the result characterizes the
limit law of the estimator of the counterfactual distribution in Theorem
4.1, which in turn determines the limit laws of the estimators of the
counterfactual quantiles and other functionals, via Theorem 4.1 and
Corollaries 4.1 and 4.2.

While proving Theorem \ref{theorem:dr}, we also establish the following
corollaries that may be of independent interest.

\begin{corollary}[Limit law and exchangeable bootstrap for DR coefficient
process]
\label{corDR} Let $\{(Y_{ji},X_{ji}): 1 \leq i \leq n_{j}\}$ be a sample of
i.i.d. copies of the random vector $(Y_{j},X_{j})$ that has probability law $%
P_{j}$ and obeys Condition DR. (1) As $n_{j} \to\infty$, the DR coefficient
process possesses the following first order approximation and limit law: 
\begin{equation*}
\sqrt{n_{j}}(\widehat\beta_{j}(\cdot) - \beta_{j}(\cdot)) = \widehat {%
\mathbb{G}}_{j}( \psi_{j,\cdot}) + o_{\Pr}(1) \rightsquigarrow{\mathbb{G}}%
_{j}(\psi_{j,\cdot}) \text{ in } \ell^{\infty}({\mathcal{Y}}_{j})^{d_x},
\end{equation*}
where ${\mathbb{G}}_j$ is a $P_{j}$- Brownian Bridge. (2) The exchangeable
bootstrap law is consistent for the limit law, namely, as $n_{j} \to\infty$, 
\begin{equation*}
\sqrt{n_{j}}(\widehat\beta^{*}_{j}(\cdot) - \widehat\beta_{j}(\cdot))
\rightsquigarrow _{\Pr} {\mathbb{G}}_{j}(\psi_{j,\cdot} ) \text{ in }
\ell^{\infty }({\mathcal{Y}}_{j})^{d_x}.
\end{equation*}
\end{corollary}

Let $\widehat F_{Y_j|X_j}(y|x):= \Lambda(x^{\prime }\widehat \beta_j(y))$ be
the DR estimator of the conditional distribution function, and $y \mapsto
\widehat F^r_{Y_j|X_j}(y|x)$ be the non-decreasing rearrangement of $y
\mapsto \widehat F_{Y_j|X_j}(y|x)$ over the region $\mathcal{Y}_j$. 
Let $\widehat Q_{Y_j|X_j}(u|x) = \widehat F^{r\leftarrow}_{Y_j|X_j}(u|x)$ be
the DR estimator of the conditional quantile function, obtained by inverting
the rearranged estimator of the distribution function over the region $%
\mathcal{U}_j$. Here, $\mathcal{U}_j \subset (0,1)$ can be any compact
interval of quantile indices such that an $\epsilon$-expansion of the region 
$\{Q_{Y_j|X_j}(u|x) : u \in \mathcal{U}_j\}$ is contained in $\mathcal{Y}_j$%
, for all $x \in \mathcal{X}_j$. Also, we use the star superscript to denote
the bootstrap versions of all these estimators, and define 
\begin{equation*}
\bar \ell_{j,u,x} (Y_j,X_j):= - \frac{1}{f_{Y_i|X_j}(Q_{Y_j|X_j}(u|x) |x)}%
\ell_{j,Q_{Y_j|X_j}(u|x) ,x}(Y_j,X_j).
\end{equation*}

\begin{corollary}[Limit law and exchangeable bootstrap for DR-based
estimators of conditional distribution and quantile functions]
\label{corDR2} Suppose that the region of interest $\mathcal{Y}_{j}\mathcal{X%
}_{j}$ is a compact subset of $\mathbb{R}^{1+d_x}$, $\mathcal{Y}_j$ is an
interval, the conditions of Corollary \ref{corDR} hold, and $%
f_{Y_j|X_j}(y|x)>0$ on $\mathcal{Y}_j\mathcal{X}_j$. Then, (1) As $n_{j}
\to\infty$, in $\ell^{\infty}({\mathcal{Y}}_{j}\mathcal{X}_{j})$, $\sqrt{%
n_{j}}(\widehat F_{Y_j|X_j}(y|x) - F_{Y_j|X_j}(y|x) ) = \widehat {\mathbb{G}}%
_{j}( \ell_{j,y,x}) + o_{\Pr}(1) \rightsquigarrow{\mathbb{G}}_{j}(
\ell_{j,y,x}),$ and $\sqrt{n_{j}}(\widehat F^r_{Y_j|X_j}(y|x) -
F_{Y_j|X_j}(y|x) ) = \widehat {\mathbb{G}}_{j}( \ell_{j,y,x}) + o_{\Pr}(1)
\rightsquigarrow{\mathbb{G}}_{j}( \ell_{j,y,x}),$ as stochastic processes
indexed by $(y,x) \in {\mathcal{Y}}_{j}\mathcal{X}_{j}$. In $\ell^{\infty}({%
\mathcal{U}}_{j}\mathcal{X}_{j})$, $\sqrt{n_{j}}(\widehat Q_{Y_j|X_j}(u|x) -
Q_{Y_j|X_j}(u|x) ) = \widehat {\mathbb{G}}_{j}( \bar \ell_{j,u,x} ) +
o_{\Pr}(1) \rightsquigarrow{\mathbb{G}}_{j}( \bar \ell_{j,u,x}),$ as a
stochastic process indexed by $(u,x) \in {\mathcal{U}}_{j}\mathcal{X}_{j}$.
(2) The exchangeable bootstrap law is consistent for the limit laws, namely,
as $n_{j} \to\infty$, in $\ell^{\infty}({\mathcal{Y}}_{j}\mathcal{X}_{j})$, $%
\sqrt{n_{j}}(\widehat F^*_{Y_j|X_j}(y|x) - \widehat F_{Y_j|X_j}(y|x) )
\rightsquigarrow_{\Pr} {\mathbb{G}}_{j}( \ell_{j,y,x}),$ and $\sqrt{n_{j}}%
(\widehat F^{r*}_{Y_j|X_j}(y|x) - \widehat F^{r}_{Y_j|X_j}(y|x) )
\rightsquigarrow_{\Pr} {\mathbb{G}}_{j}( \ell_{j,y,x}),$ as stochastic
processes indexed by $(y,x) \in {\mathcal{Y}}_{j}\mathcal{X}_{j}$. In $%
\ell^{\infty}({\mathcal{U}}_{j}\mathcal{X}_{j})$, $\sqrt{n_{j}}(\widehat
Q^*_{Y_j|X_j}(u|x) - \widehat Q_{Y_j|X_j}(u|x) ) \rightsquigarrow_{\Pr} {%
\mathbb{G}}_{j}( \bar \ell_{j,u,x}),$ as a stochastic process indexed by $%
(u,x) \in {\mathcal{U}}_{j}\mathcal{X}_{j}$.
\end{corollary}

Corollary \ref{corDR2} establishes first order approximations, functional
central limit theorems and exchangeable bootstrap validity for DR-based
estimators of the conditional distribution and quantile functions. The two
estimators of the conditional distribution function -- $\widehat F_{Y_j|X_j}$
and $\widehat F^r_{Y_j|X_j}$ -- are asymptotically equivalent. However, $%
\widehat F_{Y_j|X_j}$ is not necessarily monotone, while $\widehat
F^r_{Y_j|X_j}$ is monotone and has better finite sample properties
(Chernozhukov, Fernandez-Val, and Galichon, 2009).

The limit distribution and bootstrap consistency results in Corollaries \ref%
{corDR} and \ref{corDR2} are new. They have already been applied in several
studies (Chernozhukov, Fernandez-Val and Kowalski, 2011, Rothe, 2012, and
Rothe and Wied, 2012). Note that unlike Theorem \ref{theorem:dr} and
Corollary \ref{corDR2}, Corollary \ref{corDR} does not rely on compactness
of the region $\mathcal{Y}_{j}\mathcal{X}_{j}$.

\section{Labor Market Institutions and the Distribution of Wages}

In this section we apply our estimation and inference procedures to
re-analyze the evolution of the U.S. wage distribution between 1979 and
1988. The first goal here is to compare the methods proposed in Section 3
and to discuss the various choices that practitioners need to make. The
second goal is to provide support for the findings of DiNardo, Fortin, and
Lemieux (1996, DFL hereafter) with a rigorous econometric analysis. Indeed,
we provide confidence intervals for real-valued and function-valued effects
of the institutional and labor market factors driving changes in the wage
distribution, thereby quantifying their economic and statistical
significance. We also provide a variance decomposition of the covariate
composition effect into within-group and between-group components. 

We use the same dataset and variables as in DFL, extracted from the outgoing
rotation groups of the Current Population Surveys (CPS) in 1979 and 1988.
The outcome variable of interest is the hourly log-wage in 1979 dollars. The
regressors include a union status indicator, nine education dummy variables
interacted with experience, a quartic term in experience, two occupation
dummy variables, twenty industry dummy variables, and indicators for race,
SMSA, marital status, and part-time status. Following DFL we weigh the
observations by the product of the CPS sampling weights and the hours
worked. We analyze the data only for men for the sake of brevity.\footnote{%
Results for women can be found in Appendix C of the supplemental material.}

The major factors suspected to have an important role in the evolution of
the wage distribution between 1979 and 1988 are the minimum wage, whose real
value declined by 27 percent, the level of unionization, whose level
declined from 32 percent to 21 percent in our sample, and the
characteristics of the labor force, whose education levels and other
characteristics changed substantially during this period. Thus, following
DFL, we decompose the total change in the US wage distribution into the sum
of four effects: (1) the effect of the change in minimum wage, (2) the
effect of de-unionization, (3) the effect of changes in the characteristics
of the labor force other than unionization, and (4) the wage structure
effect. We stress that this decomposition has a causal interpretation only
under additional conditions analogous to the ones laid out in Section 2.3.

We formally define these four effects as differences between appropriately
chosen counterfactual distributions. Let $F_{Y{\langle (t,s)|(r,v)\rangle }}$
denote the counterfactual distribution of log-wages $Y$ when the wage
structure is as in year $t$, the minimum wage $M$ is at the level observed
in year $s$, the union status $U$ is distributed as in year $r$, and the
other worker characteristics $C$ are distributed as in year $v$. We use two
indexes to refer to the conditional and covariate distributions because we
treat the minimum wage as a feature of the conditional distribution and we
want to separate union status from the other covariates. Given these
counterfactual distributions, we can decompose the observed change in the
distribution of wages between 1979 (year 0) and 1988 (year 1) into the sum
of the previous four effects: {\small 
\begin{equation}
\begin{array}{lllll}
F_{Y{\langle (1,1)|(1,1)\rangle }}-F_{Y{\langle (0,0)|(0,0)\rangle }} & = & 
\underset{(1)}{[F_{Y{\langle (1,1)|(1,1)\rangle }}-F_{Y{\langle
(1,0)|(1,1)\rangle }}]}+\underset{(2)}{[F_{Y{\langle (1,0)|(1,1)\rangle }%
}-F_{Y{\langle (1,0)|(0,1)\rangle }}]} &  &  \\ 
& + & \underset{(3)}{[F_{Y{\langle (1,0)|(0,1)\rangle }}-F_{Y{\langle
(1,0)|(0,0)\rangle }}]}+\underset{(4)}{[F_{Y{\langle (1,0)|(0,0)\rangle }%
}-F_{Y{\langle (0,0)|(0,0)\rangle }}]}. &  & 
\end{array}
\label{eq: four effects}
\end{equation}%
} In constructing the decompositions (\ref{eq: four effects}), we follow the
same sequential order as in DFL.\footnote{%
The order of the decomposition matters because it defines the counterfactual
distributions and effects of interest. We report some results for the
reverse sequential order in Appendix C of the supplemental material. The
results are similar under the two alternative sequential orders.} 

We next describe how to identify and estimate the various counterfactual
distributions appearing in (\ref{eq: four effects}). The first
counterfactual distribution is $F_{Y{\langle (1,0)|(1,1)\rangle }}$, the
distribution of wages that we would observe in 1988 if the real minimum wage
was as high as in 1979. Identifying this quantity requires additional
assumptions.\footnote{%
We cannot identify this quantity from random variation in minimum wage,
since the same federal minimum wage applies to all individuals and state
level minimum wages varied little across states in the years considered.}
Following DFL, the first strategy we employ is to assume the conditional
wage density at or below the minimum wage depends only on the value of the
minimum wage, and the minimum wage has no employment effects and no
spillover effects on wages above its level. Under these conditions, DFL show
that 
\begin{equation}
F_{Y_{(1,0)}|X_{1}}(y|x)=\left\{ 
\begin{array}{ll}
F_{Y_{(0,0)}|X_{0}}\left( y|x\right) \frac{F_{Y_{(1,1)}|X_{1}}\left(
m_{0}|x\right) }{F_{Y_{(0,0)}|X_{0}}\left( m_{0}|x\right) }, & 
\hbox{if $y <
m_{0}$;} \\ 
F_{Y_{(1,1)}|X_{1}}\left( y|x\right) , & \hbox{if $y \geq m_{0}$;}%
\end{array}%
\right.  \label{eq: min wage}
\end{equation}%
where $F_{Y_{(t,s)}|X_{t}}(y|x)$ denotes the conditional distribution of
wages in year $t$ given worker characteristics $X_{t}=(U_{t},C_{t})$ when
the level of the minimum wage is as in year $s$, and $m_{s}$ denotes the
level of the minimum wage in year $s$. The second strategy we employ
completely avoids modeling the conditional wage distribution below the
minimal wage by simply censoring the observed wages below the minimum wage
to the value of the minimum wage, i.e. 
\begin{equation}
F_{Y_{(1,0)}|X_{1}}(y|x)=\left\{ 
\begin{array}{ll}
0, & \hbox{if $y < m_{0}$;} \\ 
F_{Y_{(1,1)}|X_{1}}\left( y|x\right) , & \hbox{if $y \geq m_{0}$.}%
\end{array}%
\right.  \label{eq: strategy 2}
\end{equation}

Given either (\ref{eq: min wage}) or (\ref{eq: strategy 2}) we identify the
counterfactual distribution of wages using the representation: 
\begin{equation}
F_{Y{\langle (1,0)|(1,1)\rangle }}(y)=\int
F_{Y_{(1,0)}|X_{1}}(y|x)dF_{X_{1}}(x),  \label{eq: dis 1}
\end{equation}%
where $F_{X_{t}}$ is the joint distribution of worker characteristics and
union status in year $t$. The other counterfactual marginal distributions we
need are 
\begin{equation}
F_{Y{\langle (1,0)|(0,1)\rangle }}(y)=\int \int F_{Y_{(1,0)|X_{1}}}\left(
y|x\right) dF_{U_{0}|C_{0}}(u|c)dF_{C_{1}}(c)  \label{eq: dis 2}
\end{equation}%
and 
\begin{equation}
F_{Y{\langle (1,0)|(0,0)\rangle }}(y)=\int F_{Y_{(1,0)}|X_{1}}\left(
y|x\right) dF_{X_{0}}\left( x\right) .  \label{eq: dis 3}
\end{equation}%
All the components of these distributions are identified and we estimate
them using the plug-in principle. In particular, we estimate the conditional
distribution $F_{U_{0}|C_{0}}(u|c),u\in \{0,1\},$ by logistic regression,
and $F_{X_{1}}$, $F_{C_{1}}$ and $F_{X_{0}}$ by the empirical distributions.

From a practical standpoint, the main implementation decision concerns the
choice of the estimator of the conditional distributions, $%
F_{Y_{(j,j)}|X_{j}}\left( y|x\right) ,$ for $j\in \{0,1\}$. We consider the
use of quantile regression, distribution regression, classical regression,
and duration/transformation regression. The classical regression and the
duration regression models are parsimonious special cases of the first two
models. However, these models are not appropriate in this application due to
substantial conditional heteroskedasticity in log wages (Lemieux, 2006, and
Angrist, Chernozhukov, and Fernandez-Val, 2006). As the additional
restrictions that these two models impose are rejected by the data, we focus
on the distribution and quantile regression approaches.

Distribution and quantile regressions impose different parametric
restrictions on the data generating process. A linear model for the
conditional quantile function may not provide a good approximation to the
conditional quantiles near the minimum wage, where the conditional quantile
function may be highly nonlinear. Indeed, under the assumptions of DFL the
conditional wage function has different determinants below and above the
minimum wage. In contrast, the distribution regression model may well
capture this type of behavior, since it allows the model coefficients to
depend directly on the wage levels.

A second characteristic of our data set is the sizeable presence of mass
points around the minimum wage and at some other round-dollar amounts. For
instance, 20\% of the wages take exactly 1 out of 6 values and 50\% of the
wages take exactly 1 out of 25 values. We compare the distribution and
quantile regression estimators in a simulation exercise calibrated to fit
many properties of the data set. The results presented in Appendix B of the
supplemental material show that quantile regression is more accurate when
the dependent variable is continuous but performs worse than distribution
regression in the presence of realistic mass points. Based on these
simulations and on specification tests that reject the linear quantile
regression model, we employ the distribution regression approach to generate
the main empirical results.\footnote{%
Rothe and Wied (2012) propose new specification tests for conditional
distribution models. Applying their tests to a similar dataset, they reject
the quantile regression model but not the distribution regression model.}
Since most of the problems for quantile regression take place in the region
of the minimum wage, we also check the robustness of our results with a
censoring approach. We censor wages from below at the value of the minimum
wage and then apply censored quantile and distribution regressions to the
resulting data.

We present our empirical results in Table 1 and Figures 1--3. In Table 1, we
report the estimation and inference results for the decomposition (\ref{eq:
four effects}) of the changes in various measures of wage dispersion between
1979 and 1988 estimated using logit distribution regressions. Figures 1-3
refine these results by presenting estimates and 95\% simultaneous
confidence intervals for several major counterfactual effects of interest,
including quantile, distribution and Lorenz effects. We construct the
simultaneous confidence bands using 100 bootstrap replications and a grid of
quantile indices $\left\{ 0.02,0.021,...,0.98\right\} $. 

We see in the top panels of Figures 1-3 that the low end of the distribution
is significantly lower in 1988 while the upper end is significantly higher
in 1988. This pattern reflects the well-known increase in wage inequality
during this period. Next we turn to the decomposition of the total change
into the sum of the four effects. For this decomposition we focus mostly on
quantile functions for comparability with recent studies and to facilitate
interpretation.\footnote{%
Discreteness of wage data implies that the quantile functions have jumps. To
avoid this erratic behavior in the graphical representations of the results,
we display smoothed quantile functions. The non-smoothed results are
available from the authors. The quantile functions were smoothed using a
bandwidth of 0.015 and a Gaussian kernel. The results in Table 1 have not
been smoothed.} From Figure 1, we see that the contribution of
de-unionization to the total change is quantitatively small and has a
U-shaped effect across the quantile indexes. The magnitude and shape of this
effect on the marginal quantiles between the first and last decile sharply
contrast with the quantitatively large and monotonically decreasing shape of
the effect of the union status on the conditional quantile function for this
range of indexes (Chamberlain, 1994).\footnote{%
We find similar estimates to Chamberlain (1994) for the effect of union
status on the conditional quantile function in our CPS data.} This
comparison illustrates the difference between conditional and unconditional
effects. The unconditional effects depend not only on the conditional
effects but also on the characteristics of the workers who switched their
unionization status. Obviously, de-unionization cannot affect those who were
not unionized at the beginning of the period, which is 70 percent of the
workers. In our data, the unionization rate declines from 32 to 21 percent,
thus affecting only 11 percent of the workers. Thus, even though the
conditional impact of switching from union to non-union status can be
quantitatively large, it has a quantitatively small effect on the marginal
distribution.

From Figure 1, we also see that the change in the distribution of worker
characteristics (other than union status) is responsible for a large part of
the increase in wage inequality. The importance of these composition effects
has been recently stressed by Lemieux (2006) and Autor, Katz and Kearney
(2008). The composition effect, including the de-unionization and worker
characteristics effects, is realized through two channels: between-group and
within-group inequality. To understand the effect of these channels on wage
dispersion it is useful to consider a linear quantile model $Y=X^{\prime
}\beta (U)$, where $X$ is independent of $U$. By the law of total variance,
we can decompose the variance of $Y$ into: 
\begin{equation}
\mathrm{Var}[Y]=E[\beta (U)]^{\prime }\mathrm{Var}[X]E[\beta (U)]+\mathrm{%
trace}\{E[XX^{\prime }]\mathrm{Var}[\beta (U)]\},
\end{equation}%
where between-group inequality corresponds to the first term and
within-group inequality corresponds to the second term.\footnote{%
See Aaberge, Bjerve, and Doksum (2005) for an analogous decomposition of the
pseudo-Lorenz curve. Similar within-between decompositions can also be
constructed using distribution regression models.} When we keep the
coefficients fixed, a change in the distribution of the covariates increases
inequality through the first channel if the variance of the covariates
increases and through the second channel if the proportion of high-variance
groups increases. In our case, both components increased by about 10\%
between 1979 and 1988. The increase in the proportion of college graduate
from 19\% to 23\% is an example of the observed composition changes. It
raised between-group inequality because highly educated workers earn
conditional average wages well above the unconditional average, and
within-group inequality because of the higher wage volatility faced by these
workers.\footnote{%
This is an empirical fact in our data set and not a theoretical fact.
Increasing the proportion of educated workers can in principle either
increase or decrease either component. To compute these effects, we kept the
coefficients constant at their values obtained from estimating $F_{Y\left(
1,0\right) |X}$ and changed the distribution of the covariates from $%
F_{X_{0}}$ to $F_{X_{1}}$. See Appendix D of the supplemental material for
more details on the computation of the variance decomposition.}

We also include estimates of the wage structure effect, sometimes referred
to as the price effect, which captures changes in the conditional
distribution of log hourly wages. It represents the difference we would
observe if the distribution of worker characteristics and union status, and
the minimum wage remained unchanged during this period. This effect has a
U-shaped pattern, which is similar to the pattern Autor, Katz and Kearney
(2006a) find for the period between 1990 and 2000. They relate this pattern
to a bi-polarization of employment into low and high skill jobs. However,
they do not find a U-shaped pattern for the period between 1980 and 1990. A
possible explanation for the apparent absence of this pattern in their
analysis might be that the declining minimum wage masks this phenomenon. In
our analysis, once we control for this temporary factor, we do uncover the
U-shaped pattern for the price component in the 80s.

In Figure C1 of the supplemental material , we check the robustness of the
results with respect to the link function used to implement the DR
estimator. The results previously analyzed were obtained with a logistic
link function. The differences between the estimates obtained with the
logistic, normal, uniform (linear probability model), Cauchy and
complementary log-log link functions are so modest that the lines are almost
indistinguishable. As we mentioned above, the assumptions about the minimum
wage are also delicate, since the mechanism that generates wages strictly
below this level is not clear; it could be measurement error, non-coverage,
or non-compliance with the law. To check the robustness of the results to
the DFL assumptions about the minimum wage and to our semi-parametric model
of the conditional distribution, we re-estimate the decomposition using
censored linear quantile regression and censored distribution regression
with a logit link, censoring the wage data below the minimum wage. For
censored quantile regression, we use Powell's (1986) censored quantile
regression estimated by Chernozhukov and Hong's (2002) algorithm. For
censored distribution regression, we simply censor to zero the distribution
regression estimates of the conditional distributions below the minimum wage
and recompute the functionals of interest. We find the results in Figure C2
of the Supplemental Material to be very similar for the quantile and
distribution regressions, and they are not very sensitive to the censoring.

Overall, our estimates and confidence intervals reinforce the findings of
DFL, giving them a rigorous econometric foundation. Even though the sample
size is large, the precision of some of the estimates was unclear to us a
priori. For instance, only a relatively small proportion of workers are
affected by unions. We provide standard errors and confidence intervals,
which demonstrate the statistical and economic significance of the results.
Moreover, we validate the results with a wide array of estimation methods.
The similarity of the estimates may come as a surprise because the
estimators make different parametric assumptions. However, in a fully
saturated model all the estimators we have applied would give numerically
the same results. The similarity of the results can be explained by the
flexibility of our parametric model. Finally, we give a variance
decomposition of the composition effect that shows that the increase in wage
inequality is due to both between-group and within-group inequality
components.


\section{Conclusion and directions for future work}

This paper develops methods for performing inference about the effect on an
outcome of interest of a change in either the distribution of covariates or
the relationship of the outcome with these covariates. 
The validity of the proposed inference procedures in large samples relies
only on the applicability of a functional central limit theorem and the
consistency of the bootstrap for the estimators of the covariate and
conditional distributions. These conditions hold for the empirical
distribution function estimator of the covariate distribution and for the
most common regression estimators of the conditional distribution, such as
classical, quantile, duration/transformation, and distribution regressions.
Thus, we offer valid inference procedures for several popular existing
estimators and introduce distribution regression to estimate counterfactual
distributions.

We focus on functionals of the marginal counterfactual distributions but we
do not consider their joint distribution. This joint distribution is
required to compute other economically interesting quantities such as the
distribution of the counterfactual effects. Abbring and Heckman (2007)
discuss various ways to identify the distribution of these effects. The
working paper version of this article provides inference procedures under a
rank invariance assumption.

We focus on semi-parametric estimators of the conditional distribution due
to their dominant role in empirical work (Angrist and Pischke, 2008). We
hope to extend the analysis to nonparametric estimators in future work.
Fully nonparametric estimators are in principle attractive but their
implementation in samples of moderate size might be problematic. Rothe
(2010) makes first steps in this direction and highlights some of the
difficulties.

As mentioned in Foonote 1, our general results do not require the
observability of the outcome of interest. If $F_{Y_{j}|X_{j}}\left(
y|x\right) $ is redefined as the conditional distribution of a latent
outcome and an estimator $\widehat{F}_{Y_{j}|X_{j}}\left( y|x\right) $ that
satisfies Condition D is available, then the inference results in Section 4
apply. An interesting example is given by the policy relevant treatment
effects of Heckman and Vytlacil (2005). They consider a class of policies
that affect the probability of participation in a program but do not affect
directly the structural function of the outcome in a model with endogeneity.
For instance, one may be interested in the effect of decreasing college
tuition on wages. In their model, the policy relevant treatment effect is
the conditional marginal treatment effect integrated over the covariate
distribution and the error term in the participation equation. This type of
policy effects is outside of the scope of this paper but is certainly worth
pursuing in future research.




\appendix
\vspace{-.1in}

\section{Notation}

\label{app:notation}

Given a weakly increasing function $F: \mathcal{Y} \subseteq \mathbb{R}
\mapsto \mathcal{T} \subseteq [0,1],$ we define the left-inverse of $F$ as
the function $F^{\leftarrow}: \mathcal{T} \mapsto \overline{\mathcal{Y}},$
where $\overline{\mathcal{Y}}$ is the closure of $\mathcal{Y},$ such that 
\begin{equation*}
F^{\leftarrow}(\tau) = 
\begin{cases}
\inf \{y \in \mathcal{Y} : F(y) \geq \tau \} & \text{if $\sup_{y \in 
\mathcal{Y} } F(y) > \tau$}, \\ 
\sup \{ y \in \mathcal{Y} \} & \text{otherwise}.%
\end{cases}%
\end{equation*}
Each sample from the population $k$ is defined on a probability space $%
(\Omega_k, \mathcal{A}_k, P_k)$, and there is an underlying common
probability space $(\Omega, \mathcal{A}, \Pr)$ that contains the product $%
\times_{k \in \mathcal{K}}(\Omega_k, \mathcal{A}_k, P_k)$. We write $Z_n
\rightsquigarrow Z $ in $\mathbb{E}$ to denote the weak convergence of a
stochastic process $Z_n$ to a random element $Z$ in a normed space $\mathbb{E%
}$, as defined in van der Vaart and Wellner (1996) (VW). We write $\to_{\Pr}$
to denote convergence in outer probability. We write $\rightsquigarrow_{\Pr}$
to denote the weak convergence of the bootstrap law in outer probability, as
formally defined in Section 4. Given the sequences of stochastic processes $%
Z_{m1}, ..., Z_{mn},$ $m \in \mathcal{M}$ for some finite set $\mathcal{M}$,
taking values in normed spaces $\mathbb{E}_m$, we say that $Z_{mn}
\rightsquigarrow Z_m$ jointly in $m \in \mathcal{M},$ if $(Z_{mn}: m \in 
\mathcal{M}) \rightsquigarrow (Z_m: m \in \mathcal{M}) \text{ in } \mathbb{E}%
=\times_{m \in \mathcal{M}}\mathbb{E}_m,$ where the product space $\mathbb{E}
$ is endowed with the norm $\|\cdot\|_{\mathbb{E}}= \vee_{m \in \mathcal{M}%
}\|\cdot\|_{\mathbb{E}_m}$, see Section 1.4 in VW. The space $\ell^{\infty}(%
\mathcal{F})$ represents the space of real-valued bounded functions defined
on the index set equipped with the supremum norm $\|\cdot\|_{ \ell^{\infty}(%
\mathcal{F})}$. Following VW, we use the simplified notation $\|\cdot\|_{ 
\mathcal{F}}$ to denote the supremum norm. Given a measurable subset $%
\mathcal{X}$ of $\mathbb{R}^k$, a class $\mathcal{F}$ of measurable
functions $f: \mathcal{X} \to \mathbb{R}$ is called a universal Donsker
class if for every probability measure $P$ on $\mathcal{X},$ $\sqrt{n}(P_n -
P) \rightsquigarrow \mathbb{G}$ in $\ell^{\infty}(\mathcal{F}),$ where $P_n$
is the empirical measure and $\mathbb{G}$ is a $P$-Brownian bridge (Dudley,
1987). By Dudley (1987) a sufficient condition for a uniformly bounded class
of measurable functions $\mathcal{F}$ to be universal Donsker is the
Koltchinskii-Pollard's entropy condition, which requires the uniform
covering entropy integral for $\mathcal{F}$ to be finite, and suitable
measurability, namely that $\mathcal{F}$ is an image admissible Suslin class
(Dudley, 1987). We call these conditions the Dudley-Koltchinskii-Pollard
condition, and call a class of functions $\mathcal{F}$ that obeys them a
Dudley-Koltchinskii-Pollard (DKP) class. The measurability condition,
developed by Dudley (1987), is mild and holds in most applications,
including in our analysis. We do not explicitly discuss this condition in
what follows. Finally, by a rectangle in $\overline{\mathbb{R}}^d$, we mean
any region of the form $\times_{k=1}^d R_k$, where $R_k$ is an interval of
the form $(a_k,b_k), [a_k,b_k], (a_k, b_k]$, or $[b_k,a_k)$, for $a_k, b_k
\in \overline{\mathbb{R}}$.

\section{ Tools}

We shall use the functional delta method, as formulated in VW. Let $\mathbb{D%
}_{0}$, $\mathbb{D}$, and $\mathbb{E} $ be normed spaces, with $\mathbb{D}%
_{0} \subset\mathbb{D}$. A map $\phi: \mathbb{D}_{\phi} \subset\mathbb{D}
\mapsto\mathbb{E}$ is called \textit{Hadamard-differentiable} at $\theta\in%
\mathbb{D}_{\phi}$ tangentially to $\mathbb{D}_{0}$ if there is a continuous
linear map $\phi_{\theta}^{\prime}: \mathbb{D}_{0} \mapsto\mathbb{E}$ such
that 
\begin{equation*}
\frac{\phi(\theta+ t_{n} h_{n}) - \phi(\theta)}{t_{n}} \rightarrow\phi
_{\theta}^{\prime}(h), \ \ \ n \rightarrow\infty,
\end{equation*}
for all sequences $t_{n} \rightarrow0$ and $h_{n} \rightarrow h \in \mathbb{D%
}_{0}$ such that $\theta+ t_{n} h_{n} \in\mathbb{D}_{\phi}$ for every $n$.




\begin{lemma}[Functional delta-method]
\label{lemma:delta-method} Let $\mathbb{D}_{0}$, $\mathbb{D}$, and $\mathbb{E%
}$ be normed spaces. Let $\phi: \mathbb{D}_{\phi} \subset\mathbb{D} \mapsto%
\mathbb{E}$ be Hadamard-differentiable at $\theta$ tangentially to $\mathbb{D%
}_{0}$. Let $X_{n}$ be a sequence of stochastic processes taking values in $%
\mathbb{D}_{\phi}$ such that $r_{n} (X_{n} - \theta) \rightsquigarrow X$ in $%
\mathbb{D}$, where $X$ is separable and takes its values in $\mathbb{D}_{0}$%
, for some sequence of constants $r_{n} \rightarrow\infty$. Then $%
r_{n}\left( \phi(X_{n}) - \phi(\theta) \right)
\rightsquigarrow\phi_{\theta}^{\prime}(X)$ in $\mathbb{E}$. If $%
\phi_{\theta}^{\prime}$ is defined and continuous on the whole of $\mathbb{D}
$, then the sequence $r_{n}\left( \phi(X_{n}) - \phi(\theta) \right) -
\phi_{\theta}^{\prime}\left( r_{n} (X_{n} - \theta)\right) $ converges to
zero in outer probability.
\end{lemma}

The applicability of the method is greatly enhanced by the fact that
Hadamard differentiation obeys the chain rule, for a formal statement of
which we refer to VW. We also use the following simple ``stacking rule" in
the proofs.


\begin{lemma}[Stacking rule]
\label{lemma:stacking-rule} If $\phi_1: \mathbb{D}_{\phi_1} \subset\mathbb{D}%
_1 \mapsto\mathbb{E}_{1}$ is Hadamard-differentiable at $\theta_1\in\mathbb{D%
}_{\phi_1}$ tangentially to $\mathbb{D}_{10}$ with derivative $%
\phi_{1\theta_1}^{\prime}$ and $\phi_2: \mathbb{D}_{\phi_2} \subset\mathbb{D}%
_2 \mapsto\mathbb{E}_{2}$ is Hadamard-differentiable at $\theta_2\in\mathbb{D%
}_{\phi_2}$ tangentially to $\mathbb{D}_{20}$ with derivative $%
\phi_{2\theta_2}^{\prime}$, then $\phi = (\phi_1, \phi_2): \mathbb{D}%
_{\phi_1} \times \mathbb{D}_{\phi_2} \subset \mathbb{D}_{1} \times \mathbb{D}%
_{2} \mapsto \mathbb{E}_{1} \times \mathbb{E}_{2}$ is
Hadamard-differentiable at $\theta= (\theta_1,\theta_2)$ tangentially to $%
\mathbb{D}_{01}\times \mathbb{D}_{02}$ with derivative $\phi_{\theta}^{%
\prime} = (\phi_{1\theta_1}^{\prime}, \phi_{2\theta_2}^{\prime})$.
\end{lemma}

Let $D_{n}$ denote the data vector and $M_{n}$ be a vector of random
variables, used to generate bootstrap draws or simulation draws (this may
depend on particular method). Consider sequences of random elements $V_{n} =
V_{n}(D_{n})$ and $G^{*}_{n} = G_{n}(D_{n}, M_{n})$ in a normed space $%
\mathbb{D}$, where the sequence $G_{n}=\sqrt {n}(V_{n} - V)$ weakly
converges unconditionally to the tight random element $G$, and $G^{*}_{n}$
converges conditionally given $D_{n}$ in distribution to $G,$ in
probability, denoted as $G_{n} \rightsquigarrow G$ and $G^{*}_{n}
\rightsquigarrow_{\Pr} G $, respectively.\footnote{%
This standard concept is recalled in Section 4; see also VW, Chap. 3.6.} Let 
$V^{*}_{n} = V_{n} + G^{*}_{n}/\sqrt{n}$ denote the bootstrap or simulation
draw of $V_{n}$.

\begin{lemma}[Delta-method for bootstrap and other simulation methods]
\label{lemma:delta-method-bootstrap} Let $\mathbb{D}_{0}$, $\mathbb{D}$, and 
$\mathbb{E}$ be normed spaces, with $\mathbb{D}_{0} \subset\mathbb{D}$. Let $%
\phi: \mathbb{D}_{\phi} \subset\mathbb{D} \mapsto\mathbb{E}$ be
Hadamard-differentiable at $V$ tangentially to $\mathbb{D}_{0}$, with the
derivative map $\phi_{V}^{\prime }$. Let $V_{n}$ and $V^{*}_{n}$ be maps as
indicated previously with values in $\mathbb{D}_{\phi}$ such that $\sqrt{n}%
(V_{n} - V) \rightsquigarrow G$ and $\sqrt{n}(V^{*}_{n} - V_{n})
\rightsquigarrow_{\Pr} G$ in $\mathbb{D}$, where $G$ is separable and takes
its values in $\mathbb{D}_{0}$. Then $\sqrt{n}(\phi(V^{*}_{n}) -
\phi(V_{n})) \rightsquigarrow_{\Pr} \phi_{V}^{\prime }(G) $ in $\mathbb{E}$.
\end{lemma}


Another technical result that we use in the sequel concerns the equivalence
of continuous and uniform convergence.

\begin{lemma}[Uniform convergence via continuous convergence]
\label{Lemma:Resnick} Let $\mathbb{D}$ and $\mathbb{E}$ be complete
separable metric spaces, with $\mathbb{D}$ compact. Suppose $f: \mathbb{D}
\mapsto\mathbb{E}$ is continuous. Then a sequence of functions $f_{n}: 
\mathbb{D} \mapsto\mathbb{E}$ converges to $f$ uniformly on $\mathbb{D}$ if
and only if for any convergent sequence $x_{n} \to x$ in $\mathbb{D}$ we
have that $f_{n}(x_{n}) \to f(x)$.
\end{lemma}


For the proofs of Lemmas \ref{lemma:delta-method} and \ref%
{lemma:delta-method-bootstrap}, see VW, Chap. 3.9. Lemma \ref%
{lemma:stacking-rule} follows from the definition of Hadamard derivative and
product space. For the proof of Lemma \ref{Lemma:Resnick}, see, for example,
Resnick (1987), page 2.

\section{Proof of Lemma 2.1}


First, note that $Y = \sum_{j \in \mathcal{J}} 1(J = j)Y_j^*,$ so that 
\begin{equation*}
F_{Y\mid J, X}(y \mid j, x) = F_{Y^*_j \mid J,X}(y \mid j,x).
\end{equation*}
Also, $Y_j \equiv Y \mid J=j$ and $X_k \equiv X \mid J=k$, so that $F_{Y\mid
J,X}(y\mid j,x)\equiv F_{Y_{j}\mid X_{j}}(y\mid x)$ and $F_{X\mid J}(x\mid
k)\equiv F_{X_{k}}(x)$ (by definition). Hence, by iterating expectations 
\begin{eqnarray*}
F_{Y^*_j \mid J}(y\mid k)=\int_{\mathcal{X}_{k}}F_{Y^*_j\mid J,X}(y\mid
k,x)dF_{X\mid J}(x\mid k)& = &\int_{\mathcal{X}_{k}}F_{Y^*_j \mid J,X}(y\mid
j,x)dF_{X\mid J}(x\mid k) \\
& = & \int_{\mathcal{X}_{k}}F_{Y_{j}\mid X_{j}}(y\mid x)dF_{X_{k}}(x),
\end{eqnarray*}%
where the second equality follows by conditional exogeneity (\ref{exog}),
and the last uses the facts stated above. \qed

\section{Proof of Theorems \protect\ref{theorem:main}--\protect\ref%
{theorem:bs} and Corollaries \protect\ref{cor1}--\protect\ref{cor2}.}

\subsection{Key ingredient: Hadamard differentiability of counterfactual
operator}

\label{sec:hdiff}

It suffices to consider a single pair $(j,k) \in\mathcal{J}\mathcal{K}$. In
order to keep the notation simple, we drop the indices $(j,k)$ wherever
possible.

We need some setup and preliminary observations. Let $\ell^{\infty}_m(%
\mathcal{Y}\mathcal{X})$ denote the set of all bounded and measurable
mappings $\mathcal{Y}\mathcal{X} \mapsto \mathbb{R}$. Let $\mathcal{F}$, $Z$%
, and $G$ be specified as in Condition D, with the indices $(j,k)$ omitted
from the subscripts. We consider $\mathcal{Y}\mathcal{X}$ as a subset of $%
\overline{\mathbb{R}}^{1+d_x}$, with relative topology. Let $\rho$ denote a
standard metric on $\overline{\mathbb{R}}^{1+d_x}$. The closure of $\mathcal{%
Y}\mathcal{X} $ under $\rho$, denoted $\overline{\mathcal{Y}\mathcal{X}}$,
is compact in $\overline{\mathbb{R}}^{1+d_x}$. By Condition D, a.s. $Z$
takes values in $UC(\mathcal{Y}\mathcal{X}, \rho)$, the set of functions
mapping $\mathcal{Y}\mathcal{X}$ to the real line that are uniformly
continuous with respect to metric $\rho$ , and can be continuously extended
to $\overline{\mathcal{Y}\mathcal{X}}$, so that $UC(\mathcal{Y}\mathcal{X},
\rho) \subset\ell^{\infty}_m(\mathcal{Y}\mathcal{X})$. By Condition D, $G
\in UC(\mathcal{F}, \lambda)$ a.s., where $\lambda(f,\tilde{f}) = [P (f -
\tilde f)^2]^{1/2}$ is a (semi) metric on $\mathcal{F} $, under which $%
\mathcal{F}$ is totally bounded.

\begin{lemma}[Hadamard differentiability of counterfactual operator]
\label{lemma: Hadamard dif with estimated cov} Let $\mathcal{YX} \subseteq 
\mathbb{R}^{1+d_x}$, and $\mathcal{F}$ be the class of bounded functions,
mapping $\overline{\mathbb{R}}^{d_x}$ to $\mathbb{R}$, that contains $\{
F_{Y|X}(y| \cdot): y \in\mathcal{Y}\} $ as well as the indicators of all the
rectangles in $\overline{\mathbb{R}}^{d_x}$, such that $\mathcal{F}$ is
totally bounded under $\lambda$. Let $\mathbb{D}_{\phi}$ be the product of
the space of measurable functions $\Gamma: \mathcal{Y}\mathcal{X} \mapsto[0,1%
]$ defined by $(y,x) \mapsto \Gamma(y,x)$ and the bounded maps $\Pi: 
\mathcal{F} \mapsto\mathbb{R}$ defined by $f \mapsto\int f d\Pi$, where $\Pi$
is restricted to be a probability measure on $\mathcal{X}$. Consider the map 
$\phi: \mathbb{D}_{\phi} \subset \mathbb{D} = \ell^{\infty}_m(\mathcal{Y}%
\mathcal{X}) \times\ell^{\infty }( \mathcal{F}) \mapsto\mathbb{E} =
\ell^{\infty}(\mathcal{Y})$, defined by 
\begin{equation*}
(\Gamma, \Pi) \mapsto\phi(\Gamma, \Pi) : = \int \Gamma(\cdot, x) d \Pi (x).
\end{equation*}
Then the map $\phi$ is well defined. Moreover, the map $\phi$ is
Hadamard-differentiable at $(\Gamma,\Pi) = (F_{Y|X}, F_{X})$, tangentially
to the subset $\mathbb{D}_{0} = UC(\mathcal{YX}, \rho) \times UC(\mathcal{F}%
, \lambda) $, with the derivative map $(\gamma, \pi)
\mapsto\phi^{\prime}_{F_{Y|X},F_{X}}(\gamma, \pi)$ mapping $\mathbb{D} $ to $%
\mathbb{E}$ defined by 
\begin{equation*}
\phi^{\prime}_{F_{Y|X},F_{X}}(\gamma, \pi)(y) := \int\gamma(y, x) d F_{X}(x)
+ \pi( F_{Y|X}(y| \cdot)),
\end{equation*}
where the derivative is defined and is continuous on $\mathbb{D}$.
\end{lemma}

\noindent\textbf{Proof of Lemma \ref{lemma: Hadamard dif with estimated cov}%
. } First we show that the map is well defined. Any probability measure $\Pi$
on $\mathcal{X}$ is determined by the values $\int f d \Pi$ for $f \in 
\mathcal{F}$, since $\mathcal{F}$ contains all the indicators of the
rectangles in $\overline{\mathbb{R}}^{d_x}$. By Caratheodory's extension
theorem $\Pi(A) = \Pi 1_A$ is well defined on all Borel subsets $A$ of $%
\mathbb{R}^{d_x}$. Since $x \mapsto \Gamma(y,x)$ is Borel measurable and
takes values in $[0,1]$, it follows that $\int \Gamma(y, x) d \Pi(x) $ is
well defined as a Lebesgue integral, and $\int \Gamma(\cdot, x) d \Pi (x)
\in \ell^{\infty}(\mathcal{Y}) $.

Next we show the main claim. Consider any sequence $(\Gamma^{t}, \Pi^{t}) \in%
\mathbb{D}_{\phi}$ such that for $\gamma^{t} := (\Gamma^{t} - F_{Y|X})/t,$
and $\pi^{t} (f) := \int f d(\Pi^{t}- F_{X})/t,$ 
\begin{equation*}
\begin{array}{lll}
(\gamma^{t}, \pi^{t}) \to(\gamma, \pi), \ \ \text{in } \ \ \ell_m^{\infty }(%
\mathcal{Y}\mathcal{X}) \times\ell^{\infty}(\mathcal{F}), \text{ where }
(\gamma, \pi) \in\mathbb{D}_{0}. &  & 
\end{array}%
\end{equation*}
We want to show that as $t \searrow0$ 
\begin{equation*}
\frac{\phi(\Gamma^{t}, \Pi^{t})- \phi(F_{Y|X},F_{X})}{t} - \phi^{\prime
}_{F_{Y|X},F_{X}}(\gamma, \pi) \to 0 \text{ in } \ell^{\infty}(\mathcal{Y}).
\end{equation*}
Write the difference above as 
\begin{align*}
& {\int(\gamma^{t}(y,x) - \gamma(y,x)) d F_{X}(x)}+ {(\pi^{t} - \pi)(F
_{Y|X}( y |\cdot))} + {t\pi^{t}(\gamma(y|\cdot)) } + {{t
\pi^{t}(\gamma^{t}(y|\cdot) - \gamma(y|\cdot))}} \\
& =: i (y) + ii(y) + iii(y) + iv(y).
\end{align*}
Since $\gamma^{t} \to\gamma$ in $\ell^{\infty}_m(\mathcal{Y}\mathcal{X})$,
we have that $\|i\|_{\mathcal{Y}} \leq\|\gamma^{t} - \gamma\|_{\mathcal{Y}%
\mathcal{X}} \int d F_{X} \to0,$ where $\| \cdot\|_{\mathcal{Y}\mathcal{X}}$
is the supremum norm in $\ell^{\infty}_m(\mathcal{Y}\mathcal{X})$ and $%
\|\cdot\|_{\mathcal{Y}}$ is the supremum norm in $\ell^{\infty}(\mathcal{Y})$%
. Moreover, since $\pi^{t} \to\pi$ in $\ell^{\infty}(\mathcal{F})$ and $\{
F_{Y|X}(y| \cdot): y \in\mathcal{Y} \} \subset\mathcal{F}$ by assumption, we
have $\|ii\|_{\mathcal{Y}} \leq\| \pi^{t} - \pi\|_{\mathcal{F}} \to0, $
where $\|\cdot\|_{\mathcal{F}}$ is the supremum norm in $\ell^{\infty }(%
\mathcal{F})$. Further, 
\begin{align*}
\|iv\|_{\mathcal{Y}} & = \left\| \int(\gamma^{t} -\gamma)(\cdot,x) d t
\pi^{t}(x) \right\| _{\mathcal{Y}} \leq\| \gamma^{t} - \gamma \|_{\mathcal{YX%
}} \int| d (\Pi^{t}-F_{X}) | \leq\|\gamma^{t} - \gamma \|_{\mathcal{YX}} \ 2
\to0,
\end{align*}
since $t d \pi^{t} = d (\Pi^{t}-F_{X}) \text{ and } \int| d (\Pi^{t}-F_{X})
| \leq\int d\Pi^{t} + \int d F_{X}\leq2, $ where $\int|d \mu|$ indicates the
total variation of a signed measure $\mu$.

Since $\gamma$ is continuous on the compact semi-metric space $(\overline {%
\mathcal{YX}}, \rho)$, there exists a finite partition of $\overline {%
\mathbb{R}}^{1+d_x}$ into non-overlapping rectangular regions $(R_{im}: 1
\leq i \leq m) $ (rectangles are allowed not to include their sides to make
them non-overlapping) such that $\gamma$ varies at most $\epsilon$ on $%
\mathcal{Y}\mathcal{X} \cap R_{im}$. Let $p_{m}(y,x) := (y_{im},x_{im})$ if $%
(y,x) \in \mathcal{Y}\mathcal{X} \cap R_{im}$, where $(y_{im},x_{im})$ is an
arbitrarily chosen point within $\mathcal{YX} \cap R_{im}$ for each $i$;
also let $\chi_{imy}(x) := 1\{ (y,x) \in R_{im}\}$. Then, as $t \to 0$, 
\begin{align*}
& \| iii\|_{\mathcal{Y}} = \left\| \int\gamma(\cdot, x) t d \pi
^{t}(x)\right\| _{\mathcal{Y}} \leq\left\| \int(\gamma- \gamma\circ p_{m} )
(\cdot,x) t d \pi^{t}(x) \right\| _{\mathcal{Y}} + \left\| \int(\gamma\circ{%
p_{m}}) (\cdot, x) t d \pi^{t}(x) \right\| _{\mathcal{Y}} \\
& \leq\| \gamma- \gamma\circ{p_{m}}\|_{\mathcal{YX}} \int| t d \pi^{t} | +
\sum_{i=1}^{m} |\gamma(y_{im},x_{im})| t \left \|\pi^{t}
(\chi_{im\cdot})\right\|_{\mathcal{Y}} \\
& \leq\| \gamma- \gamma\circ{p_{m}}\|_{\mathcal{YX}} 2+ t m \|\gamma\|_{%
\mathcal{YX}} \max_{1 \leq i \leq m} \left\|\pi^{t} (\chi_{im\cdot})\right
\|_{\mathcal{Y}} \leq2\epsilon+t m \|\gamma\|_{\mathcal{YX}}
\left\|\pi^t\right \|_{\mathcal{F}} \\
& \leq2\epsilon+ t m \left[ \|\gamma\|_{\mathcal{YX}} \left \| \pi \right
\|_{\mathcal{F}} + o(1) \right] \leq2\epsilon+ O(t) \to2 \epsilon,
\end{align*}
since $\{\chi_{imy}: 1\leq i \leq m, y \in \mathcal{Y}\} \subset\mathcal{F}$%
, so that $\max_i \| \pi^{t} (\chi_{im\cdot}) \|_{\mathcal{Y}} \leq \|
\pi^t\|_{\mathcal{F}} \to \|\pi\|_{\mathcal{F}} < \infty$.\footnote{%
The set $\mathcal{F}$ is allowed to include zero, the indicator of an empty
rectangle.} The constant $\epsilon$ is arbitrary, so $\| iii\|_{\mathcal{Y}}
\to 0$ as $t \to 0$.

Note that the derivative is well-defined over the entire $\mathbb{D}$ and is
in fact continuous with respect to the norm on $\mathbb{D}$ given by $%
\|\cdot\|_{\mathcal{YX}} \vee\| \cdot\|_{\mathcal{F}}$. The second component
of the derivative map is trivially continuous with respect to $\| \cdot\|_{%
\mathcal{F}}$. The first component is continuous with respect to $\|\cdot\|_{%
\mathcal{YX}}$ since 
\begin{equation*}
\left\| {\int(\gamma(\cdot,x) - \tilde\gamma(\cdot,x)) d F_{X}(x)} \right\|
_{\mathcal{Y}} \leq\| \gamma- \tilde\gamma\|_{\mathcal{Y}\mathcal{X}} \int d
F_{X}(x) .
\end{equation*}
Hence the derivative map is continuous. \qed

\subsection{Proof of Theorems \protect\ref{theorem:main} and \protect\ref%
{theorem:bs}.}

In the notation of Lemma \ref{lemma: Hadamard dif with estimated cov}, $%
\widehat F_{Y\langle j|k\rangle} (\cdot)$ $=$ $\phi(\widehat F_{Y_{j}|X_{j}}$%
, $\widehat F_{X_{k}}) (\cdot)$ $=$ $\int\widehat F_{Y_{j}|X_{j}}(\cdot|x) d
\widehat F_{X_{k}}(x)$ and $F_{Y\langle j|k\rangle} (\cdot) = \phi(
F_{Y_{j}|X_{j}}, F_{X_{k}}) (\cdot) = $ $\int F_{Y_{j}|X_{j}}(\cdot|x) d
F_{X_{k}}(x).$ The main result needed to prove the theorem is provided by
Lemma \ref{lemma: Hadamard dif with estimated cov} , which established that
the map $\phi$ is Hadamard differentiable. This result holds uniformly in $%
(j,k) \in\mathcal{J}\mathcal{K}$, since $\mathcal{J}\mathcal{K}$ is a finite
set. Moreover, under condition S, condition D can be restated as: 
\begin{equation*}
\left( \sqrt{n} ( \widehat{F}_{Y_{j}|X_{j}}(y| x) - F_{Y_{j}|X_{j}}(y| x)), 
\sqrt{n} \int f d( \widehat{F}_{X_{k}} - F_{X_{k}}) \right) \rightsquigarrow
\left( \sqrt{s_{j}} Z_{j}(y,x), \sqrt{s_{k}} G_{{k}}(f) \right) ,
\end{equation*}
as stochastic processes indexed by $(y,x,j,k,f) \in\mathcal{YXJKF}$ in the
metric space $\ell^{\infty}(\mathcal{YXJKF})^2 $. By the Functional Delta
Method quoted in Lemma \ref{lemma:delta-method}, it follows that 
\begin{eqnarray}
\sqrt{n} (\widehat F_{Y\langle j|k\rangle} - F_{Y\langle j|k\rangle} )(y) &
& = \int\sqrt{n} [ \widehat{F}_{Y_{j}|X_{j}}(y| x) - F_{Y_{j}|X_{j}}(y| x)]
d F_{X_k}(x)  \notag \\
& & + \int F_{Y_{j}|X_{j}}(y|x) \sqrt{n} d [ \widehat F_{X_{k}}(x) -
F_{X_{k}} (x)] + o_{\Pr}(1)  \label{expand} \\
& & \rightsquigarrow\bar Z_{jk}(y):= \sqrt{s_{j}} \int Z_{j}(y,x) d
F_{X_k}(x) + \sqrt{s_{k}} G_{{k}}(F_{Y_{j}|X_{j}}(y|\cdot)) ,  \notag
\end{eqnarray}
jointly in $(j,k) \in\mathcal{J}\mathcal{K}$. The first order expansion
given after the equality in (\ref{expand}) above is not needed to prove the
theorem, but it can be useful for other applications. The a.s. uniform $\rho$%
-continuity of the sample paths of $\bar Z_{jk}$ follows from the a.s.
uniform $\rho$-continuity of the sample paths of $Z_{j}(y,x)$ with respect
to $(y,x)$ and from the a.s. uniform continuity of the sample paths of $G_{{k%
}}(f)$ with respect to $f$ under the metric $\lambda_k,$ by Condition D.
Indeed, the uniform continuity of $F_{Y_{j}|X_{j}}(y|\cdot)$ with respect to 
$y$ under the metric $\lambda_k$, as stated in Condition D(b), implies a.s.
uniform continuity of the sample paths of $y \mapsto G_{{k}%
}(F_{Y_{j}|X_{j}}(y|\cdot))$. The first claim thus is proven.

In order to show the second claim, we first examine in detail the simple
case where $y \mapsto\widehat F_{Y\langle j|k\rangle}(y)$ is weakly
increasing in $y$. (For example, qr-based estimators are necessarily weakly
increasing, while dr-based estimators need not be.) In this case $\widehat
Q_{Y \langle j|k\rangle} =\widehat F^{\leftarrow}_{Y\langle j|k\rangle}$ and
Hadamard differentiability of the quantile (left inverse) operator (Doss and
Gill, 1992, VW) implies by the functional delta method: 
\begin{align}
\sqrt{n} \left( \widehat{Q}_{Y{\langle j|k\rangle}}(\tau) - Q_{Y{\langle
j|k\rangle}}(\tau) \right) & = -\frac{\sqrt{n} (\widehat F_{Y\langle
j|k\rangle} - F_{Y\langle j|k\rangle})}{ f_{Y{\langle j|k\rangle}}} (Q_{Y{%
\langle j|k\rangle}}(\tau)) + o_{\Pr}(1)  \label{eq: concl1} \\
& \rightsquigarrow - \frac{\bar Z_{jk}}{ f_{Y{\langle j|k\rangle}}} (Q_{Y{%
\langle j|k\rangle}}(\tau)),  \label{eq: concl2}
\end{align}
as a stochastic process indexed by $(\tau,j,k) \in{\mathcal{T}}\mathcal{J}%
\mathcal{K}$ in the metric space $\ell^{\infty}( {\mathcal{T}}\mathcal{J}%
\mathcal{K})$.

When $y \mapsto\widehat F_{Y\langle j|k\rangle}(y)$ is not weakly
increasing, the previous argument does not apply because the references
cited above require $\widehat F_{Y\langle j|k\rangle}$ to be a proper
distribution function. In this case, with probability converging to one we
have that $\widehat Q_{Y \langle j|k\rangle} : = \widehat F^{r
\leftarrow}_{Y\langle j|k\rangle}$, where $\widehat F^r_{Y\langle
j|k\rangle} $ is the monotone rearrangement of $\widehat F_{Y\langle
j|k\rangle}$ on the interval $[a,b]$ defined in the statement of Theorem \ref%
{theorem:main}. In order to establish the properties of this estimator, we
first recall the relevant result on Hadamard differentiability of the
monotone rearrangement operator derived by Chernozhukov, Fernandez-Val, and
Galichon (2010). Let $F$ be a continuously differentiable function on the
interval $[a,b]$ with strictly positive derivative $f$. Consider the
rearrangement map $G \mapsto G^{r}$, which maps bounded measurable functions 
$G$ on the domain $[a,b]$ and produces cadlag functions $G^{r}$ on the same
domain. This map, considered as a map $\ell^{\infty}_m([a,b])
\mapsto\ell^{\infty}_m([a,b])$, is Hadamard differentiable at $F$
tangentially to $C([a,b])$, with the derivative map given by the identity $g
\mapsto g$ which is defined and continuous on the whole $\ell^{%
\infty}_m([a,b])$. Therefore, we conclude by the functional delta method
that for all $(j,k) \in \mathcal{J}\mathcal{K} $, $\sqrt{n} (\widehat
F^{r}_{Y\langle j|k\rangle} - F_{Y\langle j|k\rangle} )(\cdot) = \sqrt{n}
(\widehat F_{Y\langle j|k\rangle} - F_{Y\langle j|k\rangle} )(\cdot) +
o_{\Pr}(1).$ Hence the rearranged estimator is first order equivalent to the
original estimator and thus inherits the limit distribution. Now apply the
differentiability of the quantile operator and the delta method again to
reach the same final conclusions (\ref{eq: concl1})- (\ref{eq: concl2}) as
above.

A.s. uniform continuity of the sample paths of $V_{jk}$ follows from the
continuity and positivity assumption on the density function, giving
continuity of the quantile function $\tau \mapsto Q_{Y\langle
j|k\rangle}(\tau)$, and from the a.s. uniform continuity of sample paths of $%
\bar Z_{jk}$ established in the first part of the theorem.

Theorem \ref{theorem:bs} follows from the application of the functional
delta method for the (generalized) bootstrap quoted in Lemma \ref%
{lemma:delta-method-bootstrap} and the chain rule for the Hadamard
derivative. \qed

\subsection{Proof of Corollaries \protect\ref{cor1}--\protect\ref{cor2}}

Corollary \ref{cor1} follows from Theorem \ref{theorem:main} by the extended
continuous mapping theorem. Corollary \ref{cor2} follows by the functional
delta method. \qed

\section{Proof of Theorem \protect\ref{theorem:qr} and \protect\ref%
{theorem:dr}}

It is convenient to organize the proof in several steps. The task is
complex: We need to show convergence and bootstrap convergence
simultaneously for estimators of conditional distributions based on QR or DR
and for estimators of covariate distributions based on empirical measures.
Since both distribution and quantile regression processes are Z-processes,
we can complete the task efficiently by showing the Hadamard
differentiability of the so called Z-maps. Hence in Section E.1 we present a
functional delta method for Z-maps (Lemma \ref{lemma:differentiability of
Z-functionals}) and show how to apply it to a generic Z-problem (Lemma \ref%
{lemma: Zexample}). The results of this section are of independent interest.
In Section E.2 we present the proofs for Section E.1. In Section E.3 we
present the results on convergence of empirical measures, which take into
account dependencies across samples in the presence of transformation
samples. Finally, with all these ingredients, we prove Theorems \ref%
{theorem:qr} and \ref{theorem:dr} and their corollaries in Sections E.4 and
E.5.


\subsection{Main ingredient: functional delta method for Z-processes}

\label{appendix:delta_zprocesses}

\label{appendix:zprocess} In our leading examples, we have a functional
parameter $u\mapsto \theta (u)$ where $u\in \mathcal{U}$ and $\theta (u)\in
\Theta \subseteq \mathbb{R}^{p}$, and, for each $u\in \mathcal{U}$, the true
value $\theta _{0}(u)$ solves the $p$-vector of moment equations $\Psi
(\theta ,u)=0. $ For estimation purposes we have an empirical analog of the
above moment functions $\widehat{\Psi }(\theta ,u)$. For each $u\in \mathcal{%
U}$, the estimator $\widehat{\theta }(u)$ satisfies 
\begin{equation*}
\Vert \widehat{\Psi }(\widehat{\theta }(u),u)\Vert ^{2}\leq \inf_{\theta \in
\Theta }\Vert \widehat{\Psi }(\theta ,u)\Vert ^{2}+{\widehat r}(u)^2,
\end{equation*}%
with $\Vert {\widehat r}\Vert _{{\mathcal{U}}}=o_{\Pr }(n^{-1/2})$.
Similarly suppose that a bootstrap or simulation method is available that
produces a pair $(\widehat{\Psi }^{\ast },{\widehat r}^{\ast })$ and the
corresponding estimator $\widehat{\theta }^{\ast }(u)$ that obeys $\Vert 
\widehat{\Psi }^*(\widehat{\theta }^{\ast }(u),u)\Vert ^{2}\leq \inf_{\theta
\in \Theta }\Vert \widehat{\Psi }^{\ast }(\theta ,u)\Vert ^{2}+ {\widehat r}%
^{\ast}(u)^2,$ with $\Vert {\widehat r}^{\ast }\Vert _{{\mathcal{U}}}=o_{\Pr
}(n^{-1/2})$.

We can represent the above estimator and estimand as 
\begin{equation*}
\widehat \theta(\cdot) = \phi( \widehat \Psi(\cdot,\cdot), \widehat r
(\cdot) ) \text{ and } \theta_0(\cdot) = \phi( \Psi(\cdot,\cdot), 0 )
\end{equation*}
where $\phi$ is a \emph{Z-map} formally defined as follows. Consider a $p$%
-vector $\psi(\theta,u) $ indexed by $(\theta,u)$ as a generic value of $%
\Psi $. An element $\theta\in\Theta $ is an $r(u)$-approximate zero of the
map $\theta\mapsto\psi(\theta,u) $ if 
\begin{equation*}
\| \psi(\theta,u) \|^{2} \leq\inf_{\theta^{\prime}\in\Theta} \|
\psi(\theta^{\prime},u) \|^{2} + r(u)^{2},
\end{equation*}
where $r(u)\in\mathbb{R}$ is a numerical tolerance parameter. Let $%
(\psi(\cdot, u), r(u)) \mapsto \phi(\psi(\cdot, u), r(u))$ be a
deterministic map from $\ell^{\infty}(\Theta)^p \times \mathbb{R}$ to $%
\Theta $ that assigns one of its $r(u)$-approximate zeroes to each element $%
\psi(\cdot, u) \in\ell^{\infty }(\Theta)^p$. Further, in our case $%
\psi(\cdot, u)$'s are all indexed by $u$, and so we can think of $\psi=
(\psi(\cdot, u): u \in \mathcal{U})$ as an element of $\ell
^{\infty}(\Theta\times \mathcal{U})^{p}$, and of $r = (r (u): u \in \mathcal{%
U})$ as an element of $\ell^{\infty}(\mathcal{U})$. Then we can define the
Z-map $(\psi, r) \mapsto \phi( \psi, r) $ as a map that assigns a function $%
u \mapsto \phi(\psi(\cdot,u), r(u))$ to each element $(\psi,r) $. This map
is from the metric space $\ell ^{\infty}(\Theta\times \mathcal{U})^{p}
\times \ell^{\infty}(\mathcal{U})$ to the metric space $\ell^{\infty}(%
\mathcal{U})^p$. The properties of the Z-processes therefore rely on
Hadamard differentiability of the Z-map 
\begin{equation*}
(\psi,r) \mapsto\phi(\psi, r)
\end{equation*}
at $(\psi,r) = (\Psi,0)$, i.e. differentiability with respect to the
underlying vector of moments function and with respect to numerical
tolerance parameter $r$.

We make the following assumption about the vector of moment functions. Let $%
B_{\delta}(\theta)$ denote a closed ball of radius $\delta$ centered at $%
\theta$.

\textsc{\ Condition Z.} \textit{\ Let $\mathcal{U}$ be a compact set of some
metric space, and $\Theta $ be an arbitrary subset of $\mathbb{R}^{p}$.
Assume (i) for each $u\in \mathcal{U}$, $\Psi (\cdot ,u):\Theta \mapsto 
\mathbb{R}^{p}$ possesses a unique zero at $\theta _{0}(u)$, and, for some $%
\delta >0$, $\mathcal{N}:=\cup _{u\in \mathcal{U}}B_{\delta }(\theta
_{0}(u)) $ is a compact subset of $\mathbb{R}^p$ contained in $\Theta $,
(ii) The inverse of $\Psi (\cdot ,u)$ defined as $\Psi ^{-1}(x ,u) := \{
\theta \in \Theta: \Psi(\theta,u) =x \}$ is continuous at $x=0$ uniformly in 
$u\in \mathcal{U}$ with respect to the Hausdorff distance, (iii) there
exists $\dot{\Psi}_{\theta _{0}(u),u}$ such that $\lim_{t \searrow
0}\sup_{u\in {\mathcal{U}},\Vert h\Vert=1}|t^{-1}[\Psi(\theta_0(u)+th,u) -
\Psi(\theta_{0}(u),u)] -\dot{\Psi}_{\theta _{0}(u),u}h|=0$, where $%
\inf_{u\in \mathcal{U}}\inf_{\Vert h\Vert =1}$ $\Vert \dot{\Psi}_{\theta
_{0}(u),u}h\Vert >0$, and (iv) the maps $u \mapsto \theta_0(u)$ and $u
\mapsto \dot{\Psi}_{\theta _{0}(u),u}$ are continuous.}






The following lemma is useful for verifying Condition Z.

\begin{lemma}[Simple sufficient condition for Z]
\label{cont inverse} Suppose that $\Theta =\mathbb{R}^{p}$, and $\mathcal{U}$
is a compact interval in $\mathbb{R}$. Let $\mathcal{I}$ be an open set
containing $\mathcal{U}$. Suppose that (a) $\Psi :\Theta \times \mathcal{I}
\mapsto \mathbb{R}^{p}$ is continuous, and $\theta \mapsto \Psi (\theta ,u) $
is the gradient of a convex function in $\theta $ for each $u\in \mathcal{U}$%
, (b) for each $u\in \mathcal{U}$, $\Psi (\theta _{0}(u),u)=0$, (c) $\frac{%
\partial }{\partial (\theta ^{\prime },u)}\Psi (\theta ,u)$ exists at $%
(\theta _{0}(u),u)$ and is continuous at $(\theta _{0}(u),u)$ for each $u\in 
\mathcal{U}$, and $\dot{\Psi}_{\theta _{0}(u),u}:=\frac{\partial }{\partial
\theta ^{\prime }}\Psi (\theta ,u)|_{\theta _{0}(u)}$ obeys $\inf_{u\in 
\mathcal{U}}\inf_{\Vert h\Vert =1}\Vert \dot{\Psi}_{\theta _{0}(u),u}h\Vert
>c_{0}>0$. Then Condition Z holds and $u \mapsto \theta_0(u)$ is
continuously differentiable.
\end{lemma}

\begin{lemma}[Hadamard differentiability of approximate Z-maps]
\label{lemma:differentiability of Z-functionals} Suppose that Condition
Z(i)-(iii) holds. Then, the map $(\psi ,r)\mapsto \phi (\psi ,r)$ is
Hadamard differentiable at $(\psi ,r)=(\Psi ,0)$ as a map $\phi :\mathbb{D}%
=\ell ^{\infty }(\Theta \times \mathcal{U})^{p}\times \ell ^{\infty }(%
\mathcal{U})\mapsto \mathbb{E}=\ell ^{\infty }(\mathcal{U})^{p}$
tangentially to $\mathbb{D}_{0}:= C(\mathcal{N}\times \mathcal{U}
)^{p}\times \{0\}$, where $C(\mathcal{N}\times \mathcal{U} )^{p}$ denotes
the subset of functions in $\ell ^{\infty }(\Theta \times \mathcal{U})^{p}$
that are continuous on $\mathcal{N}\times \mathcal{U} $. The derivative map $%
(z,0)\mapsto \phi _{\Psi ,0}^{\prime }(z,0)$ is defined by 
\begin{equation*}
\phi _{\Psi ,0}^{\prime }(z,0)=-\dot{\Psi}_{\theta _{0}(\cdot ),\cdot
}^{-1}z(\theta _{0}(\cdot ),\cdot ),
\end{equation*}%
where $(z,0)\mapsto \phi _{\Psi ,0}^{\prime }(z,0)$ is defined and
continuous over $z\in \ell ^{\infty }(\Theta \times \mathcal{U})^{p}$. If in
addition Condition Z(iv) holds, then $u \mapsto -\dot{\Psi}_{\theta
_{0}(u),u }^{-1}z(\theta _{0}(u ), u )$ is continuous.
\end{lemma}

This lemma is an alternative to Lemma 3.9.34 in VW on Hadamard
differentiability of Z-maps in general normed spaces, which we found
difficult to use in our case.\footnote{%
We also found difficult to use the version of Lemma 3.9.34 of VW given in
Theorem 13.5 of Kosorok (2008, Chap. 13.3).} (The paths of quantile
regression processes $\widehat{\theta }(\cdot )$ in the non-univariate case
are somewhat irregular and it is not apparent how to place them in an
entropically simple parameter space.) Moreover, our lemma applies to
approximate Z-estimators. This allows us to cover quantile regression
processes, where exact Z-estimators do not exist for \emph{any} sample size.
The following lemma shows how to apply Lemma \ref{lemma:differentiability of
Z-functionals} to a generic Z-problem.

\begin{lemma}[Limit distribution for approximate Z-estimators]
\label{lemma: Zexample} Suppose condition Z(i)-(iii) holds. If $\sqrt{n}(%
\widehat{\Psi }-\Psi )\rightsquigarrow Z$ in $\ell ^{\infty }(\Theta \times 
\mathcal{U})^{p},$ where $Z$ is a Gaussian process with a.s. uniformly
continuous paths on $\mathcal{N}\times \mathcal{U}$, and $\|n^{1/2}\widehat{r%
}\|_{\mathcal{U}} \to_{\Pr} 0,$ then 
\begin{equation*}
\sqrt{n}(\widehat{\theta }(\cdot )-\theta _{0}(\cdot ))=-\dot{\Psi}_{\theta
_{0}(\cdot ),\cdot }^{-1}\sqrt{n}(\widehat{\Psi }-\Psi )(\theta _{0}(\cdot
),\cdot )+o_{\Pr }(1)\rightsquigarrow -\dot{\Psi}_{\theta _{0}(\cdot ),\cdot
}^{-1}\left[ Z(\theta _{0}(\cdot ),\cdot )\right] \text{ in }\ell ^{\infty }(%
\mathcal{U})^{p}.
\end{equation*}%
If Condition Z(iv) also holds, then the paths $u \mapsto -\dot{\Psi}_{\theta
_{0}(u ),u}^{-1}\left[ Z(\theta _{0}(u),u )\right]$ are uniformly
continuous, a.s. Moreover, if $\sqrt{n}(\widehat{\Psi }^{\ast }-\widehat{%
\Psi })\rightsquigarrow _{\Pr }Z$ in $\ell ^{\infty }(\Theta \times \mathcal{%
U})^{p},$ and $\|n^{1/2}\widehat{r}^*\|_{\mathcal{U}} \to_{\Pr} 0,$ then 
\begin{equation*}
\sqrt{n}(\widehat{\theta }^{\ast }(\cdot )-\widehat{\theta }(\cdot
))\rightsquigarrow_{\Pr } -\dot{\Psi}_{\theta _{0}(\cdot ),\cdot }^{-1}\left[
Z(\theta _{0}(\cdot ),\cdot )\right] \text{ in } \ell ^{\infty }(\mathcal{U}%
)^{p}.
\end{equation*}
\end{lemma}

\begin{remark}
(Central limit theorem for exchangeable bootstrap) Primitive conditions for $%
\sqrt{n}(\widehat{\Psi }^{\ast }-\widehat{\Psi })\rightsquigarrow _{\Pr }Z$
in $\ell ^{\infty }(\Theta \times \mathcal{U})^{p}$ are given in VW for the
case of exchangeable bootstrap.
\end{remark}

\subsection{Proofs of Lemma \protect\ref{cont inverse}-\protect\ref{lemma:
Zexample}}

\noindent \textbf{Proof of Lemma \ref{cont inverse}. } To show Condition
Z(i), note that for each $u\in \mathcal{U}$, $\Psi (\cdot ,u):\Theta \mapsto 
\mathbb{R}^{p}$ possesses a unique zero at $\theta _{0}(u)$ by conditions
(a) - (c). By the Implicit Function Theorem, $\partial \theta
_{0}(u)/\partial u= - \dot{\Psi}_{\theta _{0}(u),u}^{-1} \times [\partial
\Psi(\theta _{0}(u),u)/\partial u],$ which is uniformly bounded and
continuous in $u\in \mathcal{U}$ by condition (c) and compactness of $%
\mathcal{U}$. Hence $\mathcal{N}=\cup _{u\in \mathcal{U}} B_{\delta }(\theta
_{0}(u))$ is a compact subset of $\Theta $ for any $\delta >0$. This
verifies Condition Z(i) and also implies condition Z(iv) in view of
condition (c) and continuous differentiability (and hence continuity) of $u
\mapsto \theta_0(u)$. 

To show Condition Z(iii), take any sequence $(u_{t},h_{t})\rightarrow(u,h) $
with $u\in \mathcal{U},h\in \mathbb{R}^{p}$ and then note that, for some $%
t^{\ast}\in\lbrack0,t]$, $\Delta(u_{t},h_{t})=t^{-1}\{\Psi(%
\theta_0(u_{t})+th_{t},u_{t})-\Psi(\theta_0 (u_{t}),u_{t})\}=\frac{%
\partial\Psi}{\partial \theta^{\prime }}(\theta_0(u_{t})+t^{%
\ast}h_{t},u_{t})h_{t}$$\rightarrow\frac{\partial\Psi}{\partial
\theta^{\prime }}(\theta_0 (u),u)h=\dot{\Psi}_{\theta _{0}(u),u} h$ using
the continuity characterizations of the derivative $\partial\Psi/\partial
\theta$ and the continuity of $u\mapsto\theta_0(u)$ established in the first
paragraph. Hence by Lemma \ref{Lemma:Resnick}, we conclude that $\sup_{u\in {%
\mathcal{U}},\Vert h\Vert=1}|\Delta(u,h)-\dot{\Psi}_{\theta
_{0}(u),u}h|\rightarrow0 $ as $t\searrow0$.

To show Condition Z(ii), we need to verify that for any $x_t \to 0$ such
that $x_t \in \Psi(\Theta,u)$, $d_H(\Psi^{-1}(x_t,u) , \Psi^{-1}(0, u)) \to
0 $, where $d_H$ is the Hausdorff distance, uniformly in $u \in \mathcal{U}$%
. Suppose by contradiction that this is not true, then there is $(x_t, u_t)$
with $x_t \to 0$ and $u_t \in \mathcal{U}$ such that $d_H(\Psi^{-1}(x_t,u_t)
, \Psi^{-1}(0, u_t)) \not \to 0.$ By compactness of $\mathcal{U}$, we can
select a further subsequence $(x_k,u_k)$ such that $u_k \to u$, where $u \in 
\mathcal{U}$. We have that $\Psi^{-1}(0, u)=\theta_0(u)$ is continuous in $u
\in \mathcal{U}$, so we must have $d_H(\Psi^{-1}(x_k,u_k) , \Psi^{-1}(0, u))
\not \to 0.$ Hence there is a further subsequence $y_l \in
\Psi^{-1}(x_l,u_l) $ with $y_l \to y$ in $\overline{\mathbb{R}}^p$, such
that $y \neq \Psi^{-1}(0, u)=\theta_0(u)$, and such that $x_l= \Psi(y_l,u_l)
\to 0$. If $y \in \mathbb{R}^p $, by continuity $\Psi(y_{l}, u_{l}) \to
\Psi(y,u) \neq 0$ since $y \neq \Psi^{-1}(0, u)$, yielding a contradiction.
If $y \in \overline{\mathbb{R}}^p \setminus \mathbb{R}^p$, we need to show
that $\| \Psi (y_{l}, u_{l})\| \not \to 0$ to obtain a contradiction. Note
that for $h \in \mathbb{R}^p: \|h\|=1$, $u \in \mathcal{U}$, and scalar $%
\delta \in \mathbb{R}$, the map $\delta \mapsto \Psi (\theta_0(u) + \delta
h, u)^{\prime }h$ is non-decreasing by $\theta \mapsto \Psi(\theta,u)$ being
the gradient of a convex function. Since $\Psi(\theta_0(u),u) = 0$, conclude
that $|\Psi (\theta_0(u) + \delta h, u)^{\prime }h | $ is non-decreasing in $%
|\delta|$. Moreover, $\|\Psi (\theta_0(u) + \delta h, u)\| \geq |\Psi
(\theta_0(u) + \delta h, u)^{\prime }h |$ for any $(h, u, \delta)$. Hence to
get contradiction it suffices to show that $\inf_{u \in \mathcal{U},
\|h\|=1} | \Psi (\theta_0(u) + \delta h, u)^{\prime }h| >0$ for some $%
\delta>0$. Indeed, for small enough $\delta>0$, by computation similar to
that above and condition (c), this quantity is bounded below by $(1/2)
\delta \inf_{u \in \mathcal{U}} \inf_{\|h\|=1} |h^{\prime }\dot \Psi_{\theta
_{0}(u),u} h | \geq c_0 \delta/2 >0$.\qed.

\noindent \textbf{Proof of Lemma \ref{lemma:differentiability of
Z-functionals}.} Consider $\psi _{t}=\Psi +tz_{t}$ and $r_{t}=0+tq_{t}$ with 
$z_{t}\rightarrow z\text{ in }\ell ^{\infty }(\Theta \times \mathcal{U})^{p}$
where $z\in C(\mathcal{N}\times \mathcal{U})^{p}$ and $q_{t}\rightarrow 0$
in $\ell ^{\infty }(\mathcal{U})$. Then, for $\theta _{t}(\cdot)=\phi (\psi
_{t},r_{t})$ we need to prove that uniformly in $u\in \mathcal{U}$, 
\begin{equation*}
\frac{\theta _{t}(u)-\theta _{0}(u)}{t}\rightarrow \phi _{\Psi ,0}^{\prime
}(z,0)(u)= - \dot{\Psi}_{\theta _{0}(u),u}^{-1}[z(\theta _{0}(u),u)].
\end{equation*}

We have that $\Psi(\theta_{0}(u), u) =0$ for all $u \in \mathcal{U}$. By
definition, $\theta_{t}(u)$ satisfies 
\begin{equation*}
\|\Psi(\theta_{t}(u), u) - \Psi(\theta_{0}(u),u) + t z_{t}(\theta
_{t}(u),u)\|^{2} \leq\inf_{\theta\in\Theta} \| \Psi(\theta,u) + t
z_{t}(\theta,u) \|^{2} + t^{2} q^2_{t}(u) =: t^{2} \lambda^{2}_{t}(u) +
t^{2} q^{2}_{t}(u),
\end{equation*}
uniformly in $u \in \mathcal{U}$. The rest of the proof has three steps. In
Step 1, we establish a rate of convergence of $\theta_{t}(\cdot)$ to $%
\theta_0(\cdot)$. In Step 2, we verify the main claim of the lemma
concerning the linear representation for $t^{-1}(\theta_{t}(\cdot) -
\theta_{0}(\cdot))$, assuming that $\lambda_{t}(\cdot) = o(1)$. In Step 3,
we verify that $\lambda_{t}(\cdot) = o(1)$.

\textsc{Step 1.} Here we show that uniformly in $u \in \mathcal{U}$, $%
\|\theta_{t}(u)- \theta_{0}(u)\| = O(t).$ First observe that $\sup_{(\theta,
u) \in \Theta \times \mathcal{U}}\|z_t(\theta, u)\| = O(1)$ by $z_t \to z$
and $\sup_{(\theta, u) \in \Theta \times \mathcal{U}}\|z (\theta, u)\| <
\infty$. Then note that $\lambda_{t}(u) \leq\| t^{-1}\Psi(\theta_{0}(u), u)
+ z_{t}(\theta_{0}(u), u) \| = \| z(\theta_{0}(u),u) + o(1)\| = O(1)$
uniformly in $u \in \mathcal{U}$. We conclude that uniformly in $u \in 
\mathcal{U}$, as $t \searrow0$: $t^{-1} (\Psi(\theta_{t}(u),u) -
\Psi(\theta_{0}(u),u)) = -z_{t}(\theta _{t}(u),u) + O(\lambda_{t}(u) +
q_{t}(u)) = O(1)$ and $\|\Psi(\theta_{t}(u),u) - \Psi(\theta_{0}(u),u)\| =
O(t).$ By assumption $\Psi(\cdot,u)$ has a unique zero at $\theta_{0}(u)$
and has an inverse that is continuous at zero uniformly in $u \in \mathcal{U}
$; hence it follows that uniformly in $u \in \mathcal{U}$, $\|\theta_{t}(u)
- \theta_{0}(u)\| \leq d_{H}(\Psi^{-1}(\Psi(\theta_{t}(u),u),u),
\Psi^{-1}(0,u)) \to0,$ where $d_{H} $ is the Hausdorff distance. By
condition Z(iii) uniformly in $u \in \mathcal{U}$ 
\begin{align*}
\liminf_{t \searrow0} \frac{\|\Psi(\theta_{t}(u),u) - \Psi(\theta_{0}(u),u)\|%
}{\|\theta_{t}(u) - \theta_{0}(u)\|} & \geq\liminf_{t \searrow0} %
\displaystyle \frac{\|\dot\Psi_{\theta_{0}(u),u}[\theta_{t}(u) - \theta
_{0}(u)]\|}{ \| \theta_{t}(u) - \theta_{0}(u)\|} \\
& \geq\inf_{\|h\| =1 } \| \dot\Psi_{\theta_{0}(u),u} h\| = c>0,
\end{align*}
where $h$ ranges over $\mathbb{R}^{p}$, and $c>0$ by assumption. The claim
of the step follows.

\textsc{Step 2.} (Main) Here we verify the main claim of the lemma. Using
Condition Z(iii) again, conclude $\|\Psi(\theta_{t}(u),u) -
\Psi(\theta_{0}(u),u) - \dot\Psi_{\theta_{0}(u),u}[\theta_{t}(u) - \theta
_{0}(u)] \| = o(t) $ uniformly in $u \in \mathcal{U}$. Below we show that $%
\lambda_{t}(u) = o(1)$ and we also have $q_{t}(u) = o(1)$ uniformly in $u
\in \mathcal{U}$ by assumption. Thus, we can conclude that uniformly in $u
\in \mathcal{U}$, $t^{-1} (\Psi(\theta_{t}(u),u) - \Psi(\theta_{0}(u),u)) =
-z_{t}(\theta_{t}(u),u) + o(1) = - z(\theta_{0}(u),u) + o(1)$ and 
\begin{align*}
t^{-1} [\theta_{t}(u) - \theta_{0}(u)] & = \dot\Psi_{\theta_{0}(u),u}^{-1} 
\left[ t^{-1} (\Psi(\theta_{t}(u),u) - \Psi(\theta_{0}(u),u)) + o(1) \right]
\\
& = - \dot\Psi_{\theta_{0}(u),u}^{-1} \left[ z(\theta_{0}(u),u) \right] +
o(1).
\end{align*}

\textsc{Step 3.} In this step we show that $\lambda_{t}(u) = o(1)$ uniformly
in $u \in \mathcal{U}$. Note that for $\bar\theta_{t}(u) := \theta_{0}(u) -
t \dot \Psi_{\theta_{0}(u),u}^{-1} \left[ z(\theta_{0}(u),u) \right] =
\theta _{0}(u) + O(t)$, we have that $\bar\theta_{t} (u)\in \mathcal{N}
=\cup _{u\in \mathcal{U}}B_{\delta }(\theta _{0}(u)),$ for small enough $t $%
, uniformly in $u \in \mathcal{U}$; moreover, $\lambda_{t}(u) \leq\| t^{-1}
\Psi (\bar\theta_{t}(u),u) + z_{t}(\bar\theta_{t}(u),u)\|$ which is equal to 
$\| - \dot\Psi _{\theta_{0}(u),u} \{ \dot\Psi^{-1}_{\theta_{0}(u),u}
[z(\theta_{0}(u),u) ]\} + z(\theta_{0}(u),u) + o(1)\| = o(1),$ as $t
\searrow0.$ \qed

\noindent \text{ } \noindent \textbf{Proof of Lemma \ref{lemma: Zexample}.}
We shall omit the dependence on $u$, previously signified by $(\cdot ),$ in
what follows. Then, in the notation of Lemma \ref{lemma:differentiability of
Z-functionals}, $\widehat{\theta }=\phi (\widehat{\Psi },\widehat{r})$ is an
estimator of $\theta _{0}=\phi (\Psi ,0)$. By the Hadamard differentiability
of the $\phi $-map shown in Lemma \ref{lemma:differentiability of
Z-functionals}, the weak convergence conclusion follows. The first order
expansion follows by noting that the linear map $\psi \mapsto -\dot{\Psi}%
_{\theta _{0}}^{-1}\psi $ is trivially Hadamard differentiable at $\psi
=\Psi $, and so by stacking, $(-\sqrt{n}(\widehat{\theta }-\theta _{0}),\dot{%
\Psi}_{\theta _{0}}^{-1}\sqrt{n}(\widehat{\Psi }-\Psi ))\rightsquigarrow (%
\dot{\Psi}_{\theta _{0}}^{-1}Z,\dot{\Psi}_{\theta _{0}}^{-1}Z)$ in $\ell
^{\infty }(\mathcal{U})^{2p}$, and so the difference between the terms
converges in outer probability to zero. The validity of bootstrap follows
from the delta method for the bootstrap. \qed

\subsection{Limits of empirical measures}

The following result is useful to organize thoughts for the case of
transformation sampling. Let 
\begin{equation*}
\widehat{\mathbb{G}}_{k}(f) := \frac{1}{\sqrt{n_{k}}} \sum_{i=1}^{n}\left( f
(Y_{ki},X_{ki}) - \int f d P_{k}\right) \ \ \text{ and } \widehat{\mathbb{G}}%
^{*}_{k}(f) := \frac{1}{\sqrt{n_{k}}} \sum_{i=1}^{n} (w_{ki} - \bar w_{k}) f
(Y_{ki},X_{ki})
\end{equation*}
be the empirical and exchangeable bootstrap processes for the sample from
population $k$.

\begin{lemma}
\label{lemma:empirical measures} Suppose Conditions S, SM, and EB hold. Let $%
\mathcal{F}: \mathcal{X} \mapsto \mathbb{R}$ be a DKP class (as defined in
Appendix \ref{app:notation}), where $\mathcal{X} \supseteq \cup_{k \in 
\mathcal{K}} \mathcal{X}_k$. (1) Then $\widehat{\mathbb{G}}_{k}(f)
\rightsquigarrow{\mathbb{G}}_{k}(f)$ and $\widehat {\mathbb{G}}_{k}^{*}(f)
\rightsquigarrow_{\Pr} \mathbb{G}_{k}(f)$ in $\ell^{\infty}(\mathcal{K}_{0}%
\mathcal{F})$, as stochastic processes indexed by $(k,f) \in\mathcal{K}_{0}%
\mathcal{F}$. (2) Moreover, $\widehat{\mathbb{G}}_{k}(f) \rightsquigarrow{%
\mathbb{G}}_{k}(f)$ and $\widehat {\mathbb{G}}_{k}^{*}(f)
\rightsquigarrow_{\Pr} {\mathbb{G}}_{k}(f) $ in $\ell^{\infty}(\mathcal{K}%
\mathcal{F})$, as stochastic processes indexed by $(k,f) \in\mathcal{K}%
\mathcal{F} $, where ${\mathbb{G}}_{k}(f) = {\mathbb{G}}_{l(k)}(f \circ
g_{l(k),k})$, provided that $\mathcal{F} \circ g_{l(k),k}$ continues to be a
DKP class for all $k \in \mathcal{K}_t$.
\end{lemma}

\noindent \textbf{Proof of Lemma \ref{lemma:empirical measures}.} Note that $%
\mathcal{F}$ is a universal Donsker class by Dudley (1987). Statement (1)
then follows from the independence of samples across $k\in\mathcal{K}_{0}$,
so that joint convergence follows from the marginal convergence for each $%
k\in\mathcal{K}_{0}$, and from the results on exchangeable bootstrap given
in Chapter 3.6 of VW. Let $\mathcal{F}$ be a DKP class. To show Statement
(2) we note that $\widehat{{\mathbb{G}}}_{k}(f)=\widehat{{\mathbb{G}}}%
_{m}(f\circ g_{m,k})$ for $m=l(k) \in \mathcal{K}_0$. Recall that $l(\cdot)$
denotes the indexing function that indicates the population $l(k)$ from
which the $k$-th population is created by transformation, in particular $%
l(k) = k$ if $k \in \mathcal{K}_0$. Thus, $l^{-1}(m)= \{ k \in \mathcal{K}:
l(k) = m\}$ is the set of all populations created from the $m$-th population
that includes $m$ itself. Let $\mathcal{F}^{\prime }$ consist of $\mathcal{F}
$ and $\mathcal{F}\circ g_{m,k}$ for all $k\in l^{-1}(m) =\{m,...\} \subset%
\mathcal{K}$ and all $m \in \mathcal{K}_0$. Then $\mathcal{F}^{\prime }$ is
a DKP class, since it is a finite union of DKP classes ($\mathcal{K}$ is
finite), so statement (2) follows from statement (1). In fact, this shows
that the convergence analysis is reducible to the independent case by
suitably enriching $\mathcal{F}$ into the class $\mathcal{F}^{\prime }$. \qed%
.

\subsection{Proof of Theorem \protect\ref{theorem:qr}}

(Validity of QR based Counterfactual Analysis) The proof of Lemma \ref%
{lemma:empirical measures} shows that by suitably enlarging the class $%
\mathcal{F}$, it suffices to consider only the independent samples, i.e.
those with population indices $k \in\mathcal{K}_{0}$. Moreover, by
independence across $k$, the joint convergence result follows from the
marginal convergence for each $k$ separately. It remains to examine each
case with $k \in\mathcal{J} $ separately, since otherwise for a given $k
\not \in \mathcal{J}$, the convergence of empirical measures and associated
bootstrap result are already shown in Lemma \ref{lemma:empirical measures}.
In what follows, since the proof can be done for each $k$ marginally, we
shall omit the index $k$ to simplify the notation.

\textsc{Step 1.}(Results for coefficients and empirical measures). Let $%
\mathcal{F}$ be a DKP class, as defined in Appendix \ref{app:notation}. We
use the Z-process framework described in Appendix \ref{appendix:zprocess},
where we let $\theta (u)=\beta _{{}}(u)$, $p = d_x$, and $\Theta =\mathbb{R}%
^{d_{x}}$. Lemma \ref{lemma: Zexample} above illustrates the use of the
delta method for a single Z-estimation problem, which the reader may find
helpful before reading this proof. Let $\varphi _{u,\beta
}(Y_{{}},X_{{}})=(1\{Y_{{}}\leq X_{{}}^{\prime }\beta \} - u)X_{{}},$ $\Psi
_{{}}(\beta ,u)=P_{{}}[\varphi _{u,\beta }],$ and $\widehat{\Psi }%
_{{}}(\beta ,u)=P_{n}[\varphi _{u,\beta }] $, where $P_{n}$ is the empirical
measure and $P_{{}}$ is the corresponding probability measure. From the
subgradient characterization, we know that the QR estimator obeys $\widehat{%
\beta }(u)=\phi (\widehat{\Psi }_{{}}(\cdot ,u),\widehat{r}(u)),\ \widehat{r}%
(u)=\max_{1\leq i\leq n_{{}}}\Vert X_{i}\Vert d_{x}/n_{{}},$ for each $u\in {%
\mathcal{U}}$, with $n_{{}}^{1/2}\Vert \widehat{r}\Vert _{\mathcal{U}%
}\rightarrow _{\Pr }0$, where $\phi $ is an approximate Z-map as defined in
Appendix \ref{appendix:zprocess}. The random vector $\widehat{\beta }(u)$
and $\int fd\widehat{F}_{X_{{}}}=P_{n}(f) $ are estimators of $\beta
(u)=\phi (\Psi _{{}}(\cdot ,u),0)$ and $\int fdF_{X_{{}}}=P(f).$ Then, by
Step 3 below 
\begin{equation*}
(\sqrt{n}(\widehat{\Psi }_{{}}-\Psi _{{}}),\widehat{\mathbb{G}}%
_{{}})\rightsquigarrow (W_{{}} , {\mathbb{G}}_{{}}) \text{ in } \ell
^{\infty }(\mathbb{R}^{d_{x}}\times \mathcal{U})^{d_{x}}\times \ell ^{\infty
}(\mathcal{F}),\ W_{{}}(\beta ,u)={\mathbb{G}}_{{}}\varphi _{u,\beta },
\end{equation*}%
where $W_{{}}$ has continuous paths a.s. Step 4 verifies the conditions of
Lemma E.1 for $\dot \Psi_{\theta_{0}(u),u} = J(u)$, thereby also implying
continuous differentiability of $u \mapsto \beta(u)$. Then, by Lemma \ref%
{lemma:differentiability of Z-functionals}, the map $\phi $ is
Hadamard-differentiable with derivative map $(w,0) \mapsto -J_{{}}^{-1}w$ at 
$(\Psi _{{}},0)$ (tangentially to $C(\mathcal{N}\times \mathcal{U}
)^{d_x}\times \{0\}$, where $\mathcal{N} = \cup_{u \in \mathcal{U}}
B_{\delta}(\beta(u))$). Therefore, we can conclude by the functional delta
method that $(\sqrt{n_{{}}}(\widehat{\beta }_{{}}(\cdot )-\beta _{{}}(\cdot
)),\widehat{\mathbb{G}}_{{}})\rightsquigarrow (-J_{{}}^{-1}(\cdot
)W_{{}}(\beta _{{}}(\cdot ),\cdot ),{\mathbb{G}}_{{}})$ in $\ell ^{\infty }({%
\mathcal{U}})^{d_{x}}\times \ell ^{\infty }(\mathcal{F}) $. The process $%
-J_{{}}^{-1}(\cdot )W_{{}}(\beta _{{}}(\cdot ),\cdot )$ has continuous paths
a.s. 

Similarly, for the bootstrap version, we have from the subgradient
characterization of the QR estimator that $\widehat\beta^{*} (u) =
\phi(\widehat\Psi^*_{}(\cdot,u) , \widehat{r}^{*}(u)), \ \ \widehat{r}%
^{*}(u) = \max_{1 \leq i \leq n} z_i d_x/n,$ where $z_i = w_i \|X_{i}\|$.
Moreover, ${n_{}}^{1/2}\widehat{r}^{*}_{n_{}} \to_{\Pr} 0 $, since for some $%
p = 2 + \epsilon$ with $\epsilon>0$, 
\begin{equation*}
n^{-1/2}E [\max_{1 \leq i \leq n} z_i] \leq n^{-1/2+1/p} E[( n^{-1}
\sum_{i=1}^n z_i^p )^{1/p}] \leq n^{-1/2+1/p} [ n^{-1} \sum_{i=1}^n E[z_i^p]
]^{1/p} = o(1),
\end{equation*}
where $\max_i E[z_i^p] = \max_i E \|w_{i } X_{i }\|^{p} = \max_i E |w_{i
}|^{p} E \| X_{i }\|^{p}$ is bounded uniformly in $n$; the latter holds by
the moment assumptions in Conditions QR and EB as well as independence of
the bootstrap weights $(w_i)_{i=1}^n$ from the data $(X_i)_{i=1}^n $. By
Step 3 below and Theorem 3.6.13 of VW, $(\sqrt{n}(\widehat\Psi^{*}_{} -
\widehat \Psi_{}), \widehat{\mathbb{G}}^{*}_{}) \rightsquigarrow_{\Pr}
(W_{}, {\mathbb{G}}_{})$ in $\ell^{\infty}(\mathbb{R}^{d_x}\times\mathcal{U}%
)^{d_x} \times \ell^{\infty}(\mathcal{F})$. Therefore by the functional
delta method for bootstrap $(\sqrt{n_{}} ( \widehat\beta^{*}_{}(\cdot) -
\widehat\beta_{}(\cdot)), \widehat{{\mathbb{G}}}^{*}_{})
\rightsquigarrow_{\Pr} ( -J_{}^{-1}(\cdot) W_{} ( \beta_{}(\cdot), \cdot), {%
\mathbb{G}}_{})$ in $\ell^{\infty}({\mathcal{U}})^{d_x}\times\ell^{\infty }(%
\mathcal{F})$.

\textsc{Step 2.}(Main: Results for conditional cdfs). Here we shall rely on
compactness of $\mathcal{Y}_{}\mathcal{X}_{}$. In order to verify Condition
D, we first note that $\mathcal{F}_0 = \{F_{Y_{}|X_{}}(y|\cdot): y \in%
\mathcal{Y}_{}\}$ is a uniformly bounded ``parametric" family indexed by $y
\in\mathcal{Y}_{}$ that obeys $|F_{Y_{}|X_{}}(y|\cdot) -
F_{Y_{}|X_{}}(y^{\prime}|\cdot)| \leq L| y - y^{\prime}|$, given the
assumption that the density function $f_{Y_{}|X_{}}$ is uniformly bounded by
some constant $L$. Given compactness of $\mathcal{Y}_{}$, the uniform $%
\epsilon$-covering numbers for this class can be bounded independently of $%
F_{X_{}}$ by $\text{const}/\epsilon$, and so the Pollard's entropy integral
is finite. Hence we can construct a class of functions $\mathcal{F}$
containing the union of all the families $\mathcal{F}_0$ for the populations
in $\mathcal{J} $ and the indicators of all the rectangles in $\overline{%
\mathbb{R}}^{d_x}$. Note that these indicators form a VC class. The final
set $\mathcal{F}$ therefore is a DKP class.


Next consider the mapping $\nu :\mathbb{D}_{\nu}\subset\ell^{\infty }({%
\mathcal{U}})^{d_x}\mapsto\ell^{\infty}(\mathcal{Y}_{{}}\mathcal{X}_{{}})$,
defined as $b\mapsto\nu(b),\ \ \nu(b)(y,x)=\varepsilon+
\int_{\varepsilon}^{1-\varepsilon}1\{x^{\prime}b(u)\leq y\}du.$ It follows
from the results of Chernozhukov, Fernandez-Val, and Galichon (2010) that
this map is Hadamard differentiable at $b(\cdot)=\beta_{{}}(\cdot)$
tangentially to $C({\mathcal{U}})^{d_x},$ with the derivative map given by: $%
\alpha\mapsto\nu_{\beta_{{}}(\cdot)}^{\prime}(\alpha),\ \ \nu
_{\beta_{{}}(\cdot)}^{\prime}(\alpha)(y,x)= -
f_{Y_{{}}|X_{{}}}(y|x)x^{\prime }\alpha(F_{Y_{{}}|X_{{}}}(y|x)). $ Since $%
\widehat{F}_{Y_{{}}|X_{{}}}=\nu(\widehat{\beta}_{{}}(\cdot))$ and $\int fd%
\widehat{F}_{X_{{}}}=\int fdP_{n}$ are estimators of $F_{Y_{{}}|X_{{}}}=\nu(%
\beta_{{}}(\cdot))$ and $\int fdF_{X_{{}}}=\int fdP_{{}},$ by the delta
method 
\begin{align}
& (\sqrt{n}(\widehat{F}_{Y_{{}}|X_{{}}}-F_{Y_{{}}|X_{{}}}),\widehat{\mathbb{G%
}}_{{}})\rightsquigarrow\ \
(-\nu_{\beta_{{}}(\cdot)}^{\prime}J_{{}}^{-1}(\cdot)W_{{}}(
\beta_{{}}(\cdot), \cdot),{\mathbb{G}}_{{}}) \text{ in } \ell^{\infty}(%
\mathcal{Y}_{{}}\mathcal{X}_{{}})\times\ell^{\infty }(\mathcal{F}_{{}}), \\
& (\sqrt{n}(\widehat{F}_{Y_{{}}|X_{{}}}^{\ast}-\widehat{F}_{Y_{{}}|X_{{}}}),%
\widehat{{\mathbb{G}}}_{{}}^{\ast})\rightsquigarrow_{\Pr}(-\nu_{\beta_{{}}(%
\cdot)}^{\prime}J_{{}}^{-1}(\cdot)W_{{}}(\beta_{{}}(\cdot ), \cdot),{\mathbb{%
G}}_{{}}) \text{ in } \ell^{\infty}(\mathcal{Y}_{{}}\mathcal{X}%
_{{}})\times\ell^{\infty }(\mathcal{F}_{{}}).
\end{align}

\textsc{Step 3.} (Auxiliary: Donskerness). First, we note that $\{\varphi
_{u,\beta}(Y_{{}},X_{{}}):(u,\beta)\in\mathcal{U}\times\mathbb{R}^{d_x}\}$
is $P_{{}} $-Donsker. This follows by a standard argument, which is omitted.
Second, we note that $(u,\beta)\mapsto\varphi_{u,\beta}(Y_{{}},X_{{}})$ is $%
L^{2}(P_{{}})$ continuous by the dominated convergence theorem, and the fact
that $(\beta,u)\mapsto(1(Y\leq X^{\prime}\beta) - u)X_{{}}$ is continuous at
each $(\beta,u)\in\mathbb{R}^{d_x}\times {\mathcal{U}}$ with probability one
by the absolute continuity of $F_{Y_{{}}|X_{{}}}$, and its norm is bounded
by a square integrable function $2\Vert X_{{}}\Vert$ under $P_{{}}$. Hence ${%
\mathbb{G}}_{{}}(\varphi_{u,\beta})$ has continuous paths in $(u,\beta)$ and
the convergence results follow from the convergence results in Lemma \ref%
{lemma:empirical measures}.

\textsc{Step 4. }(Auxiliary: Verification of Conditions of Lemma E.1).We
verify conditions (a)-(c) of Lemma \ref{cont inverse}. Conditions (a) and
(b) are immediate by the assumptions. To verify (c), we can compute $\frac {%
\partial}{\partial(b^{\prime}, u)} \Psi(b,u)$ $=$ $[ E[f_{Y|X}(X^{\prime}b|
X) XX^{\prime}]$,$-E X]$ for $(b, u)$ in the neighborhood of $(\beta(u), u)$%
, where the right side is continuous at $(b, u) = (\beta(u), u)$ for each $u
\in \mathcal{U}$. This computation and continuity follows from using the
dominated convergence theorem, the a.s. continuity and uniform boundedness
of the mapping $y \mapsto f_{Y|X}(y| X)$, as well as $E\|X\|^{2} < \infty$.
By assumption, the minimum eigenvalue of $J(u) =
E[f_{Y|X}(X^{\prime}\beta(u)| X) XX^{\prime}] $ is bounded away from zero
uniformly in $u \in{\mathcal{U}}$. \qed

\subsection{Proof of Corollaries \protect\ref{corQR} and \protect\ref{corQR2}%
}

Corollary \ref{corQR} is derived in Step 1 of the proof of Theorem \ref%
{theorem:qr}, where the first-order expansion of the conclusion (1) follows
by an argument similar to the proof of Lemma \ref{lemma: Zexample}. The
results for the estimators of the conditional quantile function in Corollary %
\ref{corQR2} follow from Corollary \ref{corQR} by the functional delta
method, and the Hadamard differentiability of the rearrangement operator
(uniformly with respect to an index) derived in Chernozhukov, Fernandez-Val,
and Galichon (2010). The results for the estimator of the conditional
distribution function in Corollary \ref{corQR2} are derived in Step 2 of the
proof of Theorem \ref{theorem:qr}. \qed

\subsection{Proof of Theorem \protect\ref{theorem:dr}}

(Validity of DR based Counterfactual Analysis). As in the proof of Theorem %
\ref{theorem:qr}, it suffices to show the result for each $k \in\mathcal{J} $
separately. In what follows, since the proof can be done for each $k$
marginally, we shall omit the index $k$ to simplify the notation. We only
consider the case where $\mathcal{Y}$ is a compact interval of $\mathbb{R}$.
The case where $\mathcal{Y}$ is finite is simpler and follows similarly.

\textsc{Step 1.}(Results for coefficients and empirical measures). We use
the Z-process framework described in Appendix \ref{appendix:zprocess}, where
we let $u=y,\theta(u)=\beta_{{}}(y)$, $p = d_x,$ $\Theta=\mathbb{R}^{d_x},$
and $\mathcal{U}=\mathcal{Y}$. Lemma \ref{lemma: Zexample} illustrates the
use of the delta method for a single Z-estimation problem, which the reader
may find helpful before reading this proof. Let 
\begin{equation*}
\varphi_{y,\beta}(Y_{{}},X_{{}})=[\Lambda(X^{\prime}\beta)-1(Y\leq
y)]H(X^{\prime}\beta)X,
\end{equation*}
where $H(z)=\lambda(z)/\{\Lambda (z)[1-\Lambda(z)]\}$ and $\lambda$ is the
derivative of $\Lambda$. 
Let $\Psi_{{}}(\theta,y)=P_{{}}[\varphi_{y,\beta}]$ and $\widehat{\Psi}%
_{{}}(\theta,y)=P_{n}[\varphi_{y,\beta}]$, where $P_{n}$ is the empirical
measure and $P_{{}}$ is the corresponding probability measure. From the
first order conditions, the DR estimator obeys $\widehat{\beta}(y)=\phi(%
\widehat{\Psi}_{{}}(\cdot,y),0),$ for each $y\in\mathcal{Y}$, where $\phi$
is the Z-map defined in Appendix \ref{appendix:zprocess}. The random vector $%
\widehat{\beta}(y)$ and $\int fd\widehat {F}_{X_{{}}}=P_{n}(f)$ are
estimators of $\beta(y)=\phi(\Psi_{{}}(\cdot,y),0)$ and $\int
fdF_{X_{{}}}=P(f).$ Then, by Step 3 below 
\begin{equation*}
(\sqrt{n}(\widehat{\Psi}_{{}}-\Psi_{{}}),\widehat{{\mathbb{G}}}%
_{{}})\rightsquigarrow(W_{{}},{\mathbb{G}}_{{}}) \text{ in } \ell^{\infty}(%
\mathbb{R}^{d_x}\times\mathcal{Y})^{d_x}\times\ell^{\infty}(\mathcal{F}),\
W_{{}}(y,\beta)={\mathbb{G}}_{{}}\varphi_{y,\beta},
\end{equation*}
where $W_{{}}$ has continuous paths a.s. Step 4 verifies the conditions of
Lemma E.1 for $\dot \Psi_{\theta_{0}(u),u} = J(y)$, which also implies that $%
y \mapsto \beta(y)$ is continuously differentiable on the interval $\mathcal{%
Y}$. Then, by Lemma \ref{lemma:differentiability of Z-functionals}, the map $%
\phi$ is Hadamard-differentiable with the derivative map $( w, 0)
\mapsto-J_{{}}^{-1}w$ at $(\Psi_{{}},0)$ (tangentially to $C(\mathcal{N}%
\times \mathcal{U} )^{d_x}\times \{0\}$, where $\mathcal{N} = \cup_{y \in 
\mathcal{Y}} B_{\delta}(\beta(y))$). Therefore, we can conclude by the
Functional Delta Method that 
\begin{equation*}
(\sqrt{n_{{}}}(\widehat{\beta}_{{}}(\cdot)-\beta_{{}}(\cdot)),\widehat{{%
\mathbb{G}}}_{{}})\rightsquigarrow(-J_{{}}^{-1}(\cdot)W_{{}}(\beta_{{}}(%
\cdot), \cdot),{\mathbb{G}}_{{}}) \text{ in } \ell^{\infty}({\mathcal{Y}}%
)^{d_x}\times\ell^{\infty}(\mathcal{F}),
\end{equation*}
where $-J_{{}}^{-1}(\cdot)W_{{}}(\beta_{{}}(\cdot), \cdot)$ has continuous
paths a.s.

Similarly, for the bootstrap version, we have from the first order
conditions of the DR estimator that $\widehat{\beta}^{\ast}(y)=\phi(\widehat{%
\Psi}^*_{{}}(\cdot,y),0),$ and $(\sqrt{n}(\widehat{\Psi}_{{}}^{\ast}-%
\widehat \Psi_{{}}),\widehat {{\mathbb{G}}}_{{}}^{\ast})\rightsquigarrow_{%
\Pr}(W_{{}},{\mathbb{G}}_{{}})$ in $\ell^{\infty}(\mathbb{R}^{d_x}\times%
\mathcal{Y})^{d_x}\times\ell^{\infty}(\mathcal{F})$ by Step 3 below and
Theorem 3.6.13 of VW. Therefore by the Functional Delta method for Bootstrap 
\begin{equation*}
(\sqrt{n_{{}}}(\widehat{\beta}_{{}}^{\ast}(\cdot)-\widehat{\beta}_{{}}(\cdot
)),\widehat{{\mathbb{G}}}_{{}}^{\ast})\rightsquigarrow_{\Pr}(-J_{{}}^{-1}(%
\cdot )W_{{}}(\cdot,\beta_{{}}(\cdot)),{\mathbb{G}}_{{}}) \text{ in }
\ell^{\infty }({\mathcal{Y}})^{d_x}\times\ell^{\infty}(\mathcal{F}).
\end{equation*}

\textsc{Step 2.}(Main: Results for conditional cdfs). Here we shall rely on
compactness of $\mathcal{Y}_{}\mathcal{X}_{}$. Then, $\mathcal{Y}$ is a
closed interval of $\mathbb{R}$. In order to verify Condition D, we first
note that $\mathcal{F}_0 = \{F_{Y_{}|X_{}}(y|\cdot): y \in\mathcal{Y}_{}\}$
is a uniformly bounded ``parametric" family indexed by $y \in\mathcal{Y}_{}$
that obeys $|F_{Y_{}|X_{}}(y|\cdot) - F_{Y_{}|X_{}}(y^{\prime}|\cdot)| \leq
L| y - y^{\prime}|$, given the assumption that the density function $%
f_{Y_{}|X_{}}$ is uniformly bounded by some constant $L$. Given compactness
of $\mathcal{Y}_{}$, the uniform $\epsilon$-covering numbers for this class
can be bounded independently of $F_{X_{}}$ by $\text{const}/\epsilon$, and
so the Pollard's entropy integral is finite. Hence we can construct a class
of functions $\mathcal{F}$ containing the union of all the families $%
\mathcal{F}_0$ for the populations in $\mathcal{J}$ and the indicators of
all rectangles in $\overline{\mathbb{R}}^{d_x}$. Note that these indicators
form a VC class. The final set $\mathcal{F}$ therefore is a DKP class.


Next consider the mapping $\nu:\mathbb{D}_{\nu}\subset\ell^{\infty }({%
\mathcal{Y}})^{d_x}\mapsto\ell^{\infty}(\mathcal{Y}_{{}}\mathcal{X}_{{}})$,
defined as $b\mapsto\nu(b),\ \ \nu(b)(x,y)=\Lambda(x^{\prime}b(y)).$ It is
straightforward to deduce that this map is Hadamard differentiable at $%
b(\cdot)=\beta_{{}}(\cdot)$ tangentially to $C({\mathcal{Y}})^{d_x}$ with
the derivative map given by: $\alpha\mapsto\nu_{\beta_{{}}(\cdot)}^{\prime
}(\alpha),\ \ \nu_{\beta_{{}}(\cdot)}^{\prime}(\alpha)(y,x)=\lambda
(x^{\prime}\beta(y))x^{\prime}\alpha(y).$ Since $\widehat{F}%
_{Y_{{}}|X_{{}}}=\nu(\widehat{\beta}_{{}}(\cdot))$ and $\int fd\widehat{F}%
_{X_{{}}}=\int fdP_{n}$ are estimators of $F_{Y_{{}}|X_{{}}}=\nu(\beta_{{}}(%
\cdot))$ and $\int fdF_{X_{{}}}=\int fdP_{{}},$ by the functional delta
method 
\begin{align}
& (\sqrt{n}(\widehat{F}_{Y_{{}}|X_{{}}}-F_{Y_{{}}|X_{{}}}),\widehat{\mathbb{G%
}}_{{}})\rightsquigarrow\ \
(-\nu_{\beta_{{}}(\cdot)}^{\prime}J_{{}}^{-1}(\cdot)W_{{}}(\beta_{{}}(%
\cdot), \cdot),{\mathbb{G}}_{{}}) \text{ in } \ell^{\infty}(\mathcal{Y}_{{}}%
\mathcal{X}_{{}})\times\ell^{\infty }(\mathcal{F}_{{}}), \\
& (\sqrt{n}(\widehat{F}_{Y_{{}}|X_{{}}}^{\ast}-\widehat{F}_{Y_{{}}|X_{{}}}),%
\widehat{{\mathbb{G}}}_{{}}^{\ast})\rightsquigarrow_{\Pr}(-\nu_{\beta_{{}}(%
\cdot)}^{\prime}J_{{}}^{-1}(\cdot)W_{{}}(\beta_{{}}(\cdot ), \cdot),{\mathbb{%
G}}_{{}}) \text{ in } \ell^{\infty}(\mathcal{Y}_{{}}\mathcal{X}%
_{{}})\times\ell^{\infty }(\mathcal{F}_{{}}).
\end{align}

\textsc{Step 3.} (Auxiliary: Donskerness). We verify that $%
\{\varphi_{y,\beta }(Y_{{}},X_{{}}):(y,\beta)\in\mathcal{Y}\times\mathbb{R}%
^{d_x}\}$ is $P_{{}}$-Donsker with a square integrable envelope. The
function classes $\mathcal{F}_{1}=\{X^{\prime}\beta:\beta\in\mathbb{R}%
^{d_x}\}$, $\mathcal{F}_{2}=\{1(Y\leq y):y\in\mathcal{Y}\}$, and $\{X_{q} :
q=1,...,d_x\},$ where $q$ indexes elements of vector $X$, are VC classes of
functions. The final class $\mathcal{G} = \{(\Lambda(\mathcal{F}_{1})-%
\mathcal{F}_{2})H(\mathcal{F}_{1})X_q : q=1,...,d_x\}$ is a Lipschitz
transformation of VC classes with Lipschitz coefficient bounded by $\text{%
const} \|X\|$ and envelope function $\text{const} \|X\|$, which is
square-integrable. Hence $\mathcal{G}$ is Donsker by Example 19.9 in van der
Vaart (1998). Finally, the map $(\beta,y)\mapsto(\Lambda(X^{\prime}\beta)-1%
\{Y\leq y\})H(X^{\prime}\beta)X$ is continuous at each $(\beta ,y)\in\mathbb{%
R}^{d_x}\times\mathcal{Y}$ with probability one by the absolute continuity
of the conditional distribution of $Y$ (when $\mathcal{Y}$ is not finite).

\textsc{Step 4. }(Auxiliary: Verification of Conditions of Lemma E.1). We
verify conditions (a)-(c) of Lemma \ref{cont inverse}. Conditions (a) and
(b) are immediate by the assumptions. To verify (c), a straightforward
computation gives that for $(b, y)$ in the neighborhood of $(\beta(y), y)$, $%
\frac{\partial}{\partial(b^{\prime}, y)} \Psi(b, y) = [J(b, y), R(b, y)], $
where, for $H(z) = \lambda(z)/\{\Lambda(z)[1 - \Lambda(z)]\}$ and $h(z) =
dH(z)/dz$, 
\begin{equation*}
J(b, y) := E \left[ \{h(X^{\prime}b) [\Lambda(X^{\prime}b) - 1 (Y \leq y) ]
+ H(X^{\prime}b) \lambda(X^{\prime}b) \} XX^{\prime}\right] ,
\end{equation*}
and $R(b, y) = - E \left[ H(X^{\prime}b) f_{Y|X}(y|X)X \right] .$ Both terms
are continuous in $(b, y)$ at $(\beta(y), y)$ for each $y \in\mathcal{Y}$.
The computation above as well as the verification of continuity follows from
using the dominated convergence theorem, and the following ingredients: (1)
a.s. continuity of the map $(b, y) \mapsto\frac{\partial}{\partial b^{\prime}%
} \varphi_{b, y}(Y,X)$, (2) domination of $\| \frac{\partial}{\partial
b^{\prime}} \varphi_{b,y}(X,Y)\|$ by a square-integrable function $\text{%
const} \|X\|$, (3) a.s. continuity and uniform boundedness of the
conditional density function $y \mapsto f_{Y|X}(y|X)$, and (4) $H(X^{\prime
}b)$ being bounded uniformly on $b \in\mathbb{R}^{d_x}$, a.s. By assumption $%
J(y) = J(\beta(y), y) $ is positive-definite uniformly in $y \in{\mathcal{Y}}
$. \qed

\subsection{Proof of Corollaries \protect\ref{corDR} and \protect\ref{corDR2}%
}

Corollary \ref{corDR} is derived in Step 1 of the proof of Theorem \ref%
{theorem:dr}, where the first-order expansion of the conclusion (1) follows
by an argument similar to the proof of Lemma \ref{lemma: Zexample}. The
results for $\widehat F_{Y_j|X_j}$ and its bootstrap version $\widehat
F^*_{Y_j|X_j}$ in Corollary \ref{corDR2} are derived in Step 2 of the proof
of Theorem \ref{theorem:dr}. The rest of the results in Corollary \ref%
{corDR2} follow from the functional delta method, and the Hadamard
differentiability of the rearrangement and inverse operators (uniformly with
respect to an index) derived in Chernozhukov, Fernandez-Val, and Galichon
(2010). \qed

\linespread{1}


\begin{figure}

\begin{center}

\centering\epsfig{figure=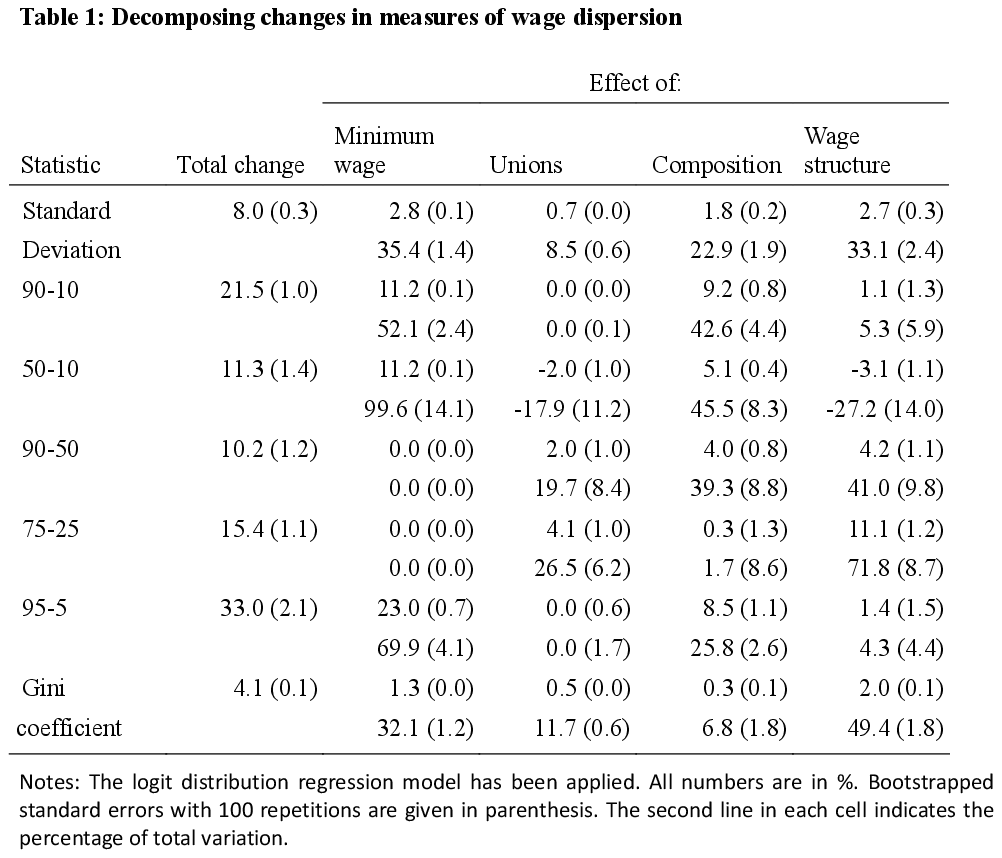,width=1.1\textwidth,height=1.1\textheight}

\end{center}

\end{figure}



\begin{figure}

\begin{center}

\centering\epsfig{figure=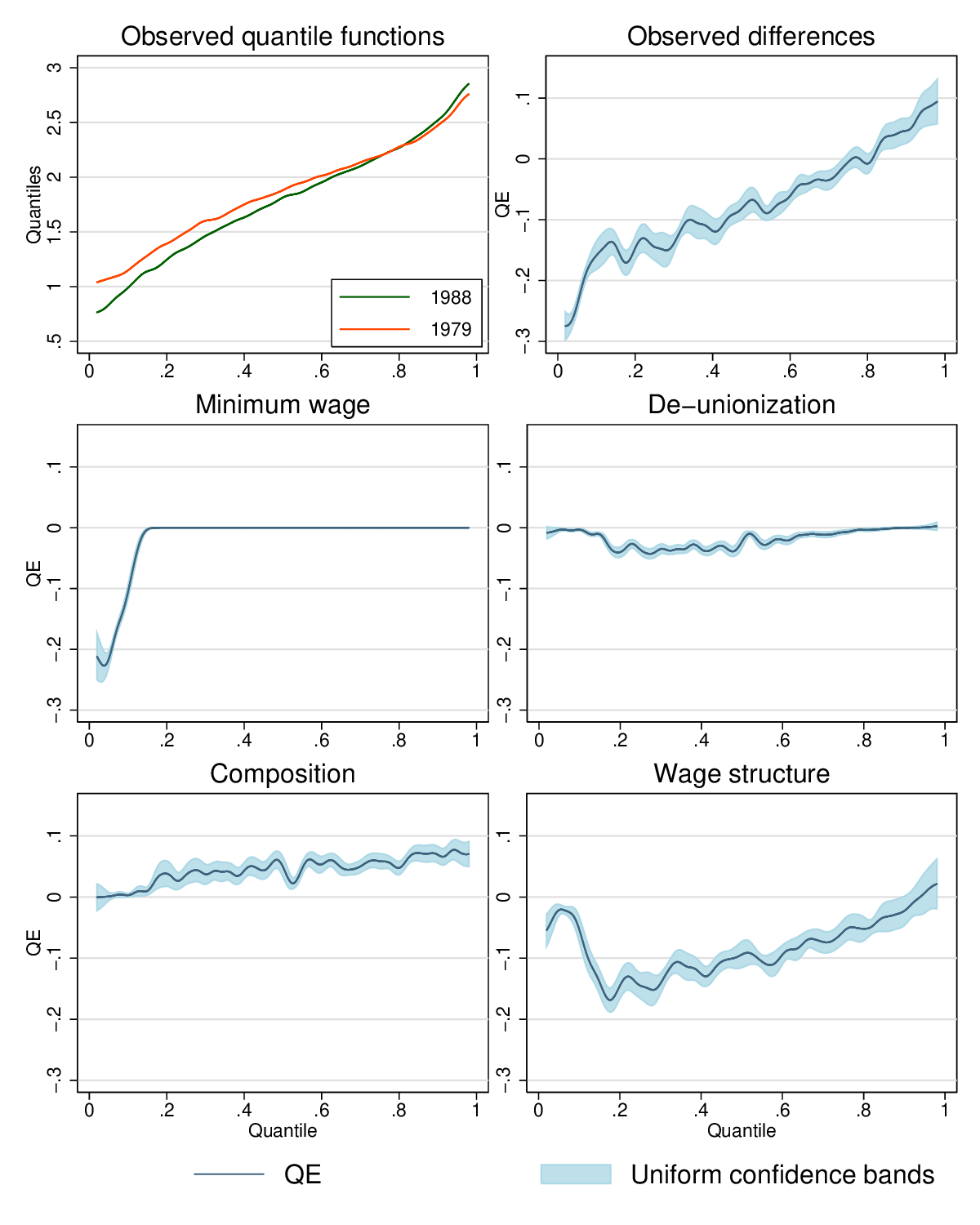,width=.9\textwidth,height=.9\textheight}

\caption{Observed quantile functions, observed differences between the quantile functions and
their decomposition into four quantile effects. The 95\% simultaneous confidence bands were
obtained by empirical bootstrap with 100 repetitions.}
\end{center}

\end{figure}



\begin{figure}

\begin{center}

\centering\epsfig{figure=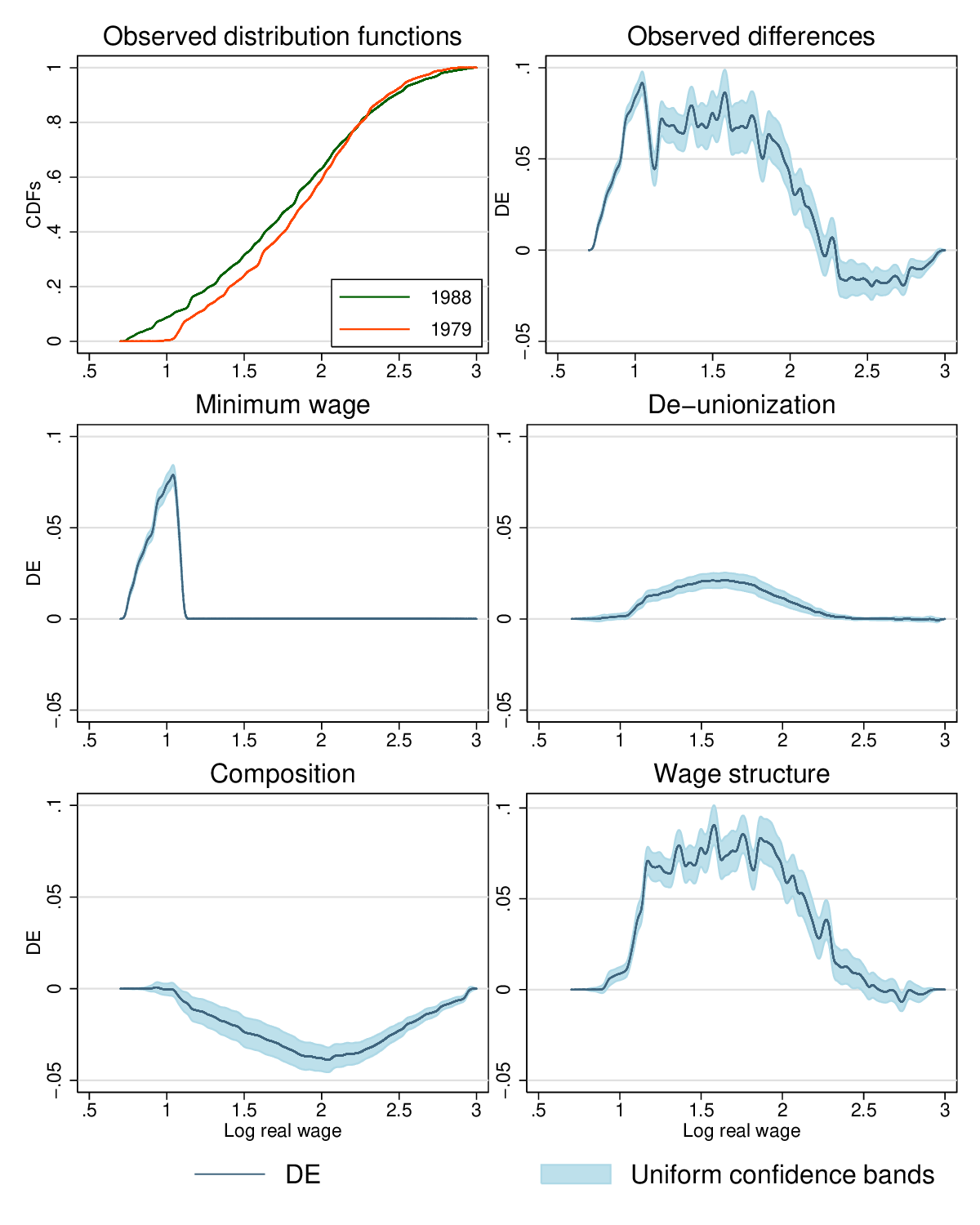,width=.9\textwidth,height=.9\textheight}

\caption{Observed distribution functions, observed differences between the distribution functions
and their decomposition into four distribution effects. The 95\% simultaneous confidence bands
were obtained by empirical bootstrap with 100 repetitions.}

\end{center}

\end{figure}



\begin{figure}

\begin{center}

\centering\epsfig{figure=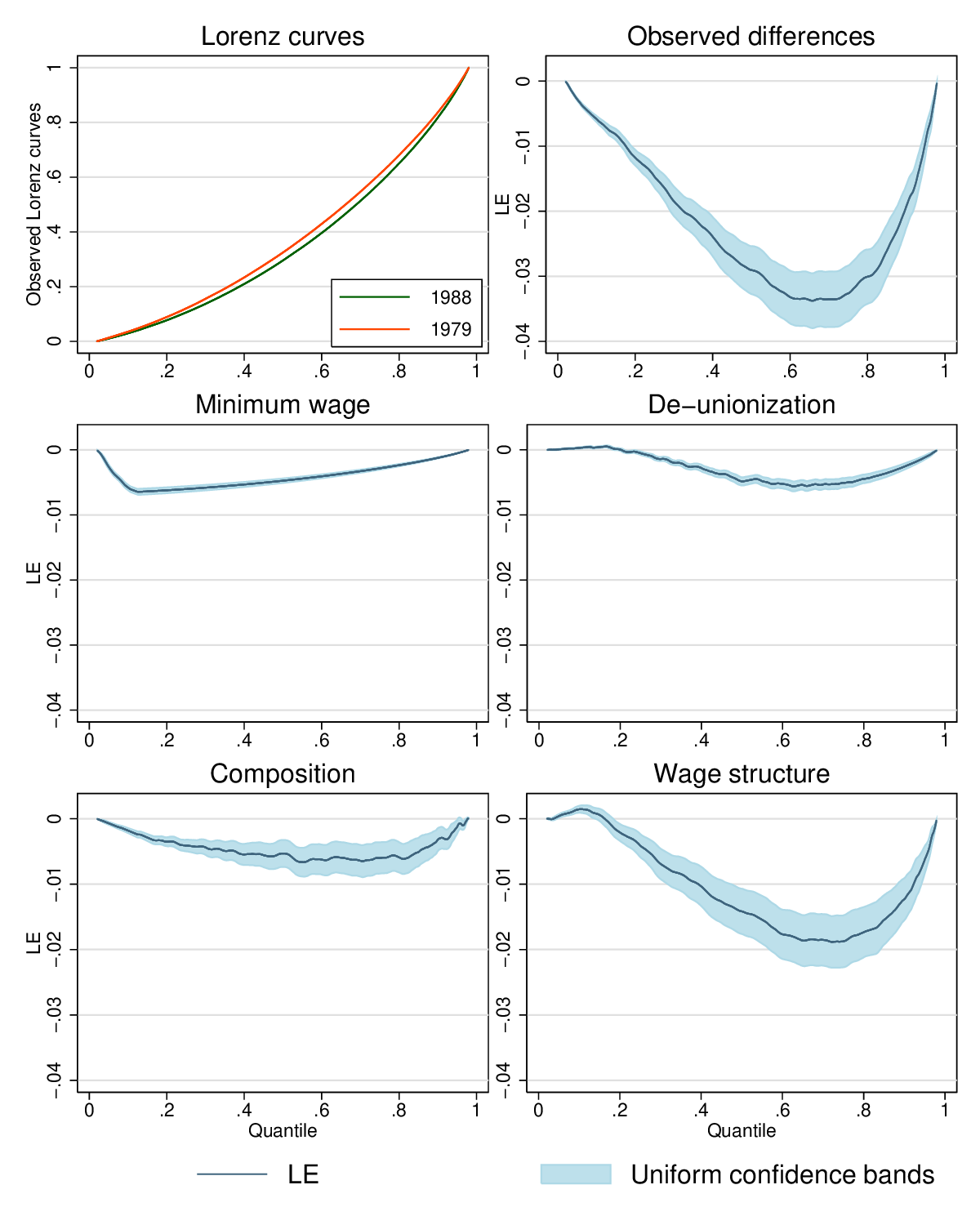, width=.9\textwidth,height=.9\textheight}

\caption{Observed Lorenz curves, observed differences between the Lorenz curves and their
decomposition into four Lorenz effects. The 95\% simultaneous confidence bands were obtained
by empirical bootstrap with 100 repetitions.}

\end{center}

\end{figure}


\end{document}